\documentclass[12pt]{article}
\pdfoutput=1

\usepackage{graphicx,psfrag,epsf,xcolor}
\usepackage{amsmath,amssymb,amsfonts}
\usepackage{array}
\usepackage{cite}
\usepackage{rotating}
\usepackage{slashed, cancel}
\usepackage{scalerel,stackengine}
\usepackage{float}
\usepackage{subcaption}
\usepackage{caption}
\usepackage{mathtools}
\definecolor{cBlue}{RGB}{0,110,191}
\usepackage[pdftitle={Threshold factorization of the Drell-Yan  
quark-gluon channel and two-loop soft function  
at next-to-leading power},hypertexnames=true]{hyperref}
\hypersetup{bookmarksnumbered,colorlinks,
    linkcolor={black},
    citecolor={cBlue},
    urlcolor={cBlue}}
\numberwithin{equation}{section}

\newcommand\Eqn[1]     {Eq.\,(\ref{#1})}
\newcommand\Eqns[2]    {Eqs.\,(\ref{#1}) and~(\ref{#2})}
\newcommand\Eqnss[2]   {Eqs.\,(\ref{#1})--(\ref{#2})}
\newcommand\eqn[1]     {eq.\,(\ref{#1})}

\newcommand{\be}{\begin{equation}}
\newcommand{\ee}{\end{equation}}
\newcommand{\bea}{\begin{eqnarray}}
\newcommand{\eea}{\end{eqnarray}}
\newcommand{\bi}{\begin{itemize}}
\newcommand{\ei}{\end{itemize}}
\newcommand{\ben}{\begin{enumerate}}
\newcommand{\een}{\end{enumerate}}
\newcommand{\bt}{\begin{tabular}}
\newcommand{\et}{\end{tabular}}

\newcommand{\nn}{\nonumber}

\newcommand{\T}{{\bf T}}
\newcommand{\eps}{\epsilon}

\newcommand{\als}{\alpha_s}

\newcommand{\ord}{{\cal O}}
\newcommand{\Lo}{{\cal L}}

\newcommand\sA{\ThisStyle{\ensurestackMath{%
			{%
\stackinset{r}{}{c}{}{\SavedStyle/}{\SavedStyle\mathcal{A}}}}}}

\newcounter{SJQ}

\begin{document}
\allowdisplaybreaks

\begin{titlepage}
		
\begin{flushright}
{\small
UWThPh 2023-11\\
IPPP/23/19\\
\today
}
\end{flushright}
		
\vskip0.7cm
\begin{center}
{\Large \bf \boldmath
Threshold factorization of the Drell-Yan \\
quark-gluon channel and two-loop soft function \\[0.18cm]
at next-to-leading power}
\end{center}

\vspace{0.5cm}
\begin{center}
{\sc Alessandro Broggio},$^{a}$
{\sc Sebastian Jaskiewicz},$^{b}$ 
and {\sc Leonardo Vernazza}$^{c}$\\[6mm]

{\it $^a$ Faculty of Physics, University of Vienna,\\
Boltzmanngasse 5, A-1090 Vienna, Austria \\[0.2cm]
}
{\it $^b$ Institute for Particle Physics Phenomenology, 
Durham University, \\
Durham DH1 3LE, United Kingdom  \\[0.2cm]
}
{\it $^c$ INFN, Sezione di Torino, 
Via P. Giuria 1, I-10125 Torino, Italy \\
and Dipartimento di Fisica Teorica, 
Universit\`a di Torino \\
}
\end{center}
		
\vspace{0.4cm}
\begin{abstract}
\vskip0.2cm\noindent
We present a factorization theorem of the 
partonic Drell-Yan off-diagonal processes 
$g\bar{q}\,(qg) \to \gamma^* + X$ in the 
kinematic threshold regime $z=Q^2/\hat{s} \to 1$   
at general subleading powers in the $(1-z)$ 
expansion. Focusing on the first order of the 
expansion (next-to-leading power accuracy with 
respect to the leading power $q \bar{q}$ channel), 
we validate the bare factorization formula up to 
$\mathcal{O}(\alpha^2_s)$. This is achieved by 
carrying out an explicit calculation of the 
generalized soft function in $d$-dimensions 
using the reduction to master integrals and 
the differential equations method.
The collinear function is a universal object which
we compute from an operator matching equation at 
one-loop level. 
Next, we integrate the soft and collinear functions over 
the convolution variables and remove the remaining initial 
state collinear singularities through PDF renormalization.
The resulting expression agrees with the known cross 
section in the literature.

\end{abstract}
\end{titlepage}

\section{Introduction}
\label{sec:introduction}

It is well documented that the perturbative 
expansion of QCD fails near the kinematic 
threshold, as the phase space for real 
emission is restricted to contain only 
low-energy (soft) radiation.
Considering the Drell-Yan (DY) process 
$A(p_A)+B(p_B) \to \gamma^*(Q) [\to\ell\bar{\ell}\,] +X$,
with $X$ being the unobserved QCD final 
state, the threshold regime of the partonic 
cross section is characterised by the 
limit $z \equiv Q^2/\hat s\to 1$, where 
$Q^2$ represents the invariant mass of the 
final state lepton pair and $\hat s$ the 
partonic centre-of-mass energy squared.
In this region, physical observables are 
expressed as a power expansion in $(1-z)\to 0$
and feature large logarithmic corrections in this variable.
Reliable results can only be obtained 
by resumming these logarithms to all 
orders in perturbation theory.
This was first achieved for the leading-power (LP) 
contribution more than thirty years ago using 
diagrammatic techniques in \cite{Sterman:1986aj,Catani:1989ne} 
and later using soft-collinear effective theory (SCET) methods 
in \cite{Idilbi:2005ky,Idilbi:2006dg,Becher:2006nr}.

Nowadays the resummation of LP threshold logarithms 
is understood in DY up to next-to-next-to-next-to-leading 
logarithmic (N$^3$LL) accuracy \cite{Moch:2005ky,Becher:2007ty,Catani:2014uta,Ajjath:2020rci}. 
However, comparatively less is known about terms which 
are suppressed by a power of $(1-z)$, 
the so-called \emph{next-to-leading power} 
(NLP) logarithms, despite the fact that studies 
of amplitudes with next-to-soft emissions have quite a long history \cite{Low:1958sn,Burnett:1967km,DelDuca:1990gz}. 
Focusing on the diagonal ($q\bar{q}$) channel 
of the Drell-Yan process, calculations of 
partonic cross sections at NLP up to 
next-to-next-to-leading order (NNLO) in the 
strong coupling expansion, and partly beyond, 
have been carried out using the 
expansion-by-regions method 
\cite{Bonocore:2014wua,Bahjat-Abbas:2018hpv} 
and diagrammatic factorization techniques 
\cite{Laenen:2008ux,Laenen:2008gt,Laenen:2010uz,Bonocore:2015esa,Bonocore:2016awd}.
Investigations of the NLP terms for the DY 
process within the effective field theory (EFT)
framework were initiated in \cite{Beneke:2018gvs} 
and \cite{Beneke:2019oqx}.
In the former, the NLP logarithms were 
resummed to leading logarithmic (LL) 
accuracy, and in the latter, the complete 
subleading power factorization theorem 
was derived. In the present work, we 
complete these investigations by 
deriving and validating up to NNLO 
the NLP factorization theorem for the 
off-diagonal $g\bar{q}\, (qg)$ channel of 
the DY production process at threshold, 
which constitutes the last missing piece up to NLP accuracy.
 
The development of this framework is timely 
as plenty of attention has recently been
given to NLP studies of the analogous 
off-diagonal channels in deep-inelastic 
scattering (DIS) at large Bjorken-$x$ 
and ``gluon thrust'' in hadronic $e^+e^-$ 
annihilation, their corresponding diagonal 
channels, and Higgs production in gluon 
fusion at threshold \cite{Bonocore:2016awd,Moult:2018jjd,Beneke:2018gvs,Moult:2019mog,Bahjat-Abbas:2019fqa,Beneke:2019mua,Moult:2019uhz,Beneke:2019oqx,Moult:2019vou,Ajjath:2020ulr,Beneke:2020ibj,Ajjath:2020sjk,vanBeekveld:2021mxn,Beneke:2022obx}.
Progress beyond LP has also been achieved 
for variables such as N-jettiness \cite{Boughezal:2016zws,Moult:2016fqy,Moult:2017jsg,Ebert:2018lzn,Boughezal:2018mvf,Boughezal:2019ggi} and the $q_T$ 
of the lepton pair or the Higgs boson \cite{Ebert:2018gsn,Cieri:2019tfv,Oleari:2020wvt}.
However, the extension of the standard 
LP factorization theorems to NLP has not 
proved straightforward. The main stumbling 
block being the ubiquitous appearance of 
endpoint divergences in the convolution 
integrals that connect the hard, (anti-) 
collinear and soft functions in NLP 
factorization theorems \cite{Beneke:2019oqx}. 
This conceptual issue has been investigated 
in a number of contexts: for instance, 
conjecture regarding the form of the 
leading double logarithms related to 
soft quark emission has been presented in 
\cite{Moult:2019uhz}, and refactorization 
ideas were developed for DIS \cite{Beneke:2020ibj} 
and for Higgs decay to two photons (and 
gluons) through bottom-quark loops 
\cite{Liu:2019oav,Liu:2020tzd,Liu:2020wbn,Liu:2022ajh}. 
Combination of standard SCET factorization 
and endpoint factorization was put forward 
in \cite{Beneke:2022obx} to arrive at a 
factorization formula for the ``gluon thrust'' 
valid in $d=4$, such that standard 
systematically improvable Renormalization 
Group (RG) methods can be applied to perform 
the resummation of large logarithms. Endpoint 
divergences have also recently been explored 
in QED \cite{Bell:2022ott} and
$B$ physics \cite{Feldmann:2022ixt,Cornella:2022ubo,Hurth:2023paz}.

Concerning the study of the all-order resummation for the $g\bar{q}$ channel of DY, results at LL accuracy have recently been obtained using diagrammatic techniques in\cite{vanBeekveld:2021mxn} and $d$-dimensional consistency relations in\cite{Beneke:2020ibjB}. 
In general, the development of a resummation framework at higher orders will require a genuine prescription for the treatment of the endpoint convolution divergences in the presence of hadronic initial states. In this respect, it is crucial to derive a consistent factorization theorem valid in $d=4$. As a first step toward this goal, however, one needs to develop a consistent factorization theorem at the level of the bare cross section, and calculate all of the contributing functions to the relevant perturbative order in dimensional regularization. While this task has been completed for the DY 
$q\bar q$ channel in \cite{Beneke:2019oqx,Broggio:2021fnr}, the $g\bar{q}$ channel of DY has not been systematically address yet. The purpose of this paper is to fill this gap:
in particular, starting from time-ordered power suppressed operators 
in SCET, we derive the factorization of the partonic cross section in terms of short-distance coefficients times the convolution of collinear and soft matrix elements, valid at any subleading power. 
We then focus on the NLP contribution, and evaluate the collinear and soft functions appearing in the factorization theorem respectively at $\mathcal{O}(\alpha_s)$ and $\mathcal{O}(\alpha^2_s)$, which is the required accuracy to reproduce the known inclusive DY cross section result at NNLO. This provides a strong sanity check of our framework and allows us to reach the same
accuracy as for the diagonal $q\bar q$ channel in \cite{Beneke:2019oqx,Broggio:2021fnr}.

The collinear function has a purely virtual origin and we evaluate it at one-loop order. Its appearance is the consequence of the insertion of the power suppressed interaction which couples a soft quark with collinear gluons and a collinear quark. This function is practically equivalent (with just some minor differences in its definition) to the radiative collinear function computed to $\mathcal{O}(\alpha^2_s)$ in \cite{Liu:2021mac}.
The calculation of the collinear function which appears in the factorization formula for the $g \bar{q}$ channel turns out to be much simpler in comparison to the evaluation of the collinear functions which appear in \cite{Beneke:2019oqx} since the derivative terms acting on the momentum conserving delta function are not present in this case.
The soft function depends on the total energy of the soft emissions as well as two additional convolution variables.
We evaluate the soft contributions at NNLO accuracy in exact $d$-dimensions by employing the reduction to master integrals and the differential equations methods that were also employed in \cite{Broggio:2021fnr}. Other soft functions which appear in the $h\to \gamma\gamma$ decay NLP factorization formula were computed in \cite{Liu:2020eqe,Bodwin:2021cpx} at $\mathcal{O}(\alpha_s)$.

Working with $d$-dimensional expressions allows us to carry out the convolution integrals at fixed-order accuracy.
This is the first step needed for the application of the refactorization ideas developed in \cite{Beneke:2022obx,Liu:2020wbn}. 
We also study asymptotic limits of the obtained functions which is useful for any such program.
However, additional insights concerning the operatorial structure of the incoming hadronic states are required to carry out this program, which we leave to a future investigation.

The paper is organised as follows: in section~\ref{sec:NLPfactorization} we derive 
the general subleading power factorization 
formula for the partonic off-diagonal channels 
of the DY process and we specialize it to NLP. 
Then, in sections~\ref{sec:CollinearFunctions} 
and~\ref{sec:SoftFunctions} we calculate the 
required NLP collinear and generalized soft 
functions up to $\mathcal{O}(\alpha_s)$ and 
$\mathcal{O}(\alpha_s^2)$, respectively.  As a last 
step in section~\ref{sec:SoftFunctions}, we consider asymptotic limits of the soft function. 
The obtained accuracy of the soft and collinear functions enables the validation of the 
factorization formula derived in 
section~\ref{sec:NLPfactorization} to 
NNLO, which is carried out in 
section~\ref{sec:Validation}. 
We summarize and discuss possible future 
developments in section~\ref{sec:summary}.

\section{Factorization near threshold}
\label{sec:NLPfactorization}

We consider the Drell-Yan invariant 
mass distribution 
\begin{equation}
\frac{d\sigma_{\rm DY}}{dQ^2} = 
\frac{4\pi\alpha_{\rm em}^2}{3 N_c Q^4}
\sum_{a,b} \int_{\tau}^1 dz \int dx_a dx_b \, 
\delta\bigg(z- \frac{\tau}{x_a x_b}\bigg)
f_{a/A}(x_a)f_{b/B}(x_b)\,\hat{\sigma}_{ab}(z) 
+ \mathcal{O}\left(\frac{\Lambda}{Q}\right),
\label{eq:dsigsq0}
\end{equation}
where $f_{a/A}(x_a)$ and $f_{b/B}(x_b)$ 
are the usual parton distribution functions 
(PDFs) of partons $a$, $b$, with momentum 
fractions $x_a$, $x_b$; $\hat{\sigma}_{ab}(z)$
represents the partonic cross section for 
the process $a(p_a) b(p_b) \to \gamma^*(q) [\to\ell\bar{\ell}\,]+X$; 
$\tau = Q^2/s$ and $z = Q^2/\hat s$ are respectively 
the hadronic and partonic threshold variables, with 
$s = (p_A + p_B)^2$ and $\hat s = (p_a + p_b)^2 
= x_a x_b \, s$ the hadronic and partonic centre 
of mass energy; $Q^2 \equiv q^2$ is the invariant 
mass of the final state photon; last, $\Lambda$ 
is the confinement scale of QCD. In the present 
analysis, we do not consider power corrections 
in $\Lambda/Q$.  

Near threshold the partonic cross 
section has the following power 
expansion
\be\label{xspowerexpansion}
\hat{\sigma}_{ab}(z) = 
\hat{\sigma}^{\,{\rm{LP}}}_{ab}(z)\,
+ {\hat{\sigma}^{\,{\rm{NLP}}}_{ab}(z)} 
+ \mathcal{O}(\lambda^2),
\ee
where the power-counting parameter is 
defined as $\lambda = \sqrt{1-z} \ll 1$; 
the leading power term in the equation 
above is ${\cal O}(\lambda^{-2})$ and the 
next-to-leading power term is 
${\cal O}(\lambda^{0})$. Only the diagonal 
production channel $ab = q\bar q$ contributes
at LP. At NLP more channels open up: in 
addition to the NLP correction to the 
$q\bar q$ channel, which has been studied 
at length in \cite{Beneke:2019oqx,Broggio:2021fnr}, 
there are also contributions from the 
gluon-antiquark $g\bar q$ and the 
quark-gluon $qg$ channels, which are the 
topic of the present work. 
In this section we derive their
factorization structure near threshold, 
valid at general subleading powers. 
Given that the two channels $g\bar q$ 
and $qg$ give identical contributions, 
in what follows we focus for simplicity 
on the gluon-antiquark $g\bar q$ channel.

We derive the factorization theorem for the 
bare cross section in $d = 4 - 2\eps$ dimensions: 
in this case, dimensional regularization 
regulates the endpoint divergences arising 
in convolution integrals, hence calculations 
at fixed orders are well defined.
In order to facilitate comparison with literature, we 
will consider an equivalent form of \Eqn{eq:dsigsq0},
given by
\begin{equation}
\frac{d\sigma_{\rm DY}}{dQ^2} = \sigma_0 
\sum_{a,b} \int_{\tau}^1\, \frac{dz}{z}\, 
{\cal L}_{ab}\bigg(\frac{\tau}{z}\bigg)\, \Delta_{ab}(z) 
+ \mathcal{O}\left(\frac{\Lambda}{Q}\right),
\qquad 
\sigma_0 = 
\frac{4\pi \alpha_{\rm em}^2}{3 N_c Q^2 s},
\label{eq:dsigsqDelta}
\end{equation}
where the parton luminosity function ${\cal L}_{ab}(y)$ 
is defined as 
\be
{\cal L}_{ab}(y) = \int_y^1\frac{dx}{x} \, f_{a/A}(x) \, 
f_{b/B}\bigg(\frac{y}{x}\bigg),
\ee
and $\Delta_{ab}(z)$ is related to the partonic cross
section $\hat{\sigma}_{ab}(z)$ in \Eqn{eq:dsigsq0} by 
\begin{eqnarray} \label{DeltaDef}
\Delta_{ab}(z) = \frac{1}{(1-\epsilon)} 
\frac{\hat{\sigma}_{ab}(z)}{z}\,. 
\end{eqnarray} 
In order to obtain the factorization theorem
for $\Delta_{g\bar q}(z)$ let us start from 
the standard expression for the hadronic 
$d$-dimensional differential cross section.
After integration over the phase-space 
for the final state leptons one has
\begin{equation}\label{eq:comb_lept_2bc}
 d\sigma = \frac{4\pi\alpha_{{\rm{em}}}^2}{3s\,q^2}
 \frac{d^{d}q}{(2\pi)^d} \,\big(-g^{\mu\rho}
 \, W_{\mu\rho}\big)\,,
\end{equation}
where $W_{\mu\rho}$ is the hadronic tensor 
\begin{eqnarray}\label{eq:ampTrans}
 g^{\mu\rho} W_{\mu\rho} &=& 
 \int d^dx \,e^{-iq\cdot x}\,
 \langle A(p_A)B(p_B)|J^{\dagger\,\rho}(x)
 J_{\rho}(0)|A(p_A)B(p_B) \rangle \nonumber \\ 
 &=& \sum_X \, (2 \pi)^d 
 \delta^{(d)}\left(p_A +p_B-q-p_{X_s}-p_{X_c^{\rm PDF}}-p_{X_{\bar c}^{\rm PDF}}\right)
 \nonumber \\[-1ex]
 &&\hspace{0cm} \times\,\langle A(p_A)B(p_B)|
 J^{\dagger}_{\rho}(0)| X\rangle 
 \langle X|J^{\rho}(0)|A(p_A)B(p_B)\rangle \,,
\end{eqnarray}
and in turn $J^{\rho}=\sum_q e_q \bar{\psi}_q 
\gamma^{\rho} \psi_q$ represents the 
electromagnetic quark current. 
In \Eqn{eq:ampTrans}, $
p_{X_s}+p_{X_c^{\rm PDF}}+p_{X_{\bar c}^{\rm PDF}} 
= p_X$ is the total momentum of the final state 
radiation $X$. In order to not obscure the 
derivation of the factorization theorem, in 
what follows we work with a single quark 
flavour and set the electromagnetic 
charge $e_q=1$.

The partonic kinematic threshold corresponds 
to the case where almost all of the energy in 
the interaction between the two incoming partons 
is carried away by the intermediate off-shell 
boson,~$\gamma^*$. 
The relevant modes consist thus of PDF-collinear and 
PDF-anticollinear modes, threshold-collinear and 
threshold-anticollinear modes, and soft modes.
Momenta are decomposed along two light-like 
directions $n_{\pm}^{\mu}$, defined by 
$n_-^{\mu} = 2 p_A^{\mu}/\sqrt{s}$,
$n_+^{\mu} = 2 p_B^{\mu}/\sqrt{s}$, such that 
a generic momentum $l^{\mu}$ has decomposition 
$l^{\mu} = n_+l \, n_-^{\mu}/2 
+ n_-l \, n_+^{\mu}/2 + l^{\mu}_{\perp}$.
The scaling of a given momentum $l^{\mu}$ 
is thus indicated by the scaling of the 
components $(n_+ l, l_{\perp}, n_-l)$. 
The PDF-collinear and anticollinear modes 
scale respectively as $p_{X_{c}^{\rm PDF}}^{\mu} 
\sim (Q,\Lambda, \Lambda^2/Q)$ and
$p_{X_{\bar c}^{\rm PDF}}^{\mu} \sim 
(\Lambda^2/Q,\Lambda,Q)$, where 
$\Lambda \ll Q\lambda = Q(1-z)^{1/2}$ 
represents the confinement scale of QCD, 
while threshold collinear, anticollinear, 
and soft modes scales respectively as 
$p_{c}~\sim~Q(1,\lambda,\lambda^2)$,
$p_{\bar c}~\sim~Q(\lambda^2,\lambda,1)$
and $p_{s}~\sim~Q(\lambda^2,\lambda^2,\lambda^2)$.
We work in position-space SCET
\cite{Beneke:2002ph,Beneke:2002ni} and describe 
each mode by its own set of fields. We have therefore 
PDF- and threshold-collinear quarks and gluons, 
and soft quarks and gluons. These fields are 
expressed in terms of gauge-invariant building 
blocks, after the soft-collinear decoupling 
transformation has been applied. For instance, 
we express collinear fields modes in terms of 
the gauge-invariant building blocks 
\bea\label{coll-fields-def} 
\chi^{(0)}_c(z) &=& Y_+^{\dagger}(z_-) \chi_c(z), 
\qquad
\mathcal{A}^{(0) \mu}_c(z) =
Y_+^\dagger(z_-)\mathcal{A}^{\mu}_c(z)Y_+(z_-),
\eea
where $\chi_c = W_c^\dagger \xi_c$ 
and $\mathcal{A}_c^\mu = 
W_c^{\dagger}\left[i D_c^\mu\,W_c\right]$ 
represent the original non-decoupled 
collinear fields. In turn, the 
covariant derivative is defined as 
$i D_c^\mu = i \partial^\mu 
+ g_s  A^\mu_c$. In what follows we 
will always work with decoupled fields, 
and drop the index~$(0)$. In 
\Eqn{coll-fields-def} the soft Wilson 
lines are defined as 
\bea
Y_{\pm}\left(x\right)&=&\mathbf{P}
\exp\left[ig_s\int_{-\infty}^{0}ds\,n_{\mp}
A_{s}\left(x+sn_{\mp}\right)\right],
\eea
while the collinear Wilson lines reads 
\bea 
W_{c}\left(x\right)&=&\mathbf{P}
\exp\left[ig_s\int_{-\infty}^{0}ds\,n_+
A_{c}\left(x+sn_+\right)\right].
\eea
In both of the above equations $\mathbf{P}$
is the path ordering operator. Notice also 
that in \eqn{coll-fields-def} the soft Wilson
lines are evaluated at $z_-^{\mu} \equiv 
(n_+ z) n_-^{\mu}/2$, in accordance with 
the fact that soft fields are multipole 
expanded when multiplying collinear fields
\cite{Beneke:2002ni}. Concerning soft fields, 
we write them in terms of the 
gauge invariant building blocks 
\be\label{eq:softFields}
{q}^{\pm} = Y_{\pm}^{\dagger}\, q_s \,.
\qquad 
\mathcal{B}_{\pm}^{\mu} = Y_{\pm}^{\dagger}
\left[ i\,D^{\mu}_s\,Y_{\pm}\right] \,,
\ee
where the soft covariant derivative 
is $i D_s^\mu = i \partial^\mu + g_s A^\mu_s$.
The power-suppressed soft-collinear 
Lagrangian $\mathcal{L}^{(i)} = \mathcal{L}^{(i)}_{\xi}
+ \mathcal{L}^{(i)}_{\xi q} +\mathcal{L}^{(i)}_{\rm YM}$ 
is then expressed in terms of these 
fields, as listed in appendix A of 
\cite{Beneke:2019oqx}. We refer to 
\cite{Beneke:2002ph,Beneke:2002ni}
for a thorough derivation of the
SCET Lagrangian and SCET fields 
definition.  

\begin{figure}[t]
\begin{center}
\includegraphics[width=0.45\textwidth]{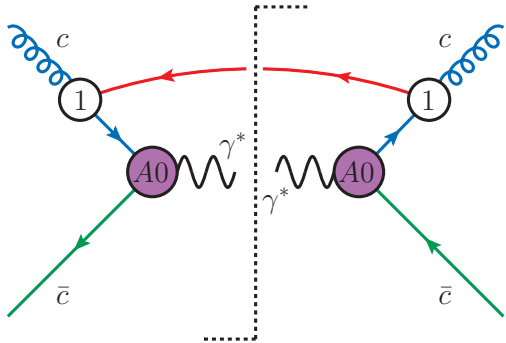} 
\end{center}
\caption{Schematic representation of the lowest-power 
contribution to the gluon-antiquark channel in Drell-Yan. 
It involves the conversion of a (PDF) collinear gluon 
into a (threshold) collinear quark by emission of a 
soft anti-quark. The emission proceeds first at NLP 
by means of a time-ordered product, involving the 
leading power SCET current $J^{A0}$ times the insertions 
of the power suppressed Lagrangian ${\cal L}_{\xi q}^{(1)}$, 
indicated by the index ``1'' in the picture.
The matching of threshold modes onto PDF modes 
is trivial for the anti-collinear sector, as there 
is no power suppression on this leg \cite{Beneke:2019oqx}.}
\label{fig:XSnondecv}
\end{figure}

Given that the final state is forced to 
only contain soft radiation, the hard 
matching to SCET fields can be performed 
at amplitude level, since there are no 
contributions to the hadronic tensor where 
the currents at positions $0$ and $x$ are 
connected by hard partons. For the process 
$g\bar{q}\,\rightarrow\,\gamma^*+X$ to 
take place in the $z\to 1$ limit, the 
incoming $c$-PDF gluon must be converted 
to a threshold collinear quark through 
the emission of a soft antiquark. The 
threshold-collinear quark retains almost 
all of the momentum of the incoming 
$c$-PDF gluon. The soft-quark interaction 
with collinear fields is inherently a 
subleading power effect. This can be 
deduced from the fact that soft quarks 
at leading power only appear in the soft 
term, $\bar{q}_{s}i\slashed{D}_sq_s$, 
in the SCET Lagrangian; the soft quarks 
first appear in soft-collinear interaction 
terms at $\mathcal{O}(\lambda)$ in 
$\mathcal{L}^{(1)}_{\xi q}$, 
\be 
\mathcal{L}^{(1)}_{\xi q} = 
\bar{q}^{\,+}\sA_{c\perp}\chi_c+ {\rm h.c.}.
\ee
Therefore, unlike in the case of the 
$q\bar{q}$-channel discussed in 
\cite{Beneke:2019oqx}, the contribution 
to the Drell-Yan cross-section from the 
$g\bar{q}$-channel begins at NLP (i.e., 
at ${\cal O}(\lambda^2)$), through a 
time-ordered product insertion of 
$\mathcal{L}^{(1)}_{\xi q}$ into 
the amplitude and its complex 
conjugate, as represented in 
Fig. \ref{fig:XSnondecv}.

\subsection{Factorization at general subleading powers}
\label{sec:genSLPfact}

The formal factorization formula for the 
$g\bar{q}$-channel at general subleading 
can be obtained following closely 
the derivation for the $q\bar{q}$-channel 
given in \cite{Beneke:2019oqx}, therefore 
we provide here a rather concise discussion 
and draw attention to the differences between 
the two cases. Our discussion below follows
the derivation given in \cite{Jaskiewicz:2021cfw}.
In general, one needs to take
into account the contribution of time-ordered 
products involving the (multiple) insertions 
of $\mathcal{L}^{(i)} = \mathcal{L}^{(i)}_{\xi}
+ \mathcal{L}^{(i)}_{\xi q} +\mathcal{L}^{(i)}_{\rm YM}$
Lagrangian terms and operators arising from the 
expansion of the Drell-Yan current, with the 
constraint that the time-ordered products 
must involve a collinear gluon and at least 
one soft quark emission. Omitting the index 
structure for clarity, the general, all power, 
hard matching of the vector current is given by 
\bea \label{eq:subleadingmatching}
\bar{\psi} \gamma_\rho \psi(0)
= \sum_{m_1, {m_2} }\, 
\int \{dt_k\} \, 
\{d\bar{t}_{\bar{k}}\}  \,\widetilde{C}^{\,m_1,m_2}
\left(\,\{t_k\} , \,\{\bar{t}_{\bar{k}}\}\right)\, J_s(0)
\,J_\rho^{\,m_1,m_2}\left(\,\{t_k\} , \,\{\bar{t}_{\bar{k}}\}
\right)\,,\quad
\eea
where the symbols $\{dt_k\}$, $\{t_k\}$ 
(and the barred ones for the anticollinear direction)
indicate sets of convolution variables associated to
the number of fields in each collinear direction. 
The indices $m_1$ and $m_2$ label the basis of SCET 
operators (and their corresponding short-distance 
matching coefficients $\widetilde{C}^{\,m_1m_2} $),
following the formalism and notation developed
in \cite{Beneke:2017ztn,Beneke:2018rbh,Beneke:2019kgv}\footnote{
See e.g.~\cite{Marcantonini:2008qn,Kolodrubetz:2016uim,Feige:2017zci,Moult:2017rpl}
for the construction of power suppressed operator 
bases in the label formulation of SCET.}.
The function $J_s$ involves only soft fields
and starts at $\mathcal{O}(\lambda^3)$ 
\cite{Beneke:2017ztn}. Given that the DY 
process involves a collinear and an anticollinear 
direction, the currents $J_\rho^{\,m_1,m_2}$ 
explicitly read
\bea \label{eq:DYcurrent}
J_\rho^{\,m_1,m_2}\left(\,\{t_k\} , \,\{\bar{t}_{\bar{k}}\}
\right) =  J^{\,m_1}_{\bar{c}}\left(\{\bar{t}_{\bar{k}}\}
\right)\,\Gamma_{\rho}^{m_1,m_2}\,J^{\,m_2}_c\left(\{t_k\}
\right)\,,
\eea 
where for instance the terms $J^{\,m_2}_c\left(\,\{t_k\}
\right)$ are constructed using collinear-gauge-invariant 
collinear building blocks given in \Eqn{coll-fields-def}.  
In this general construction, the indices $m_1$, $m_2$ 
consist of letters $A$, $B$, $C$ etc., labelling the number 
of fields in a particular collinear direction, and numbers 
0, 1, 2 etc., denoting the overall power of $\lambda$ 
of the current with respect to the LP, which is 
labelled $0$. The function $\Gamma_{\rho}^{m_1,m_2}$ 
in \Eqn{eq:DYcurrent} gives the appropriate spinor 
and Lorentz structure of the operator. For instance, 
the structure of the LP operator reads 
$\Gamma^{\,A0,A0}_{\rho}=
\gamma_{\perp\rho}\,$; at $\mathcal{O}(\lambda)$ 
one has $\Gamma^{\,A0,A1}_{\rho} = n_{+\rho}$, etc. 

Hard modes are integrated out in the matching between 
QCD and SCET, and appear as short-distance coefficients 
$\widetilde{C}^{\,m_1,m_2}$ in \Eqn{eq:subleadingmatching}. 
In the next step, one integrates out threshold-collinear 
and anticollinear modes. As discussed in \cite{Beneke:2018gvs,Beneke:2019oqx},
this gives rise to a matching equation, which involves 
on the left-hand-side the SCET Lagrangian with 
threshold-collinear and soft fields, and on the 
right-hand side PDF-collinear and soft fields, 
times a short-distance coefficient, defined as 
``collinear function'', which arises as a consequence
of integrating out threshold-collinear modes. 
The collinear function is the analogue of 
the ``radiative jet'' discussed in 
\cite{DelDuca:1990gz,Bonocore:2015esa,Bonocore:2016awd},
and we refer to section 2 of \cite{Beneke:2019oqx}
for a detailed discussion. The specific form of 
the matching equation is dictated by gauge 
invariance, momentum and Fermion number 
conservation. For the $g\bar q$ channel, 
the general collinear matching equation reads
\begin{eqnarray}
\label{eq:genMatch}
&& i^{\,m}\int \{d^dz_j\}\, T
\left[\,\{\psi_{c}(t_kn_+)\}
\times\left\{\mathcal{L}^{(l)}_{}(z_j) \right\}\right] 
\nonumber \\ 
&&\hspace{1.0cm}
= 2\pi \sum_i \int du \int \{dz_{j-}\}\,
\tilde{G}_{i}\left(\{t_k\},u;\{z_{j-}\} \right)
\,\frac{1}{g_s} {\cal A}^{\text{\scriptsize PDF}}_{c}(u n_{+})
\,\mathfrak{s}_{i}(\{z_{j-}\})\,,
\end{eqnarray}
where $\{d^dz_j\} =\prod^m_{j=0} \, d^dz_j$ and  
$\{dz_{j-}\}=  \prod^m_{j=0} \frac{dn_+z_j}{2}$.  
$\{z_{j-}\}$  denotes the set of $m$ positions
at which the soft building block insertions are 
located, and $T$ represents the time-ordering operator.
On the left-hand side, 
$\left\{\mathcal{L}^{(l)}_{}(z_j) \right\}$ 
is a set of $m$ $\mathcal{O}(\lambda^l)$-suppressed 
Lagrangian insertions and $\{\psi_c(t_kn_+)\}$ 
denotes a set of $n$ collinear fields, 
$\psi_{c} = \chi_c$ or $\mathcal{A}^{\mu}_{c\perp}$, 
each dependent on  one variable from the $n$-sized 
set $\{t_k\}$, originating from the collinear part 
of the Drell-Yan current operators. On the right-hand side, 
the function $\tilde{G}_{i}$ represents the NLP
collinear function resulting from the matching 
of threshold collinear fields 
onto the initial state PDF-collinear gluon 
${\cal A}_c^{\text{\scriptsize PDF}}$ and 
is analogous to the NLP collinear function 
$\tilde{J}_{i}$ in \cite{Beneke:2018gvs,Beneke:2019oqx}, 
representing the matching 
onto an initial state PDF-collinear quark, 
which appears in the factorization of the
$q\bar q$ channel. Last, as in case of 
the $q\bar q$ channel, the soft operator 
$\mathfrak{s}_{i}(\{z_{j-}\})$ represents
a series of soft structures, containing 
the soft fields originating from the 
power-suppressed soft-collinear interaction. 
In case of the $g\bar q$ channel 
$\mathfrak{s}_{i}(\{z_{j-}\})$ must contain
at least one soft quark, i.e. 
\begin{eqnarray}\label{eq:softSet}
\mathfrak{s}_{i}(\{z_{j-}\})\hspace{+0.1cm} 
&\in& \hspace{+0.15cm} \bigg\{
\frac{g_s}{in_-\partial_z} q^{+}(z_-),\,
\ldots\, \bigg\}\,,
\end{eqnarray}
where the index $i$ labels the soft structures, 
and the ellipsis refers to additional structures 
appearing beyond NLP.  Note that, for consistency 
with the $q\bar q$ channel, we normalise the collinear 
function such that at tree level it contributes to 
${\cal O}(g_s^0)$, and we absorb a factor of $g_s$ 
into the soft structures, as can be seen explicitly 
in \Eqn{eq:softSet}. Notice also that the collinear 
gauge invariant building block itself contains a 
factor of $g_s$, which is compensated by the 
explicit factor of $g_s^{-1}$ in \Eqn{eq:genMatch}. 
The matching for the anticollinear leg, with an 
incoming antiquark, is analogous to \Eqn{eq:genMatch}, 
and involves the anticollinear function 
$\bar{\tilde{J} }_{\,\bar{i}}$, as for the $q\bar q$ 
case, and the soft structures given in Eq. (3.45) of 
\cite{Beneke:2019oqx}. After the second matching onto 
collinear and anticollinear PDF modes has been 
performed, the derivation of the factorization 
theorem follows very closely that of the $q\bar q$
channel. Taking into account the factorization 
of the state $|X \rangle = |X_{c} \rangle \otimes 
|X_{\bar c} \rangle \otimes |X_s \rangle$, 
the $g\bar q$ matrix element reads
\begin{eqnarray}
\label{eq:genrl_2b}
\langle X|\bar{\psi} \gamma_\rho \psi(0)|A(p_A)B(p_B)\rangle
&=&\sum_{  	m_1, m_2}\,\sum_{i,\bar{i}}  \, \int\, \{dt_k\} \,
\{d\bar{t}_{\bar{k}}\}  \, \widetilde{C}^{ \, 	m_1, m_2}
\left(\,\{t_k\},\,\{\bar{t}_{\bar{k}}\} \right) \nonumber\\
&& \hspace*{-4cm}\times\,
2\pi \int d\bar{u} \int \{d\bar{z}_{\,\bar{j}+}\}\,
\bar{\tilde{J} }_{\,\bar{i}}^{\,m_1}\left(\,
\{\bar{t}_{\bar{k}}\},\bar{u};
\{\bar{z}_{\,\bar{j}+}\} \right) \, \langle 
X^{{\rm PDF}}_{\bar{c}}|\bar{\chi}^{{\rm PDF}}_{\bar{c}}
(\bar{u} n_-)|B(p_B)\rangle  \nonumber\\ 
&& \hspace*{-4cm}\times \,
2\pi  \int du   \int \{dz_{j-}\}\,
\tilde{G}^{\,m_2}_{i}\left(\,\{t_k\},u;\{z_{j-}\} \right)
\, \langle X^{{\rm PDF}}_{{c}}| \frac{1}{g_s}
\mathcal{A}^{{\rm PDF}}_{c\perp}(un_+)
|A(p_A)\rangle  \nonumber   \\ 
&& \hspace*{-4cm}\times  \,
\Gamma^{m_1,m_2}_{\rho}\,\langle X_s| \, T\left(
\bar{\mathfrak{s}}_{\,\bar{i}}\,(\,\{\bar{z}_{\,\bar{j}+}\})\,
\big[ Y_{-}^\dagger\, J_s \, Y_{+} \big](0)\,
\mathfrak{s}_{i}\,(\{z_{j-}\})
\right)|0 \rangle   \,,
\end{eqnarray} 
where explicit Dirac, Lorentz, and color 
indices are suppressed. The index $k$ $(\bar{k})$ 
counts the number of collinear (anticollinear) building 
block fields in a  given current, and we sum over all 
of the possible currents. The index $j$ $(\,\bar{j}\,)$ 
counts the number of insertions of the subleading power 
Lagrangians into each collinear (anticollinear) sector, 
and again we sum over all of the occurrences.

In order to take the square of the matrix element 
in \Eqn{eq:genrl_2b} it proves useful to write it 
in terms of the momentum-space representation of 
the various functions, which are obtained by means 
of Fourier transformation. For the short-distance 
coefficients $\widetilde{C}^{\,m_1,\,m_2}$ one has
\be
\widetilde{C}\left(\{t_k\},\{ 
\bar{t}_{\bar{k}} \}\, \right) = 
\int \left\{\frac{dn_+p_{k} }{2\pi} \right\}
\left\{ \frac{dn_-\bar{p}_{\,\bar{k}}}{2\pi} \right\}  
\, e^{i  \,( n_+ p_k )\, t_k}
e^{ i  \,(n_- \bar{p}_{\,\bar{k}})\,\bar{t}_{\bar{k}}} 
\,C(\{ n_+p_k\}, \{ n_-\bar{p}_{\,\bar{k}}\}),
\ee
then we need to introduce the momentum-space 
representation of the PDF-collinear and anticollinear 
fields. For instance, for the $c$-PDF gluon field we 
use 
\begin{equation}\label{Ainvft2} 
{\mathcal{A}^{\rm{PDF}}_{c\perp }(un_+)
=\int \frac{dn_+p_a}{2\pi} e^{-i(n_+p_a)u}
\hat{\mathcal{A}}^{\rm{PDF}}_{c\perp }(n_+p_a)},
\end{equation}
with analogous definition for the $\bar c$-PDF
antiquark field. Furthermore, the Fourier transform 
of the collinear function is given as\footnote{Let 
us notice here that, in order to keep equations 
compact, we use the same symbol to indicate 
$z_-^{\mu} \equiv n_+ z 
\, \frac{n_-^{\mu}}{2} \equiv z_- 
\, n_-^{\mu}$ and $z_- \equiv \frac{n_+ z}{2}$.
Therefore, when $z_-$ appears as argument 
of functions, as in $\tilde{G}^{m_2}_{\,{i}}
\left(\,\{t_k\},\,u;\,\{\, z_{j-}\}\,
\right)$, or in scalar products $z_-\cdot k$, 
we refer to the \emph{vector} $z_-^{\mu}$, 
while when $z_-$ appears in exponents or integration 
measures, such as in $e^{-i \omega_j z_{j-}}$ or $dz_{j-}$, 
we refer to the \emph{scalar} $z_{j-} = \frac{n_+ z_j}{2}$ 
or $dz_{j-}=\frac{d(n_+z_j)}{2}$, respectively.}
\begin{eqnarray} \label{eq:anticolFTgluon}
\int \{dt_k\}\int du \,  \tilde{G}^{m_2}_{\,{i}}
\left(\,\{t_k\},\,u;\,\{\, z_{j-}\}\,
\right)\, e^{i ( n_+ p_k ) \,t_k}
e^{-i(n_+p_a)\,u}
\nonumber \\= \int \,\left\{
\frac{d\omega_j}{2\pi}\right\}\,
\,  e^{-i \omega_j z_{j-}} \, 
{G}^{m_2}_{\,{i}}\left(\,\{ n_+p_k \} ,n_+p_a;\,
\{\omega_{j}\}\right)\,,
\end{eqnarray}
with an analogous definition for the 
anticollinear function. Here the set 
$\{\omega_{j}\}$ denotes the variables 
with a soft scaling that are conjugate 
to $\{\, z_{j-}\}$ and in the exponents 
Einstein's summation convention is used. 
Also, $\left\{\frac{d\omega_j}{2\pi}\right\}
= \frac{d\omega_1}{2\pi}\times...\times
\frac{d\omega_m}{2\pi}$. 

Making use of these definitions, \Eqn{eq:genrl_2b}
becomes
\begin{eqnarray}
\label{eq:5.10cQG}
\langle X|\bar{\psi} \gamma_\rho \psi(0)|A(p_A)B(p_B)\rangle
&=& \sum_{	m_1, m_2} \,\sum_{i,\bar{i}}\,
\int \left\{\frac{dn_+p_{k} }{2\pi} \right\}
\, \left\{ \frac{dn_-\bar{p}_{\,\bar{k}}}{2\pi} \right\}
\nonumber \\ 
&& \hspace{-3cm} \times \int \,d(n_+p_a) \, 
\,d(n_-p_b)\, 
C^{\,m_1,m_2}(\{ n_+p_k\}, \,\{ n_-\bar{p}_{\bar{k}}\})
\nonumber \\ 
&& \hspace{-3cm}\times \, \int 
\left\{ \frac{d\bar{\omega}_{\,\bar{j}}}{2\pi}\right\} 
\, \bar{J}^{\,m_1}_{\,\bar{i}}
\left(\, \{ n_-\bar{p}_{\bar{k}}\},-n_-p_b;\,
\{ \bar{\omega}_{\,\bar{j}}\}\right) \,
\,\langle X^{{\rm PDF}}_{\bar{c}}|
\hat{\bar{\chi}}^{{\rm PDF}}_{\bar{c}}(n_-p_b)|B(p_B)\rangle 
\nonumber\\ 
&& \hspace{-3cm}\times \,  \int   
\left\{\frac{d\omega_j}{2\pi}\right\}
\, {G}^{\,m_2}_{\,{i}}\left(\,\{ n_+p_k \} ,n_+p_a;\,
\{\omega_{j}\}\right)  \, \langle X^{{\rm PDF}}_{{c}}| 
\frac{1}{g_s} \hat{\mathcal{A}}^{{\rm PDF}}_{c\perp}(n_+p_a)
|A(p_A)\rangle  \nonumber \\ 
&& \hspace{-2cm} \times  \,
\Gamma^{m_1,m_2}_{\rho} \int \{d\bar{z}_{\,\bar{j}+}\,\}
\int \{dz_{j-}\}\,e^{-i  \bar{\omega}_{\,\bar{j}}
	\,\bar{z}_{\,\bar{j}+}}\,
\, e^{-i \omega_j \,z_{j-}}\nonumber \\ 
&&  \hspace{-2cm} \times\,\langle X_s| \, T\left(
\bar{\mathfrak{s}}_{\,\,\bar{i}}\,(\{
\bar{z}_{\,\bar{j}+}\})\left[ Y_{-}^\dagger\, J_s\, Y_{+} \right](0)\,
\mathfrak{s}_{i}(\{z_{j-}\})
\right)|0 \rangle   \,.
\end{eqnarray} 
The matrix element can be written 
in a more compact form by introducing 
the following amplitude level coefficient 
functions
\begin{eqnarray}\label{eq:5.95}
\mathcal{D}^{m_1,m_2\,\rho}_{\,{i}\,\bar{i}}
(n_+p_a,- n_-{p}_b;\,\{\omega_j\}, \,\{\bar{\omega}_{\bar{j}}\}) 
&=& (2\pi)^2 \int  \left\{\frac{dn_+p_{k} }{2\pi} \right\}
\, \left\{ \frac{d(n_-\bar{p}_{\bar{k}})}{2\pi} \right\} 
\nonumber \\ && \hspace{-3.0cm}\times \,
C^{\,m_1,m_2}(\{ n_+p_k\}, \,\{ n_-\bar{p}_{\bar{k}}\}) 
 \times \bar{J}^{m_1}_{\,\bar{i}}
\left(\, \{ n_-\bar{p}_{\bar{k}}\},-\,n_-p_b;\,
\{ \bar{\omega}_{\bar{j}}\}\right)
\nonumber \\ && \hspace{-3.0cm}
\times  \,{\Gamma}^{\,m_1, m_2\,\rho} \,  
{G}^{m_2}_{\,{i}}\left(\,\{ n_+p_k \} ,\,n_+p_a;\,
\{\omega_{j}\}\right) \,,
\end{eqnarray}
such that the matrix element reads
\begin{eqnarray}\label{eq:5.11cqg}
\langle X|\bar{\psi} \gamma^\rho \psi(0)|A(p_A)B(p_B)\rangle
&=& \frac{1}{(2\pi)^2} \sum_{m_1, m_2} \,\sum_{i,\bar{i}}\,
\int \,d(n_+p_a) \, \,d(n_-p_b) \int 
\left\{ \frac{d\bar{\omega}_{\bar{j}}}{2\pi}\right\} 
\int\left\{\frac{d\omega_j}{2\pi}\right\}\,
\nonumber \\
&& \hspace{-3cm}
\times  
\int dg\,d\bar{g} \,e^{i(n_+p_a)g}e^{-i(n_-p_b)\bar{g}}\, 
\mathcal{D}^{m_1,m_2\,\rho}_{\,{i}\,\bar{i}}
(n_+p_a,- n_-{p}_b;\,\{\omega_j\}, \,\{\bar{\omega}_{\bar{j}}\})
\nonumber \\[1ex] && \hspace{-3cm}
\times \,\, \langle X^{{\rm PDF}}_{\bar{c}}|
{\bar{\chi}}^{{\rm PDF}}_{\bar{c}}(\bar{g}\,n_-)|B(p_B)\rangle 
\,\,  \langle X^{{\rm PDF}}_{{c}}| \frac{1}{g_s}
{\mathcal{A}}^{{\rm PDF}}_{c\perp}(g\,n_+) |A(p_A)\rangle 
\nonumber \\[1ex] && \hspace{-2cm}
\times \int \{d\bar{z}_{\bar{j}+}\,\}
\int \{dz_{j-}  \,\}\,e^{-i  \bar{\omega}_{\,\bar{j}}
\,\bar{z}_{\,\bar{j}+}}\, \, e^{-i \omega_j \,z_{j-}} 
\nonumber \\[1ex] && \hspace{-2cm}
\times \,\, \langle X_s| T
\left( \bar{\mathfrak{s}}_{\,\,\bar{i}}(\,\{\bar{z}_{\bar{j}+}\}) 
\left[ Y_{-}^\dagger\,J_s\, Y_{+} \right](0)\,
\mathfrak{s}_{i}(\,\{z_{j-}\})
\right)|0 \rangle   \,.
\end{eqnarray}

We can now insert this expression into the 
definition of the hadronic tensor $W_{\rho\mu}$ 
in \Eqn{eq:ampTrans}, with an equivalent expression 
for the complex conjugate matrix element, and obtain
the factorized expression at general subleading 
powers for the cross section given in 
\Eqn{eq:comb_lept_2bc}. To this end one still 
needs to identify the PDF-collinear and anticollinear
matrix elements respectively with the gluon and 
anti-quark parton distribution functions. For 
the former we use the following relation from  
appendix A of \cite{Beneke:2010da}\footnote{Note 
the different normalization of the gauge-invariant 
gluon field building block used in this work, 
compared to \cite{Beneke:2010da}: here 
$\mathcal{A}_c^\mu = 
W_c^{\dagger}\left[i D_c^\mu\,W_c\right]$, while 
in Eq. (2.26) of \cite{Beneke:2010da} one defines
$\mathcal{A}_c^\mu = 
g_s^{-1} \, W_c^{\dagger}\left[i D_c^\mu\,W_c\right]$. 
This difference is at the origin of the factor $g_s^{-2}$
in \Eqn{eq:gluonPDF}.}
\begin{eqnarray}\label{eq:gluonPDF}
\langle A(p_A)| \frac{1}{g_s^2}\, \mathcal{A}^{A', 
\,{\rm{PDF}}}_{c {\perp} {\eta'} }(x+g'n_+)\,  
\mathcal{A}^{A,\, {\rm{PDF}}}_{c{\perp}\eta  }(gn_+)
|A(p_A)\rangle &=& \frac{-g_{\perp \eta\eta'}}{(d-2)} 
\,\frac{\delta^{AA'}}{(N^2_c-1)} \nonumber
\\  &&\hspace{-2.8cm}\times  \int^1_{0} \frac{dx_a}{x_a}  
f_{{g}/A}(x_a)e^{ix_a(x+g'n_+-gn_+)\cdot p_A }\,,
\end{eqnarray}
while for the anticollinear matrix element we use
\begin{eqnarray}\label{bpdf}
\langle B(p_B) |{\bar{\chi}}^{{\rm PDF}}_{\bar{c},\alpha a}
(\bar{g} n_-){{\chi}}^{{\rm PDF}}_{\bar{c},\delta j}
(x+\bar{g}' n_-)|B(p_B)\rangle  && \nn \\&& \hspace{-5cm}=  
-\frac{\delta_{ja}}{N_c}\left(\frac{\slashed{n}_+}{4}
\right)_{\delta\alpha}(n_-p_B)\int_0^1 dx_b \,
e^{-i(\bar{g}n_--\bar{g}'n_--x)\cdot p_B x_b}f_{b/B}(x_b). 
\end{eqnarray}
Let us remark that, although we do not make indices 
explicit in this general derivation, it is understood 
that the indices appearing in~\Eqn{eq:gluonPDF} are 
absorbed by the collinear functions. Integrating over 
the auxiliary variables $g$, $\bar{g}$, $g'$, $\bar{g}'$ 
and the momenta $n_+p_a$, $n_+p_b$, $n_+p_a'$, $n_+p_b'$,
after some elaboration we obtain the $g\bar{q}$-channel 
Drell-Yan cross-section as defined in \Eqn{DeltaDef}:
\begin{eqnarray}\label{eq:allpowerresultQG}
\Delta_{g\bar q}(z) &=&  
\sum_{ \substack{m'_1, {m}'_2,\\m_1, m_2 }}\,
\sum_{\substack{i',\bar{i}'\\i,\bar{i}}}\, 
\int \left\{ \,\frac{d\bar{\omega}'_{\,\bar{j}'}}{2\pi}\right\}
\left\{ \,\frac{d\omega'_{j'}}{2\pi}\right\}\,
\left\{ \frac{d\bar{\omega}_{\bar{j}}}{2\pi}\right\}
\left\{\frac{d\omega_j}{2\pi}\right\} \nonumber \\ 
&& \times \frac{1}{2z(1-\epsilon)^2} \bigg[
\mathcal{D}^{m_1,m_2\,\rho}_{\,{i}\,\bar{i}}
(x_an_+p_A, x_bn_-{p}_B;\,\{\omega_j\},
\,\{\bar{\omega}_{\bar{j}}\})\,
\left(\frac{\slashed{n}_+}{4} \right)\, \nonumber \\ 
&&\hspace{2.5cm} \times  
\mathcal{D}^{*\,m_1',m_2'\,}_{\,{i}'\,\bar{i}'\,\rho}
(x_an_+p_A, x_bn_-{p}_B;\,\{\omega'_{j'}\},
\,\{\bar{\omega}'_{\bar{j}'}\}) \bigg]\, \nonumber \\ 
&& \times \int 
\frac{d^{d-1}\vec{q}}{(2\pi)^3\,2\sqrt{Q^2+\vec{q}^{\,2}}}
\, \frac{Q}{2\pi} \int d^dx \, e^{i(p_A x_a+p_B x_b- q)\cdot x}
\nonumber \\ && \hspace{2.5cm} \times
\widetilde{S}_{{{g\bar{q}}};\,{i}\,\,\bar{i}\,\,{i}'\,\,\bar{i}'}
(x; \{\omega_j\};\,\{\bar{\omega}_{\bar{j}}\};
\,\{{\omega}'_{{j}'}\};\,\{\bar{\omega}'_{\bar{j}'}\}).
\end{eqnarray}
This is the result for the general form of the 
power-suppressed $g\bar{q}$-induced partonic 
cross-section in the $z\to1$ limit. The notation 
with bars $(\,\bar{}\,)$ and tildes 
$(\,\,\widetilde{}\,\,)$ is used here in the same 
way as in the derivation of the $q\bar{q}$-induced 
partonic cross-section of \cite{Beneke:2019oqx}. 
They refer to the anticollinear direction and 
objects with dependence on the coordinate 
variables respectively. Also, the contributions 
from the complex conjugate amplitude are denoted 
with a prime $(\,'\,)$ symbol. 

In the last line of equation \eqref{eq:allpowerresultQG} 
we have introduced the generalised multi-local soft 
function for the $g\bar{q}$-channel. It is given by 
\begin{eqnarray}\label{eq:3.17QG}
\widetilde{S}_{{{g\bar{q}}};\,{i}\,\,\bar{i}\,\,{i}'\,\,\bar{i}'}
(x; \{\omega_j\};
\,\{\bar{\omega}_{\bar{j}}\};\,\{{\omega}'_{{j}'}\};
\,\{\bar{\omega}'_{\bar{j}'}\}) &= & 
\int \{ d\bar{z}'_{\bar{j}'+}\} \int \{dz'_{j'-}\}
\int \{d\bar{z}_{\bar{j}+}\,\}
\int \{dz_{j-}  \,\} \nonumber \\ 
&& \hspace{-3.5cm}\times\,e^{+i\left( 
\bar{\omega}'_{\,\bar{j}'} \bar{z}'_{\,\bar{j}'+}\right)}\, 
e^{+i\left(\omega'_{j'} z'_{j'-}\right)}\,
\,e^{-i \left( \bar{\omega}_{\,\bar{j}}
\bar{z}_{\,\bar{j}+}\right)}\,
\, e^{-i\left(\omega_j z_{j-}   \right)}
\nonumber\\ && \hspace{-3.5cm}\times  
\frac{1}{N_c^2 - 1}
\langle 0|   \bar T\left(
\bar{\mathfrak{s}}'_{i'}(\,\{x+z'_{j'-}\})\,
\left[ Y_{+}^\dagger(x)J^{\dagger}_s Y_{-}(x) \right]\,
{\mathfrak{s}}'_{\,\,\bar{i}'}(\,\{
x+ \bar{z}'_{\bar{j}'+}\}) \right)  \nonumber \\ && \hspace{-3.5cm}\times \, T\left(
\bar{\mathfrak{s}}_{\,\,\bar{i}}(\,\{\bar{z}_{\bar{j}+}\})
\left[ Y_{-}^\dagger(0)J_s Y_{+}(0) \right]\,
\mathfrak{s}_{i}(\,\{z_{j-}\})
\right)|0 \rangle \,.
\end{eqnarray}
Let us also recall that, in the same fashion 
as the $q\bar{q}$-channel factorization theorem 
in \cite{Beneke:2019oqx}, the result in  
\Eqn{eq:allpowerresultQG} is formally valid
in $d$-dimensions.  

\subsection{Factorization at next-to-leading power}
\label{sec:NLPfact}

Let us now specialise the factorization theorem in 
\Eqn{eq:allpowerresultQG} to next-to-leading power. 
As discussed above, near threshold the production of 
an off-shell photon from an initial $g\bar q$ state 
involves at least the emission of a soft quark into
the final state. Soft quarks can arise from the factor 
$J_s(0)$ in the expansion of the vector current, but 
this term is at least of ${\cal O}(\lambda^3)$, 
therefore any such contribution is beyond NLP. 
Hence, the only remaining possibility to achieve a
power suppression is provided by time-ordered products 
with insertions of the Lagrangian terms 
$\mathcal{L}^{(i)}_{\xi q}$. 
Up to NLP only the two terms with $i = 1$, $2$ can 
appear. The $\mathcal{L}^{(2)}_{\xi q}$ Lagrangian 
insertion, however, can also be dropped, as the 
corresponding amplitude would have to be interfered 
with a leading power amplitude in order to yield an
$\mathcal{O}(\lambda^2)$ power suppressed cross-section,
and such contribution vanishes. This leaves solely the 
contribution due to $\mathcal{L}^{(1)}_{\xi q}$
multiplied with the leading power $J^{A0}$ current. 
Therefore, using the Fourier transforms in
\Eqns{eq:anticolFTgluon}{Ainvft2}, at NLP 
the collinear matching in \Eqn{eq:genMatch} 
reduces to
\begin{eqnarray}\label{m1qg}
i \int d^dz \,{T}\Big[ 
\chi_{c,\gamma f}\left(tn_+\right)
\,\mathcal{L}^{(1)}_{\xi q}(z) \Big]
= 2\pi  \int \frac{d\omega}{2\pi}
\int \frac{dn_+p}{2\pi} \,e^{-i(n_+p)t} 
\int \frac{dn_+p_a}{2\pi} && \nonumber \\
&& \hspace{-11.5cm}
\times \, {G}^{\eta,A}_{\xi q;\gamma\alpha,fa}
\left(n_+p,n_+p_a; \omega \right)\, \frac{1}{g_s}
\hat{\mathcal{A}}^{{\rm PDF}\,A}_{c \perp \eta}(n_+p_a)
\int dz_-\, e^{-i\omega\,z_-}
\,\mathfrak{s}_{\xi q;\alpha,a}(z_-),
\end{eqnarray}
where we have written indices explicitly: 
$\alpha$ and $\gamma$ are  Dirac  indices, $\eta$ is a Lorentz 
index, $a,f$ and $A$ are a fundamental and adjoint 
color indices respectively. This is the analogue 
of Eq.~(3.23) in \cite{Beneke:2019oqx} for the 
$q\bar{q}$-channel at NLP. Here we do not sum 
over the soft structures, as at NLP there is only 
one, originating from $\mathcal{L}^{(1)}_{\xi q}$,
given by the term written in \Eqn{eq:softSet}. 
With explicit indices it reads 
\begin{eqnarray}\label{qgsoftstructure}
\mathfrak{s}_{\xi q;\alpha,a}(z_-) = \frac{g_s}{in_-\partial_z}q^{+}_{\alpha,a}(z_-)\,.
\end{eqnarray}
In order to simplify the next-to-leading power 
factorization formula as much as possible, we 
make use of generic properties of the collinear 
function ${G}^{\eta,A}_{\xi q;\gamma\alpha,fa}
\left(n_+p,n_+p_a; \omega \right)$ in the matching 
equation \eqref{m1qg}. Firstly, as in case of the 
$q\bar q$ channel, the collinear function must be 
proportional to the delta function in the collinear 
momenta, $\delta \left(n_+p-n_+p_a\right)$, since 
the kinematic set-up does not allow for threshold 
collinear radiation into the final state. Therefore 
the incoming $c$-PDF momentum is the same as the 
outgoing threshold collinear momentum. Further 
simplification arises from the fact that 
$\mathcal{L}^{(1)}(z)$ does not contain 
explicit factors of the position, such 
as for instance $n_- z$, thus there are 
no momentum-space derivatives acting on 
the collinear momentum delta function, 
as it happens for  
the collinear function $J_1$ in case 
of the $q\bar q$ channel, cf. Eq.~(3.40) 
in \cite{Beneke:2019oqx}. Concerning the 
color structure, we note that 
${G}^{\eta,A}_{\xi q;\gamma\alpha,fa}
\left(n_+p,n_+p_a; \omega \right)$ carries 
one adjoint color index $A$ and 
two fundamental color indices $f a$, 
therefore we can extract a 
$\mathbf{T}^{A}_{fa}$ color generator 
and transfer it into the definition of 
the soft function. The collinear function 
also carries a single Lorentz index $\eta$ 
and two Dirac indices $\gamma,\alpha$. From 
the matching equation in~\eqref{m1qg} we see 
that the Lorentz index is contracted with a 
$\perp$ structure. Therefore, in the collinear 
function a $\gamma^{\eta}_{\perp}$ must appear, 
as the only other possible single Lorentz index 
carrying structures  are $n^{\eta}_{\pm}$, 
which would vanish upon contraction with 
$\hat{\mathcal{A}}^{{\rm PDF}\,A}_{c \perp\eta}$. 
Based on these considerations, we expect the 
collinear function ${G}_{\xi q,\gamma\alpha,fa}^{\eta,A}$ 
to have the following structure at all orders: 
\be
\label{collFunction4}
{G}_{\xi q,\gamma\alpha,fa}^{\eta,A}(n_+p,  n_+p_a;\omega) = 
{G}_{\xi q}(n_+p;\omega)\,\delta\big(n_+p-n_+p_a\big)\, 
\mathbf{T}^{A}_{fa}\,\left[\slashed{n}_{-} \gamma^{\eta}_{\perp}\right]_{\gamma\alpha}\,,
\ee
where we introduced the scalar collinear function 
${G}_{\xi q}$. At tree level the factor $\slashed{n}_{-}$
arises due to the collinear quark propagator from 
the point $t n_+$ to $z$. 
At higher orders the function 
${G}_{\xi q,\gamma\alpha,fa}^{\eta,A}(n_+p,  n_+p_a;\omega)$
is given in terms of loops corrections to the 
collinear quark propagator, therefore it is 
constrained by gauge- and reparametrization 
invariance to have the form on the right-hand side 
of \Eqn{collFunction4}. This can be seen 
explicitly by noticing that additional 
factors of $\gamma_{\perp}$ cannot appear, 
because they would be contracted with  
the collinear momentum, and thus would be 
power suppressed. The only possible remaining 
structures are given by factors of $\slashed{n}_+$ 
and $\slashed{n}_-$. Given an odd number of these 
factors, one has $\slashed{n}_- \slashed{n}_+ 
\slashed{n}_- = 4 \slashed{n}_-$,
$\slashed{n}_+ \slashed{n}_+ \slashed{n}_-
= 0$, and $\slashed{n}_- \slashed{n}_- 
\slashed{n}_- = 0$. Since at least
one factor of $\slashed{n}_-$ always 
appears due to the tree level quark 
propagator, the rules above can be 
used to systematically reduce any
loop correction to the form in 
\Eqn{collFunction4}. We will show this 
explicitly at one loop in 
section~\ref{sec:CollinearFunctions}, and 
this structure was verified at two loops 
in \cite{Liu:2021mac}.

In order to simplify further \Eqn{eq:allpowerresultQG},
we note now that the time-ordered product at NLP,
both in the amplitude and the complex conjugate 
amplitude involves the leading power current $J^{A0}$,
therefore the structure $\Gamma^{m_1,m_2\,\rho}_{}$
in \Eqn{eq:5.95} (and the corresponding 
$\bar{\Gamma}^{m_1',m_2'}_{\rho}$ in the 
complex conjugate matrix element) reduce 
respectively to $\gamma_{\perp}^{\rho}$ 
and $\gamma_{\perp\rho}$. Using this 
information, and the structure of the 
collinear function in \Eqn{collFunction4}, 
we find that the spin structure implicit 
in the factor $\mathcal{D}^{\,\rho}_{\,}\,
\left(\frac{\slashed{n}_+}{4}\right)\,
\mathcal{D}^{*\,}_{\rho}$ in 
\Eqn{eq:allpowerresultQG} 
takes the form  
\be\label{spin2}
\left[\gamma_{\perp \eta }{\slashed{n}_{-}}\right]_{\sigma\beta} 
\left(\gamma_{\perp\rho }\right)_{ \beta\delta}
\left(\frac{\slashed{n}_+}{4}\right)_{\delta\lambda}
\left(\gamma^{\rho}_{\perp }\right)_{ \lambda \gamma}
\left[{\slashed{n}_{-}} \gamma^{\eta}_{\perp}\right]_{\gamma\alpha}
= 4 \, \frac{\slashed{n}_{-\sigma\alpha}}{4} (d-2)^2\,.
\ee
In the following we will absorb the factor of 
${\slashed{n}_{-\sigma\alpha}}/{4}$ into the 
definition of the soft function, which we give 
below.

\begin{figure}[t]
\begin{center}
\includegraphics[width=0.8\textwidth]{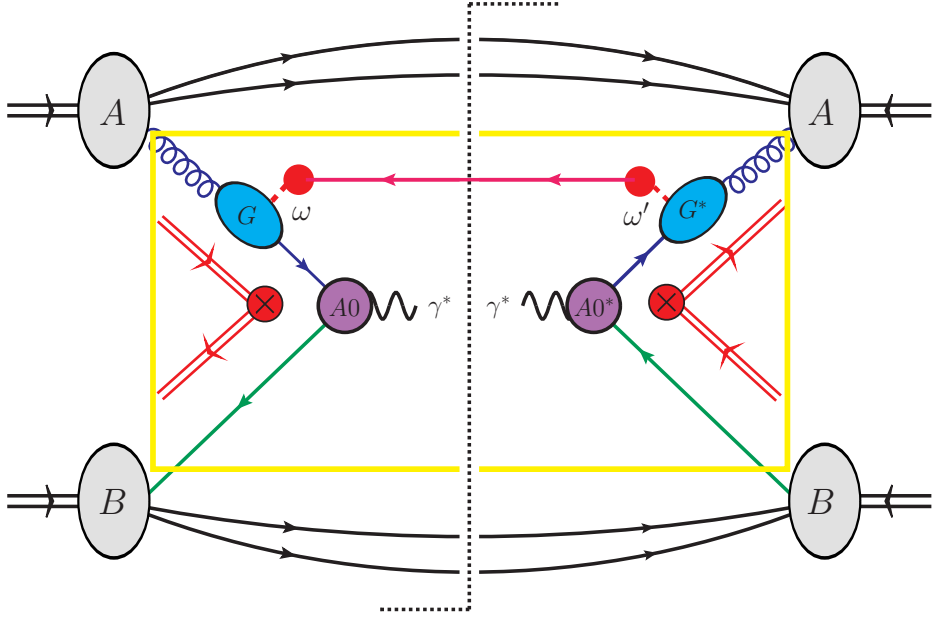} 
\end{center}
\caption{Graphical representation of the 
factorization theorem in \Eqn{eq:QGnlpfact}, 
valid for the $g\bar q$ channel in Drell-Yan 
near threshold at NLP. In this picture 
red lines represent soft fields, blue lines 
represent collinear fields, and green lines 
are anti-collinear fields. The purple circles 
denoted by ``$A0$'' and ``$A0^*$'' represent the leading power 
short-distance coefficient $C^{A0}(Q^2)$ and its complex conjugate, respectively.  
The blue ovals denoted by $G$, $G^*$ 
represents the (momentum-space representation 
of the) collinear functions, introduced in the 
matching equation \Eqn{eq:genMatch}. The red double lines represent the soft Wilson lines and the red filled circles are insertions of the soft quark building blocks.
Last, $\omega$ and $\omega'$ are the 
convolution variables between the 
collinear functions and the soft 
function.}
\label{fig:XSfactorized}
\end{figure} 
The last simplification which can be 
applied when considering \Eqn{eq:allpowerresultQG}
up to NLP concerns the phase space integration. 
As discussed for the $q\bar q$ channel in 
\cite{Beneke:2019oqx}, also the kinematic factors
in \Eqn{eq:allpowerresultQG} need to be expanded
consistently to NLP: this implies that the matrix 
element squared expanded to NLP needs to be integrated
against the LP expansion of the phase space, while
the LP matrix element needs to be multiplied by the
phase space expanded up to NLP. These two contributions
are identified respectively as ``dynamical'' and 
``kinematic'' terms: $\Delta_{\rm{NLP}}(z)
= \Delta^{dyn }_{\rm NLP}(z) + \Delta^{kin }_{\rm NLP}(z)$.
In case of the $g\bar q$ channel the matrix element
squared starts at NLP, therefore there is no kinematic
contribution:
\be\label{dynNLPqg}
\Delta_{g\bar q}(z)|_{\rm NLP} 
= \Delta^{dyn}_{g\bar q}(z)|_{\rm NLP}\,. 
\ee
Taking the LP approximation of the phase 
space implies the following replacement 
in \Eqn{eq:allpowerresultQG}:
\be\label{PSLP}
\frac{1}{2z(1-\epsilon)^2} 
\int \frac{d^{d-1}\vec{q}}{(2\pi)^3\,2\sqrt{Q^2+\vec{q}^{\,2}}}
\frac{Q}{2\pi} \int d^dx \, e^{i(p_A x_a+p_B x_b- q)\cdot x} 
\to \frac{1}{8\pi(1-\epsilon)^2} 
\int dx^0 \, e^{i \frac{\Omega}{2} x^0}\,,
\ee
where $\Omega = Q(1-z)$. As a consequence of this 
approximation, we can set $\vec{x}=0$ in the argument 
of the position space soft function. Last, we observe 
that we can carry out a further simplification
by setting the anticollinear function to its LP 
expression, $\bar{J}_{\bar{i}}
\left(n_-\bar{p}, x_b n_-p_B; \bar \omega\right) 
= \delta(n_-\bar{p} - x_b n_-p_B)$,
given that the suppression to NLP is already 
provided by the two insertions of 
$\mathcal{L}^{(1)}_{\xi q}$.

Applying all these simplifications to 
\Eqn{eq:allpowerresultQG} we finally get
\be\label{eq:QGnlpfact}
\Delta_{g\bar{q}}|_{\rm NLP}(z) = 
8 H(Q^2)\,\int {d\omega}\, d\omega' \,  
G^*_{\xi q}(x_an_+p_A;\omega') \,
G_{\xi q}(x_an_+p_A;\omega) 
\, S(\Omega,\omega, \omega') \,,
\ee
where the hard function is also given by the 
LP approximation of the LP short-distance 
coefficient squared: $H(\hat{s})
=|C^{A0,A0}(x_an_+p_A,x_bn_-p_B)|^2 
= H(Q^2)+\mathcal{O}(\lambda^2)$, and the 
soft function ${S}(\Omega,\omega,\omega')$ 
is given by
\begin{eqnarray}\label{softoperator2}
S_{g\bar{q}}(\Omega,\omega, \omega') &=&  
\int \frac{dx^0}{4\pi} \int \frac{dz_{-}}{2\pi} 
\int \frac{dz'_{-}}{2\pi}e^{-i\omega z_{-}}
e^{+i\omega' z'_{-}}e^{+i \,\Omega \, x^0/2} 
\nonumber\\ &&\times \frac{1}{N_c^2-1}
\langle 0| \bar T \Big(
\frac{g_s}{in_-\partial_{z'}}
\bar{q}^{\,+}(x^0+z'_{-})\,\mathbf{ T}^{A}\,
 \big\{Y^{\dagger}_{+}(x^0)Y_{-}(x^0)\big\}\Big)
\nonumber\\&& \times  \frac{\slashed{n}_{-}}{4} 
T\Big(\left\{Y^{\dagger}_{-}(0)Y_{+}(0)\right\} 
\, \mathbf{ T}^{A}\,\frac{g_s}{in_-\partial_{z}} 
q^{+}(z_{-}) \Big) |0\rangle \,.
\end{eqnarray}
The $\mathcal{L}^{(1)}_{\xi q}$ insertion 
in the amplitude is at position $z_{-}$, 
whereas in the conjugate amplitude we place 
the same insertion at position $z'_{-}$. 
The conjugate variables to these coordinate-space
variables are $\omega$ and $\omega'$ respectively.
A graphical representation of the factorization 
theorem in \Eqn{eq:QGnlpfact} is given in Fig. 
\ref{fig:XSfactorized}. We note that the 
factorization formula in \Eqn{eq:QGnlpfact}, 
in the same way as the results for the 
$q\bar{q}$-channel Drell-Yan cross-section, 
is formally valid in $d$-dimension. The 
objects appearing in the factorization 
formula, ${G}_{\xi q}(x_an_+p_A;\omega)$, 
${G}^*_{\xi q}(x_an_+p_A;\omega')$, 
and ${S}_{g\bar{q}}(\Omega,\omega,\omega')$,
should not be treated as renormalized objects, 
because the convolution linking the collinear 
and soft functions must be performed in 
$d$-dimensions. We stress that the 
factorization formula in \Eqn{eq:QGnlpfact} 
is valid at the level of the partonic cross 
section. In particular, in the context of SCET, it 
was argued in \cite{Becher:2007ty} that Glauber 
modes give scaleless contributions at partonic 
level and therefore Glauber fields are not included in the construction of the effective 
theory.  In the next two sections 
we compute the collinear and soft functions, 
respectively to ${\mathcal O}(\alpha_s)$ and 
${\cal O}(\alpha_s^2)$, which is necessary 
to achieve NNLO accuracy for the invariant
mass distribution.

\section{Collinear functions}
\label{sec:CollinearFunctions}

We now proceed to calculate the collinear 
function $G_{\xi q}(x_an_+p_A;\omega)$
up to one loop, by means of the matching 
equation \eqref{m1qg}. Before continuing, 
let us notice that an equivalent collinear
function has been defined in the context 
of $H \to gg$ in \cite{Liu:2021mac}, 
which has been calculated up to two 
loops. We provide here an independent
calculation and check that our result 
for the one-loop collinear function 
defined in \Eqn{m1qg} is indeed 
equivalent to the one considered in 
\cite{Liu:2021mac}.
In order to set-up the calculation 
it proves useful to introduce the 
short-hand notation 
\begin{align}
\label{eqn:matching4.1qg} 
\tilde{\mathcal{T}}_{\gamma f}(t)
&\equiv i \int d^4z \, T \Big[ 
\chi_{c,\gamma f}\left(tn_+\right) 
\mathcal{L}^{(1)}_{\xi q}(z)\Big]\,,
\end{align}
for the left-hand side of \eqref{m1qg}. 
We then consider its Fourier transform
\begin{align}
\label{eqn:matching4.1FTqg}
\mathcal{T}_{\gamma f}(n_+q) 
&=  \int dt \, e^{i (n_+q)\,t}
\,\tilde{\mathcal{T}}_{\gamma f}(t)\, ,
\end{align}
and rewrite the matching equation 
\eqref{m1qg} in momentum space as follows
\begin{eqnarray}\label{m1qg2}
\mathcal{T}_{\gamma f}(n_+q)
&=& 2\pi \int \frac{dn_+p_a}{2\pi} 
\int du \, e^{i\, (n_+p_a)\, u}\,
\int \frac{d\omega}{2\pi} 
\nonumber \\&& \hspace{-1cm}
\times  {G}^{\,\eta,A}_{\xi q;\gamma\alpha,fa}
\left(n_+q,n_+p_a; \omega \right)\,  \frac{1}{g_s} 
{\mathcal{A}}^{{\rm PDF}\,A}_{c \perp  \eta}(un_+)
\int dz_-\, e^{-i\omega\,z_-}
\,\mathfrak{s}_{\xi q;\alpha,a}(z_-)\,.
\end{eqnarray}
Starting from this definition, we extract 
the perturbative collinear functions by 
considering suitable partonic matrix elements 
of the operator matching equation. In this case, 
the relevant matrix element involves an incoming 
$c$-PDF gluon and an outgoing soft quark 
$\langle q^+(k)| ... | g(p)\rangle$. We 
compute both sides of the matching equation 
with the leading power decoupled Lagrangian, 
considering the soft fields as external. Hence, 
we only need the soft matrix element $\langle q^+(k)| 
\mathfrak{s}_{\xi q;\alpha,a}(z_-) | 0\rangle$ at tree 
level. Similarly, for the $c$-PDF matrix element 
$\langle 0| {\mathcal{A}}^{{\rm PDF}\,A}_{c  \perp \eta}(un_+) 
| g(p)\rangle$ the loop corrections are scaleless 
and $\langle 0| 
{\mathcal{A}}^{{\rm PDF}\,A}_{c\perp \eta }(un_+) 
| g(p)\rangle $ contributes only at tree level.
The calculation of the left-hand side of 
\Eqn{m1qg2} is done using the momentum-space 
Feynman rules given in appendix~A of 
\cite{Beneke:2018rbh}. Before moving 
to the description of the actual 
calculation, we point out that this 
calculation is far more straightforward 
compared to the case of the corresponding 
collinear functions in the $q\bar q$ channel, 
see \cite{Beneke:2019oqx}. This is because, 
as discussed above, we need to consider the 
single soft structure given in 
\Eqn{qgsoftstructure}; furthermore, the 
Lagrangian insertion $\mathcal{L}^{(1)}_{\xi q}$ 
which induces the power suppression does not 
contain any explicit position variables, 
which means that, in momentum space, no 
derivatives on the momentum conserving 
delta functions appear at subleading power 
soft-collinear interaction vertices. This 
reduces drastically the number of terms 
which appear in each diagram and the number
of integrals involved in the calculation.

\subsection{Tree-level collinear function}

We first consider the right-hand side of \eqref{m1qg2} 
and evaluate the tree-level matrix element
\begin{eqnarray}\label{m1qg2matrixelement}
\langle q^+_{b}(k)|  \mathcal{T}_{\gamma f}(n_+q)
|g^B(p)\rangle &=& 2\pi \int \frac{dn_+p_a}{2\pi} 
\int du \, e^{i\, (n_+p_a)\, u}\,
\int \frac{d\omega}{2\pi}  \int dz_-\, e^{-i\omega\,z_-}
\nonumber \\&& \hspace{-4cm}
\times {G}^{\,\eta,A}_{\xi q;\gamma\alpha,fa}
\left(n_+q,n_+p_a; \omega \right)\, 
\langle 0| \frac{1}{g_s}  
{\mathcal{A}}^{{\rm PDF}\,A}_{c \perp \eta}(un_+)|g^B(p)\rangle 
\, \langle q^+_{b}(k)| \mathfrak{s}_{\xi q;\alpha,a}(z_-)|0\rangle\,,
\end{eqnarray}
where $B$ is the adjoint color index and $b$ 
is the fundamental color index of the external 
final state. The $c$-PDF matrix element in
\eqref{m1qg2matrixelement} evaluates to 
the following
\be\label{eq:qgcol}
\langle 0| \frac{1}{g_s}
{\mathcal{A}}^{{\rm PDF}\,A}_{c\perp\eta}(un_+)|g^B(p)\rangle 
= \delta^{AB}\,\sqrt{Z_{g,{\rm{PDF}}}}\,
\epsilon_{ {\perp}\eta}(p)\,e^{-i(n_+p)u}.
\ee
The factor $\sqrt{Z_{g,{\rm{PDF}}}}$ is the on-shell 
renormalization factor of the $c$-PDF gluon field. 
The soft matrix element on the right-hand side of 
\eqref{m1qg2matrixelement} becomes 
\be\label{eq:qgsoft}
\langle q^+_{b}(k)|\mathfrak{s}_{\xi q;\alpha,a}(z_-)|0\rangle 
= \langle q^+_{b}(k)|\frac{g_s}{in_-\partial_z}
q^{+}_{\alpha,a}(z_-)|0 \rangle 
= - \delta_{ba}\,\frac{g_s}{n_-k}
\,v_{\alpha}(k)\, e^{iz_-\cdot k}\,.
\ee
In the second step we have used the form of the 
soft structure as given in \Eqn{qgsoftstructure}.
As originally discussed below Eq. (4.3) of 
\cite{Beneke:2019oqx} for the case of collinear 
functions involving PDF and threshold quarks, 
the matrix elements in \Eqns{eq:qgcol}{eq:qgsoft}
do not have loop corrections. This is true in 
\Eqn{eq:qgcol} because the loop corrections are 
scaleless, and it also holds for \Eqn{eq:qgsoft} 
because here the soft field is treated as external.
Next, we substitute the matrix elements evaluated 
in Eqs.~\eqref{eq:qgcol} and \eqref{eq:qgsoft} 
into~\eqref{m1qg2matrixelement} and find
\begin{eqnarray}\label{m1qg2matrixelementresult}
\langle q^+_{b}(k)| \mathcal{T}_{\gamma f}(n_+q)
|g^B(p)\rangle &=& - (2\pi) \frac{g_s}{n_-k} \,  
{G}^{\,\eta,B}_{\xi q;\gamma\alpha,fb} 
\left(n_+q,n_+p; n_-k \right)\, \nonumber \\ 
&& \times \,v_{\alpha}(k)\, \sqrt{Z_{g,{\rm{PDF}}}}
\,\epsilon_{{\perp}\eta}(p) \,.
\end{eqnarray}
This result constitutes the final expression 
for the matrix element with the chosen partonic 
external states of the right-hand side of 
the matching equation in \eqref{m1qg2}. 
Since the matrix elements in Eqs.~\eqref{eq:qgcol} 
and \eqref{eq:qgsoft} have no loop corrections, 
the expression in \Eqn{m1qg2matrixelementresult} 
is valid to all orders in $\alpha_s$. 
  
\begin{figure}[t]
	\begin{centering}
		\hskip-1cm
		\includegraphics[width=0.32\textwidth]{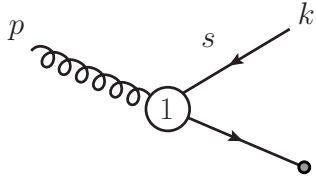}
		\par\end{centering}
	\caption{\label{fig:qgtree} 
	Tree-level EFT diagram in the $g\bar{q}$-channel.}
\end{figure}  
Having obtained an expression for the right-hand 
side of the matching equation~(\ref{m1qg2matrixelement}), 
we now compute the left-hand side, which 
corresponds to the diagram in Fig.~\ref{fig:qgtree}.
The Lagrangian insertion in \Eqn{eqn:matching4.1qg} 
can be computed by means of the Feynman rule given 
in Eq.~(A.36) of \cite{Beneke:2018rbh} and we obtain
\begin{eqnarray}\label{eq:qgtree}
\langle q^+_{b}(k)|  
\mathcal{T}_{\gamma f}(n_+q)
|g^B(p)\rangle &=& (2\pi) \delta(n_+q-n_+p)
\frac{1}{n_-k}\, g_s \mathbf{T}^B_{fb} \nonumber \\ 
&&\hspace{-1cm}\times \,
\left(\frac{\slashed{n}_{-}}{2}
\gamma_{\perp{\eta}}\right)_{\gamma\alpha}
\, v_{\alpha}(k) \sqrt{Z_{g,c}}\rvert_{{\rm{tree}}}
\epsilon^{\eta}_{\perp}(p) + \mathcal{O}(\alpha_s)\,.
\end{eqnarray}
The tree-level value of the on-shell wave 
function renormalization factor of the gluon 
field is $\sqrt{Z_{g,c}}\rvert_{{\rm{tree}}}=1$ 
in the EFT. 
  
Since at next-to-leading power in the $g\bar{q}$-channel 
only a single soft structure is relevant, no additional 
manipulations relating to the use of equation-of-motion 
identity, or on-shell and transversality conditions are 
necessary here, in contrast to the calculation of
the collinear functions for the $q\bar{q}$-channel, 
see section~4.1 of \cite{Beneke:2019oqx}. At this point, 
we simply compare the result for the left-hand side of 
the matching equation given in \Eqn{eq:qgtree} to the 
right-hand side in \Eqn{m1qg2matrixelementresult} and 
read off the tree-level result for the collinear function
\begin{eqnarray}\label{eq:qgtreecollfunc}
{G}^{\,\eta,B}_{\xi q;\gamma\alpha,fb}
\left(n_+q,n_+p; \omega \right) = 
-\,\delta(n_+q-n_+p)\,\mathbf{T}^B_{fb}\,
\left(\frac{\slashed{n}_{-}}{2}
\gamma_{\perp{\eta}}\right)_{\gamma\alpha}
+\mathcal{O}(\alpha_s)\,.
\end{eqnarray}
Using the decomposition introduced in 
\Eqn{collFunction4} we can extract the scalar 
collinear function ${G}_{\xi q}(n_+p;\omega)$
which appears in the factorization formula in 
\eqref{eq:QGnlpfact}. Namely, we find 
\begin{eqnarray}\label{eq:qgtreecollfuncscalar}
{G}^{(0)}_{\xi q}(n_+p;\omega) =-\frac{1}{2}\,,
\end{eqnarray}
where the superscript $(0)$ denotes the 
tree-level result.

\subsection{One-loop collinear function}
\label{sec:gqoneloopcoll}

\begin{figure}[t]
\begin{centering}	
\subcaptionbox{}{
\includegraphics[width=0.30\textwidth]{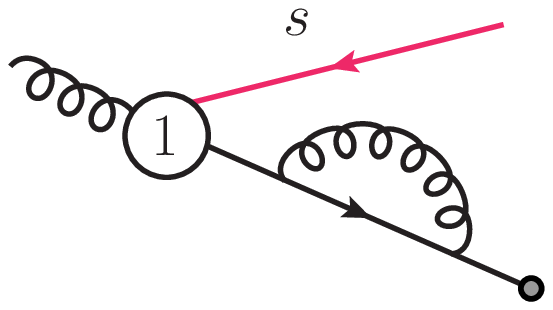}}
\subcaptionbox{}{
 \includegraphics[width=0.30\textwidth]{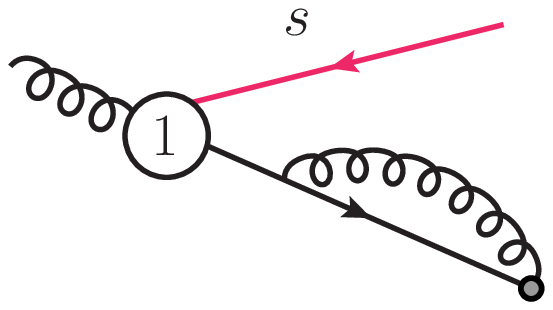}}
\subcaptionbox{}{
\includegraphics[width=0.30\textwidth]{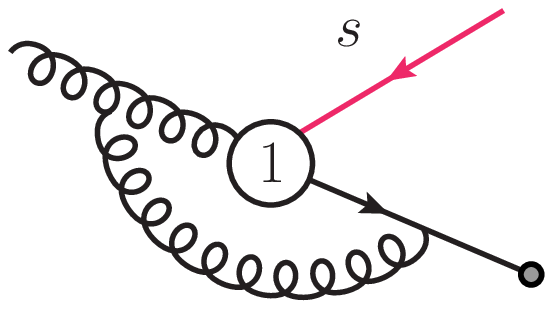}}
\subcaptionbox{}{
\includegraphics[width=0.30\textwidth]{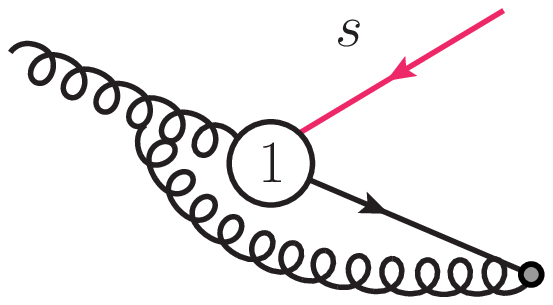}}
\subcaptionbox{}{
\includegraphics[width=0.30\textwidth]{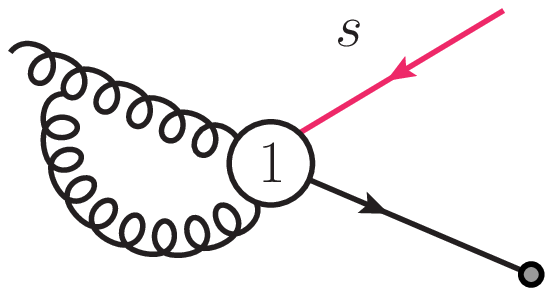}}
\subcaptionbox{}{
\includegraphics[width=0.30\textwidth]{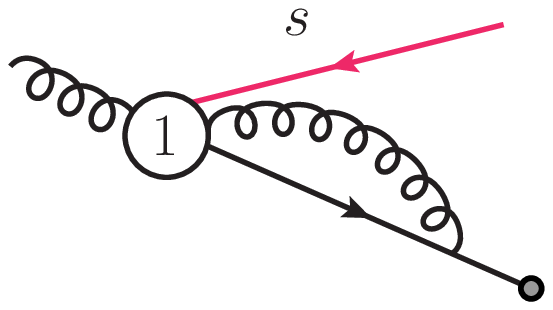}}
\subcaptionbox{}{
\includegraphics[width=0.30\textwidth]{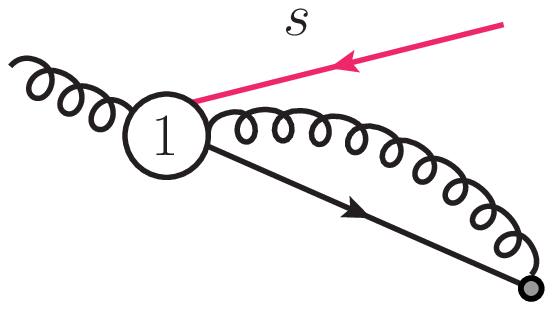}}
\subcaptionbox{}{
\includegraphics[width=0.30\textwidth]{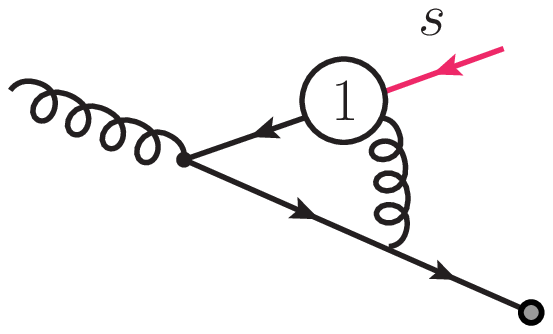}}
\subcaptionbox{}{
\includegraphics[width=0.30\textwidth]{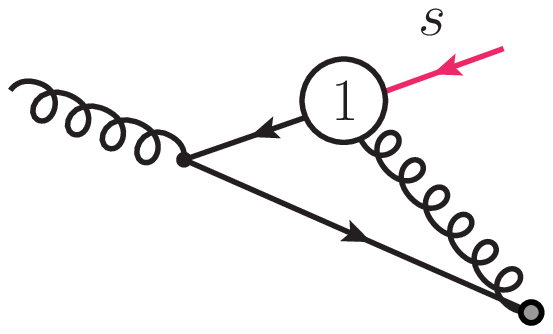}}
\par\end{centering}	
\caption{\label{fig:collinearoneloop} 
One-loop diagrams contributing to the collinear function.  
}\end{figure}
We are now ready to consider the 
$\mathcal{O}(\alpha_s)$ correction. 
For the right-hand side of the matching we 
use \Eqn{m1qg2matrixelementresult}. Here the 
on-shell wave function renormalization factor 
is unity to all orders in perturbation theory: 
$\sqrt{Z_{g,{\rm{PDF}}}}=1$. This holds in 
dimensional regularization, which we use to 
treat infrared and ultraviolet 
divergences, since the loop corrections 
are scaleless. The situation is 
the same for the $\sqrt{Z_{g,c}}$ factor on 
the left-hand side of the matching equation, 
see \Eqn{eq:qgtree}. Given that the calculation 
is relatively straightforward, we skip the details 
of the computation and present the results for 
the one-loop matrix element $\langle q^+_{b}(k)| 
\mathcal{T}_{\gamma f}(n_+q)|g^B(p)\rangle$ 
on the left-hand side of the matching 
equation~\eqref{m1qg2}, which is given 
by the sum over the one loop diagrams 
represented in Fig. 
\ref{fig:collinearoneloop} (a) -- (i). 
Let us only notice that diagram (e), (g) 
and (i) are zero, the latter because 
the collinear loop does not have a scale.
Furthermore, diagrams (a) and (b) contribute 
with a color factor $C_F$, (c) and (d)
with a color factor $(-C_A/2)$, (h) and (f)
are proportional to $(C_F-C_A/2)$. We obtain 
\begin{eqnarray}\label{qgfinal}
\langle q^+_{b}(k)| 
\mathcal{T}_{\gamma f}(n_+q)
|g^B(p)\rangle^{(1)} &=& (2\pi) \delta(n_+q-n_+p) \,
\frac{\alpha_s}{4\pi} \, g_s\,\mathbf{T}^B_{fb}
\big( C_F - C_A \big) 
\bigg( \frac{n_+p\, n_-k}{\mu^2}\bigg)^{-\eps} 
\nonumber \\ 
&& \hspace{-3.0cm} \times \, 
\frac{e^{\eps \gamma_{E}} 
\Gamma[1+\eps]\Gamma[1-\eps]^2}{\Gamma[2-2\eps]}
\,\frac{2-4\eps - \eps^2}{\epsilon^2}\,
\frac{1}{n_-k}\left(\frac{\slashed{n}_{-}}{2}
\gamma_{\perp{\eta}}\right)_{\gamma\alpha} 
v_{\alpha}(k) \, \epsilon^{\eta}_{\perp}(p) \,,
\end{eqnarray}
where the superscript (1) on the left-hand side 
indicates that the matrix element is considered at order $\alpha_s$.
We match this result to the right-hand side 
of \Eqn{m1qg2matrixelementresult}, from which 
we extract the perturbative matching coefficient, 
the collinear function, at one-loop order
\begin{eqnarray}\label{eq:qgoneloopcollfunc}
{G}^{\,\eta,B\,(1)}_{\xi q;\gamma\alpha,fb}
\left(n_+q,n_+p; \omega\right) &=& 
-\, \frac{\alpha_s}{4\pi} \,\mathbf{T}^B_{fb}
\big(C_F - C_A \big)
\left(\frac{n_+p\, \omega}{\mu^2}\right)^{-\epsilon} 
\frac{2-4\eps-\eps^2}{\epsilon^2} \nonumber \\
&&\hspace{0.0cm}\times 
\frac{e^{\epsilon \gamma_E}\Gamma[1+\epsilon]
\Gamma[{1-\epsilon}]^2}{\Gamma[{2-2\epsilon}]}
\left(\frac{\slashed{n}_{-}}{2}
\gamma_{\perp}^{\eta}\right)_{\gamma\alpha} 
\delta(n_+q-n_+p) \,.
\end{eqnarray}
Using the decomposition introduced in 
\Eqn{collFunction4}, the scalar collinear 
function ${G}_{\xi q}(n_+p;\omega)$ at 
one loop reads 
\be\label{eq:qgoneloopcollfuncscalar}
{G}^{(1)}_{\xi q}\left(n_+p; \omega \right) 
= -\, \frac{\alpha_s}{4\pi} \big(C_F - C_A \big)
\left(\frac{n_+p\, \omega}{\mu^2}\right)^{-\epsilon} 
\frac{2-4\eps-\eps^2}{2\epsilon^2}
\frac{e^{\epsilon \gamma_E}\Gamma[1+\epsilon]
\Gamma[{1-\epsilon}]^2}{\Gamma[{2-2\epsilon}]} \,.
\ee
This is the $\mathcal{O}(\alpha_s)$ correction 
to the tree-level collinear function presented in 
\Eqn{eq:qgtreecollfuncscalar}, valid to all orders 
in $\epsilon$. As anticipated it agrees with the 
result given in \cite{Liu:2021mac}, see in particular 
Eqs. (1.4), (2.1) and (2.3) there, apart from
an overall normalization factor $\propto 1/\omega$,
that by definition we shuffle into the soft 
function such as to define a dimensionless 
collinear function. Expanding 
\Eqn{eq:qgoneloopcollfuncscalar} in 
$\epsilon=(4-d)/2$ we arrive at
\begin{eqnarray}
\label{eq:qgoneloopcollfuncExpandedscalar} \nonumber
{G}^{(1)}_{\xi q}\left(n_+p; \omega\right) 
&=&-\,\frac{\alpha_s}{4\pi} \big(C_F - C_A \big)
\bigg[\frac{1}{\epsilon ^2} -\frac{1}{\epsilon}
\ln \left(\frac{n_+p \,\omega}{\mu^2}\right) \\
&&\hspace{3.0cm}-\,\frac{1}{2} -\frac{\pi^2}{12}
+ \frac{1}{2}\ln^2\left(\frac{n_+p\, \omega}{\mu^2}\right) 
+\mathcal{O}(\epsilon) \bigg]\,.
\end{eqnarray}
It is interesting to compare the above result 
for the collinear function appearing in the 
$g\bar{q}$-channel to the collinear functions 
in the $q\bar{q}$-channel, which have been 
given in their expanded form in equations
(4.31) and (4.34) of \cite{Beneke:2019oqx}.
We note that here, in contrast to $J_1^{(1)}$ 
and $J_6^{(1)}$, the collinear function 
${G}^{(1)}_{\xi q}$ exhibits $1/\epsilon^2$ poles, 
and finite logarithms  $\propto \,
\ln^2(n_+p\,\omega/\mu^2)$. 
At cross-section level, as we will show in 
section~\ref{sec:Validation} below, these 
correspond to leading logarithmic contributions 
appearing in the collinear sector. This fact  
complicates the adaptation of the resummation 
treatment developed for the $q\bar{q}$-channel
\cite{Beneke:2018gvs} to the off-diagonal 
$g\bar{q}$-channel. We provide additional 
details in section~\ref{sec:Validation}. 
It is noteworthy that the leading pole 
structure in the above equation is 
proportional to $C_F-C_A$. As discussed 
in several instances in literature, cf.~\cite{Vogt:2010cv,Liu:2017vkm,Liu:2018czl,Moult:2019uhz,Beneke:2020ibj,Beneke:2022obx}, this structure characterizes the 
conversion of a collinear gluon into a 
collinear quark by means of an emission 
of a soft-quark; its factorization into 
soft and collinear parts typically 
involves divergent convolution integrals.

\section{Soft functions}
\label{sec:SoftFunctions}

The last ingredient needed to reproduce the 
partonic cross section up to NNLO in perturbation 
theory are the one and two-loop soft function defined in
\Eqn{softoperator2}, that we calculate in this 
section. To this end it is useful to insert a 
complete set of states between the ${ \bar T}$ 
and ${ T}$ products, and use the momentum 
operator to translate the fields in the 
anti-time-ordered part. Performing the 
integration over $x^0$ in \Eqn{softoperator2} 
gives rise to an energy conserving delta function. 
Writing spinor and color indices explicitly one has 
\begin{eqnarray}\label{softoperator_scalar} \nonumber
S_{g\bar{q}}(\Omega,\omega, \omega') &=& 
\sum_{X_s} \int \frac{dz_{-}}{2\pi} \int \frac{dz'_{-}}{2\pi}
e^{-i z_{-}\omega}e^{+iz'_{-}\omega'}
\delta(\Omega - 2E_{X_s}) \\ \nonumber
&& \hspace{0.0cm}\times \, \frac{1}{N_c^2 -1}
\langle 0| \bar T \Big(
\frac{g_s}{in_-\partial_{z'}}\bar{q}^{\,+}_{\sigma k}(z'_{-})
(\T^{D})_{kl} \big\{Y^{\dagger}_{+}(0)Y_{-}(0)\big\}_{la}\Big) 
| X_s \rangle
\\&& \hspace{0.0cm}\times\,
\frac{\slashed{n}_{-\sigma\beta}}{4}  \langle X_s |
T \Big(\big\{Y^{\dagger}_{-}(0)Y_{+}(0)\big\}_{af}
(\T^{D})_{fj}\frac{g_s}{in_-\partial_{z}} 
q^{+}_{\beta j}(z_{-}) \Big) |0 \rangle.
\end{eqnarray}
In what follows we evaluate the soft function at 
first and second order in $\als$, which is indicated 
below by a superscript $(1)$ and $(2)$, respectively.

\subsection{First order}

\begin{figure}[t]
\begin{centering}
	\hskip-1cm 
	\includegraphics[width=0.32\textwidth]{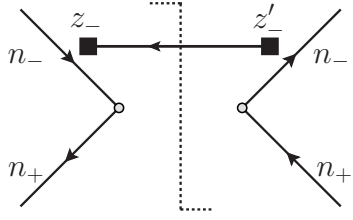}
	\par
\end{centering}
\label{fig:qg9}
\caption{Diagram contributing to the soft 
	function at first order (NLO in $\alpha_s$).}
\end{figure}  
At one loop the final state contains 
a single soft quark and the matrix 
element reads
\be\label{softoperator_scalar_2}
\langle q_{s,e}(k) |
T \Big(\big\{Y^{\dagger}_{-}(0)Y_{+}(0)\big\}_{af}
(\T^{D})_{fj} \frac{g_s}{in_-\partial_{z}} 
q^{+}_{\beta j}(z_{-}) \Big) |0 \rangle 
= \frac{g_s(\T^{D})_{ae}}{-n_- k} \, 
v_{s,\beta}(k) \, e^{i  z_-\cdot k} +\mathcal{O}(g_s^3),
\ee
The complex-conjugate matrix element 
can be derived accordingly. Inserting 
these into \Eqn{softoperator_scalar}
and introducing the notation
\be
\int [dk] \equiv 
\left(\frac{\mu^2 e^{\gamma_E}}{4\pi}\right)^{\eps}
\int \frac{d^d k}{(2\pi)^d},
\ee
we easily get
\begin{eqnarray}\label{eq:nlo-matrix} \nonumber 
S_{g\bar{q}}^{(1)}(\Omega,\omega,\omega')
&=&  2\pi \, g_s^2 \, T_F   
\int[dk] \, \frac{\delta^+(k^2)}{n_-k}  
\delta(\omega- n_-k)\delta(\omega'- n_-k) 
\delta(\Omega-n_+k-n_-k) \\
&=& \frac{\alpha_s \, T_F}{4\pi}  
\frac{e^{\epsilon \gamma_E}}{
\Gamma[1-\epsilon]} \frac{1}{\omega} 
\bigg(\frac{\mu^{2}}{\omega\,
(\Omega-\omega)}\bigg)^{\eps}
\delta(\omega- \omega') \,
\theta(\Omega-\omega) \theta(\omega),
\end{eqnarray}
which we express in terms of the 
color factor ${\rm Tr}[\T^A \T^B] 
= T_F\,\delta^{AB} = \delta^{AB}/2$,
characteristic of the quark-gluon 
channel.

\subsection{Second order: virtual-real contribution}
\label{sec:twoloopsoftvirtualreal}

\begin{figure}[t]
\begin{centering}	
\subcaptionbox{}{
\includegraphics[width=0.32\textwidth]{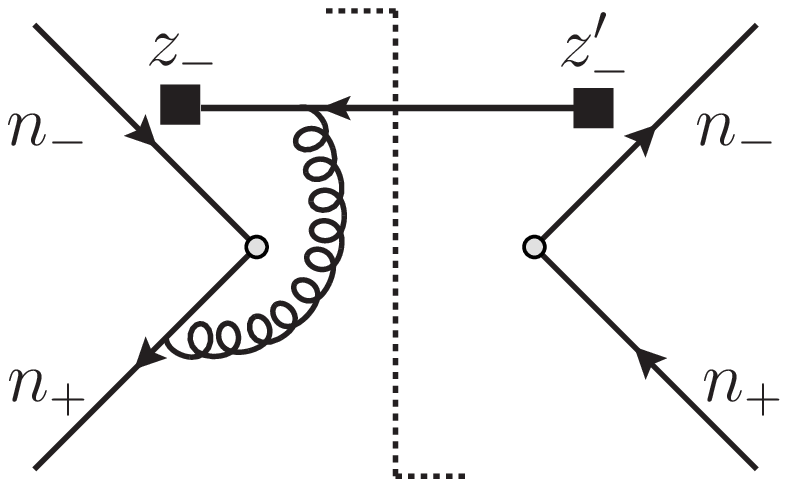}}
\subcaptionbox{}{
\includegraphics[width=0.32\textwidth]{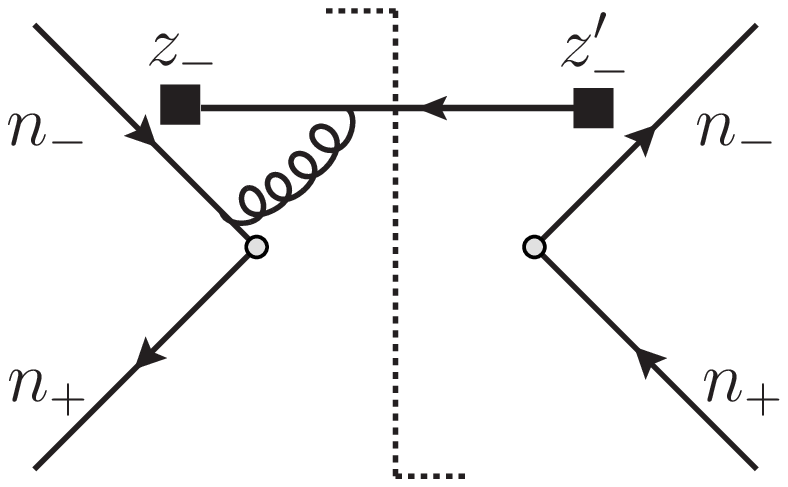}}
\subcaptionbox{}{
\includegraphics[width=0.32\textwidth]{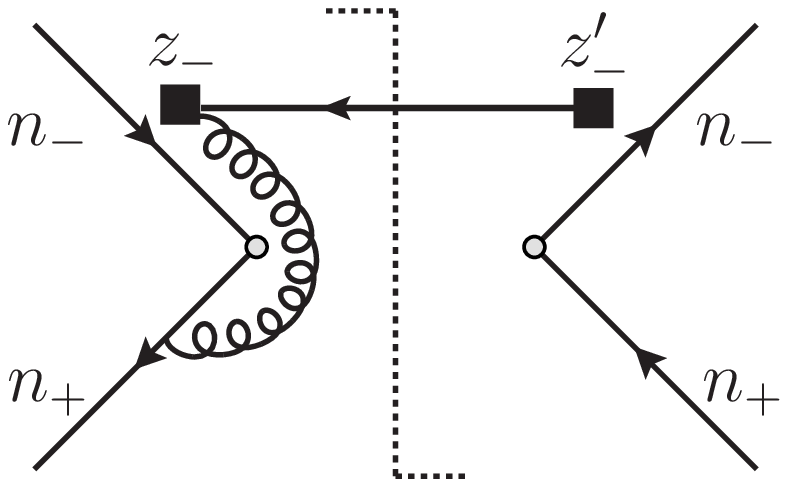}}
\subcaptionbox{}{
\includegraphics[width=0.32\textwidth]{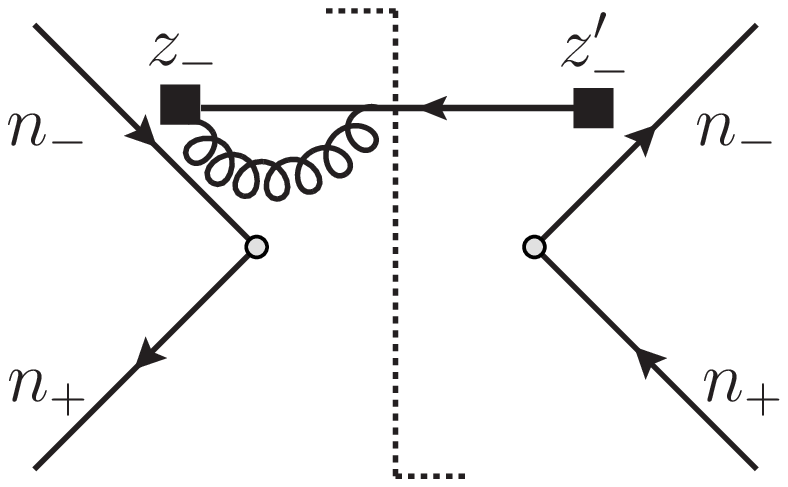}}
\subcaptionbox{}{
\includegraphics[width=0.32\textwidth]{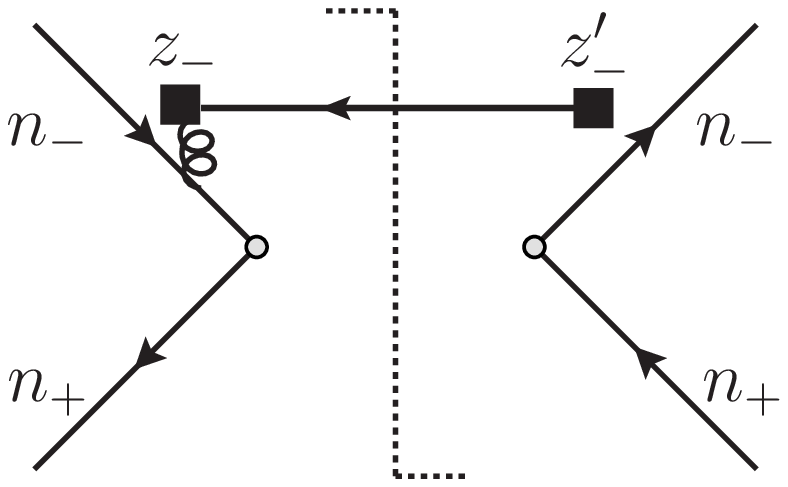}}
\subcaptionbox{}{
\includegraphics[width=0.32\textwidth]{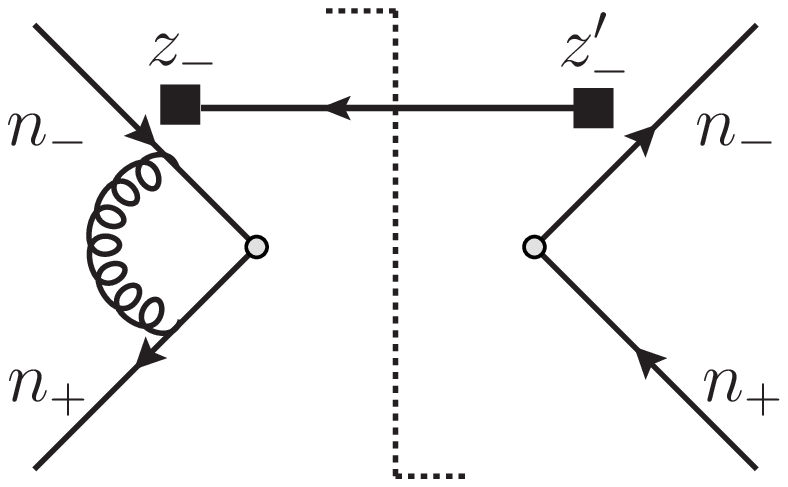}}
\par\end{centering}	
\caption{\label{fig:SQGVR}
Diagrams contributing to the virtual-real part 
of the soft function at second order in $\alpha_s$.
The part to the left (right) of the cut corresponds 
to the time-ordered (anti-time-ordered) part of the 
diagram, and lines labeled by $n_{\pm}$ with in 
(out)-going arrow correspond to soft Wilson lines 
$Y_{\mp}$($Y_{\mp}^{\dagger}$). The filled square 
in this figure stands for the soft quarks and the 
Wilson lines contained in  $q^+=   Y^{\dagger}_+q_s$.}
\end{figure}
At second order in $\alpha_s$ we need to take
into account two different types of corrections: 
the virtual-real contribution, involving a soft 
gluon loop in the matrix element (or in the 
complex conjugate one) and the real-real 
contribution, involving the emission of a 
soft gluon in addition to the soft anti-quark. 
We start by considering the virtual-real 
contribution. The one-loop matrix element 
is given by 
\bea\label{qgSoftOneEmission1L} \nonumber
\langle q_{s,e}(k_1) |
T \Big(\big\{Y^{\dagger}_{-}(0)Y_{+}(0)\big\}_{af}
(\T^{D})_{fj} \frac{g_s}{in_-\partial_z} 
q^{+}_{\beta j}(z_-) \Big) |0 \rangle && \\  \nonumber
&&\hspace{-9.0cm}=\, i \, g_s^3 \, (\T^{D})_{ae} 
\int [dk] \bigg\{ C_F \, \frac{2}{k^2 (n_+k) 
(-n_-k) (-n_- k_1)} e^{i z_-\cdot k_1 } \\ \nonumber
&&\hspace{-7.0cm}-\,\bigg( C_F - \frac{C_A}{2}\bigg) 
\frac{2}{k^2 (n_+k) 
(-n_-k) (-n_- k - n_- k_1)} e^{i z_-\cdot k_1 } \\ \nonumber
&&\hspace{-7.0cm}-\,\bigg( C_F - \frac{C_A}{2}\bigg) 
\frac{(\slashed{k} + \slashed{k}_1)\slashed{n}_+}{k^2 
(k+k_1)^2 (n_-k) (-n_- k - n_- k_1)} e^{i z_-\cdot(k_1+k) }  
\\ \nonumber
&&\hspace{-7.0cm}+\,\bigg( C_F - \frac{C_A}{2}\bigg) 
\frac{(\slashed{k} + \slashed{k}_1)\slashed{n}_-}{k^2 
(k+k_1)^2 (n_-k) (-n_- k - n_- k_1)} e^{i z_-\cdot(k_1+k) } \\ 
&&\hspace{-6.0cm}-\,  C_F \, 
\frac{(\slashed{k} + \slashed{k}_1)\slashed{n}_-}{k^2 
(k+k_1)^2 (n_- k)(-n_-k_1)} e^{iz_-\cdot k_1 }
\bigg\} \,v_{s,\beta}(k_1) \, +\mathcal{O}(g_s^5),
\eea
where linear propagators are written such 
that they carry a small positive imaginary 
factor $+i 0_+$, and the terms on the right-hand side 
represent respectively Fig. \ref{fig:SQGVR} 
(f), (c), (a), (b), (d), while (e) is immediately 
zero. Among these, it turns out that only diagram 
(c) contributes. We focus on this term and insert
the matrix element in \Eqn{softoperator_scalar} 
with the complex conjugate matrix element taken 
at lowest order; furthermore, we recall that 
we need to take into account also the complex 
conjugate one-loop matrix element times the tree 
level matrix element, such that the full 
contribution to the virtual-real part 
of the soft function reads 
\bea\label{soft-function-momentum-1r1v-NNLOb} \nonumber
S^{(2)1r1v}_{g\bar{q}}(\Omega,\omega,\omega') &=&  
{\rm Re}\bigg\{ i \, g_s^4\, T_F (2C_F - C_A)
\int [dk_1] \int [dk]  \\ \nonumber
&&\hspace{-1.5cm}\times\, 
(2\pi) \delta(k_1^2) \theta(k_1^0)\,\delta(\Omega - 2k_1^0) 
\, \delta(\omega' - n_- k_1)
\delta(\omega - n_-k - n_- k_1) \\ 
&&\hspace{-1.5cm}\times\,  
\frac{(k+k_1)^2 - k^2 + n_-k_1(n_+k + n_+ k_1)
+ n_+k_1(n_-k + n_- k_1)}{k^2(k + k_1)^2
(n_+ k) (-n_- k - n_- k_1) (-n_- k_1)}\, \bigg\}.
\eea
We evaluate the integrations over $k$ and $k_1$ 
following two independent approaches: within the
first, we integrate directly the terms appearing 
in \Eqn{soft-function-momentum-1r1v-NNLOb}; within
the second we first reduce such terms to a basis of master integrals, in which case 
\Eqn{soft-function-momentum-1r1v-NNLOb} 
reads
\begin{equation}
\label{soft-function-momentum-1r1v-NNLOc}
S^{(2)1r1v}_{g\bar{q}}(\Omega,\omega,\omega')  
= {\rm Re} \bigg\{- i \, \frac{\alpha_s^2\, T_F}{(4\pi)^2} 
(2C_F-C_A) \bigg[\frac{1}{\omega \omega'} \hat{J}_1
+\frac{(\omega+\omega')(\Omega-\omega')}{\omega \omega'} 
\hat{J}_2 \bigg] \bigg\},
\end{equation}
where the master integrals $\hat{J}_i$ are 
defined in \Eqn{eq:MISVR} and calculated in 
Eqs.~\eqref{eq:MISVR1} and \eqref{eq:MISVR2}
of appendix \ref{app-RV-MIs}. Using the results 
given there we get
\bea \nonumber
\label{soft-function-momentum-1r1v-NNLOd}
S^{(2)1r1v}_{g\bar{q}}(\Omega,\omega,\omega')  
&=& \frac{\alpha_s^2\, T_F}{(4\pi)^2} (2C_F-C_A) 
\frac{e^{2\eps \gamma_E}\,\Gamma[1+\eps]}{\eps\,
\Gamma[1-\eps]} \\ \nn
&&\hspace{-3.0cm} \times \,
{\rm Re}\bigg\{ \frac{1}{(-\omega)\omega'}
\bigg[\frac{\omega+\omega'}{\omega'}
\, _2F_1 \bigg(1,1+\eps,1-\eps,\frac{\omega}{\omega'}\bigg)
- 1 \bigg] \bigg(\frac{\mu^4}{(-\omega) 
\omega'(\Omega - \omega')^2}\bigg)^{\eps}
\theta(-\omega)   \\
&&\hspace{-2.7cm}+\, \frac{2(\omega+\omega')}{\omega\omega'(\omega'-\omega)} 
\bigg(\frac{\mu^4}{(\omega'-\omega)^2(\Omega - \omega')^2}\bigg)^{\eps}
\frac{\Gamma[1-\eps]^2}{\Gamma[1-2\eps]}\theta(\omega'-\omega)
\bigg\} \theta(\omega')\theta(\Omega-\omega').
\eea

\subsection{Second order: real-real contribution}
\label{sec:twoloopsoftrealreal}

\begin{figure}
\begin{centering}	
\subcaptionbox{}{
\includegraphics[width=0.32\textwidth]{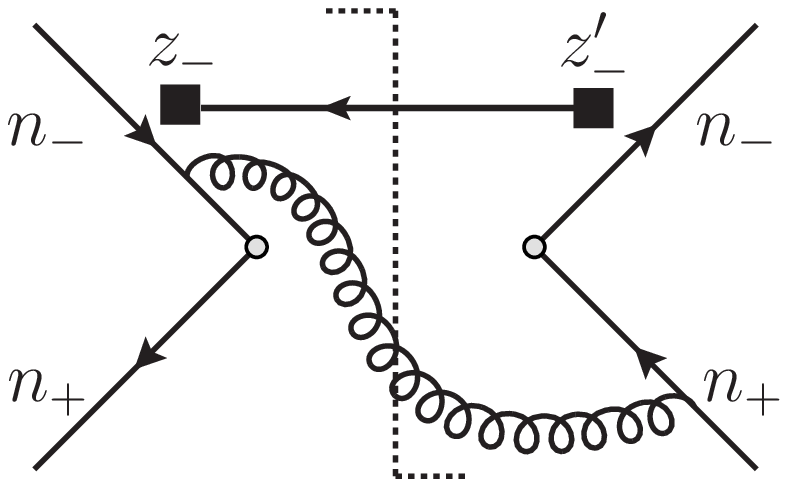} }
\subcaptionbox{}{
\includegraphics[width=0.32\textwidth]{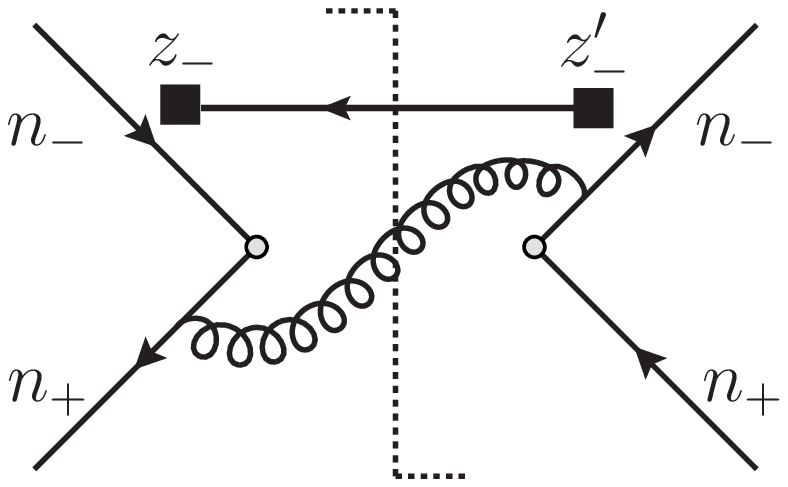}}
\subcaptionbox{}{
\includegraphics[width=0.32\textwidth]{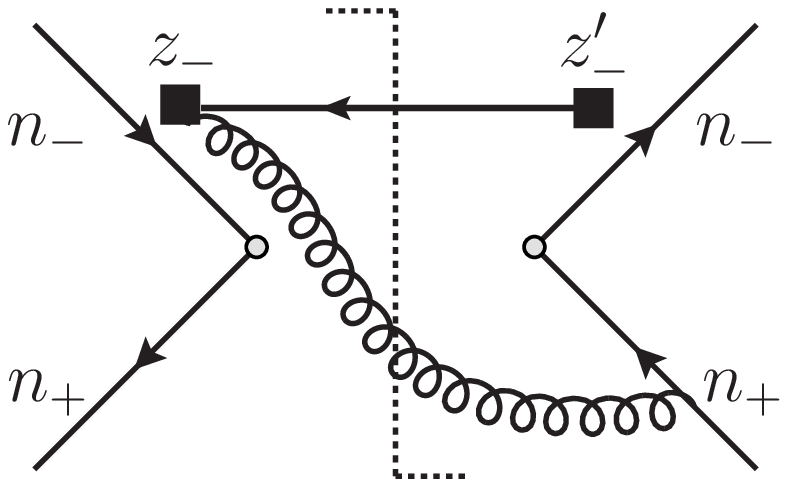}}
\subcaptionbox{}{\includegraphics[width=0.32\textwidth]{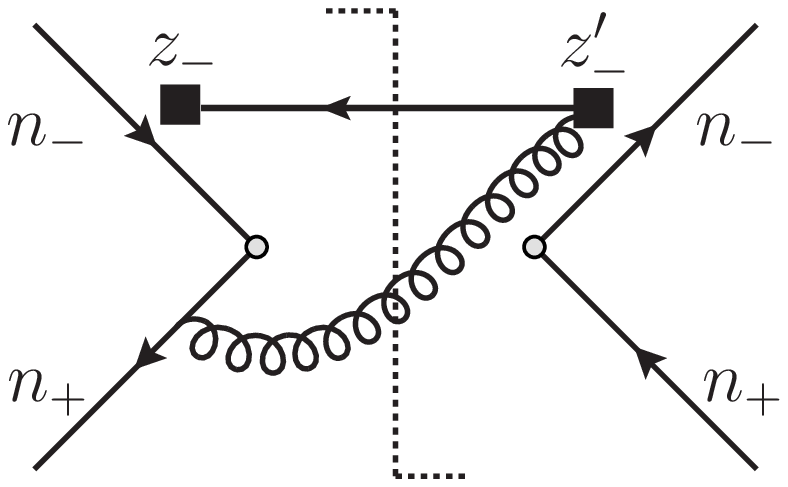}}
\subcaptionbox{}{\includegraphics[width=0.32\textwidth]{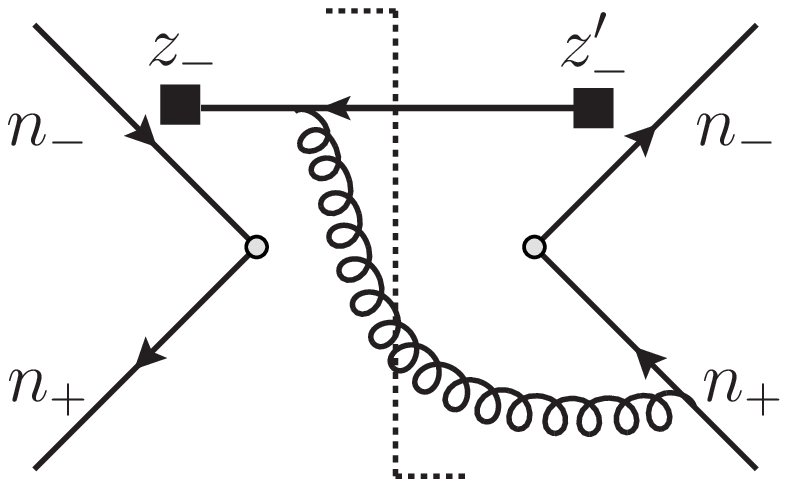}}
\subcaptionbox{}{\includegraphics[width=0.32\textwidth]{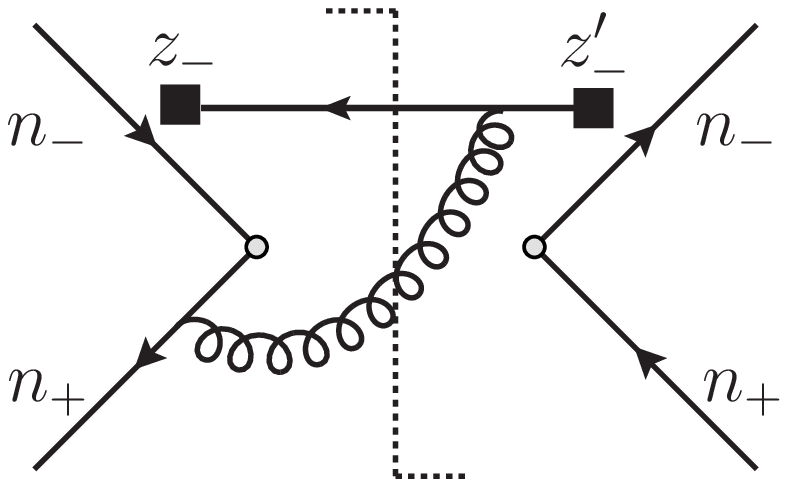}}
\subcaptionbox{}{\includegraphics[width=0.32\textwidth]{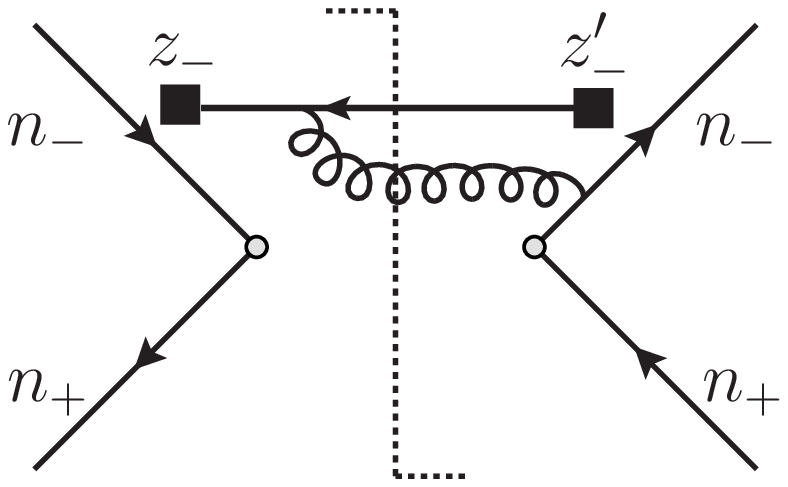}}
\subcaptionbox{}{\includegraphics[width=0.32\textwidth]{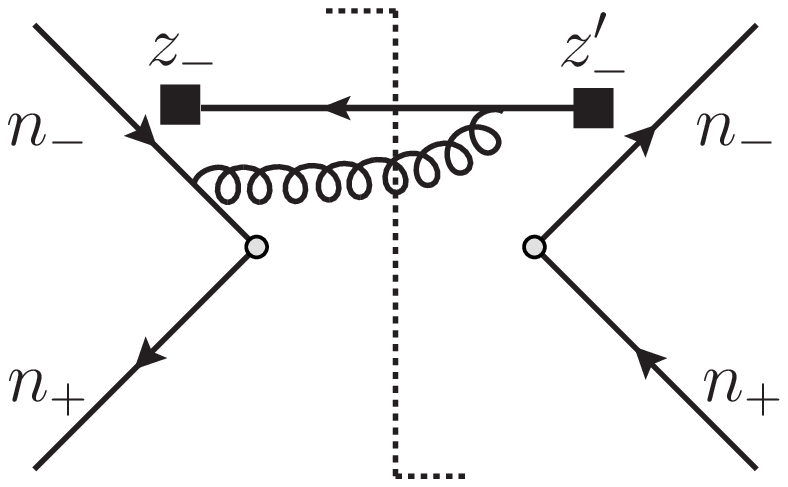}}
\subcaptionbox{}{\includegraphics[width=0.32\textwidth]{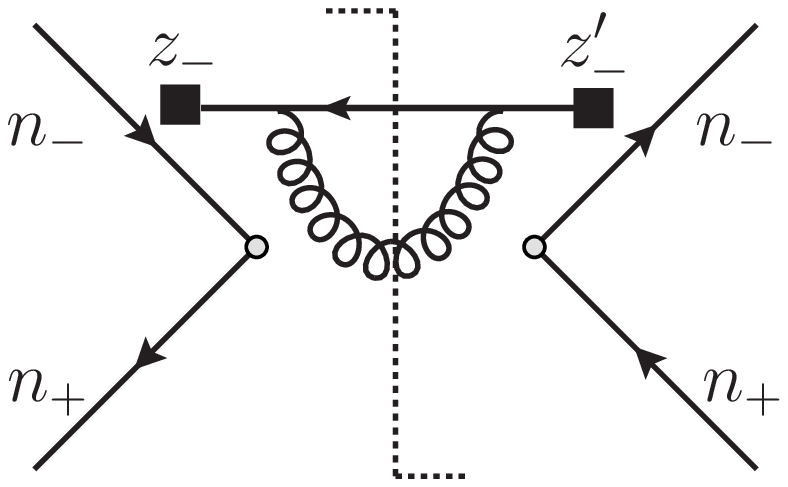}}
\par\end{centering}	\caption{\label{fig:SQG} 
Diagrams contributing to the real-real part 
of the  soft function. The part to the left 
(right) of the cut corresponds to the 
time-ordered (anti-time-ordered) part of the 
diagram, and lines labeled by $n_{\pm}$ with 
in (out)-going arrow correspond to soft Wilson 
lines $Y_{\mp}$($Y_{\mp}^{\dagger}$). The filled 
square in this figure stands for the soft quarks 
and the Wilson lines contained in  
$q^+ = Y^{\dagger}_+q_s$.
}\end{figure}
We consider now the final state emission of a soft 
gluon in addition to the soft-antiquark. 
The matrix element reads
\bea\label{qgSoftTwoEmissions} \nonumber
\langle q_{s,e}(k_1) g^{H}_{s}(k_2) |
T \bigg(\big\{Y^{\dagger}_{-}(0)Y_{+}(0)\big\}_{af}
(\T^{D})_{fj} \frac{g_s}{in_-\partial_z} 
q^{+}_{\beta j}(z_-) \bigg) |0\rangle && \\ 
&&\hspace{-11.0cm}=\, 
g_s^2 \bigg[ (\T^H \T^D)_{ae} \frac{1}{n_- k_1} 
\bigg(\frac{n_+ \varepsilon(k_2)}{n_+ k_2} 
- \frac{n_- \varepsilon(k_2)}{n_- k_2} \bigg) 
e^{i \, z_- \cdot k_1} \\ \nonumber
&&\hspace{-11.0cm}+\, 
(\T^D \T^H)_{ae} \frac{1}{n_-(k_1+k_2)} 
\bigg(\frac{n_+ \varepsilon(k_2)}{n_- k_2} 
- \frac{(\slashed{k_1}+\slashed{k_2})
\slashed{\varepsilon}(k_2)}{(k_1 + k_2)^2}\bigg) 
e^{i \, z_- \cdot (k_1+k_2)}\bigg] v_{s,\beta}(k_1) + \mathcal{O}(g_s^4).
\eea
In practice we have four contributions: three 
diagrams where the gluon is taken from one of 
the three Wilson lines in \Eqn{qgSoftTwoEmissions},
counting also the Wilson line implicit in 
$q^+ = Y^{\dagger}_+q_s$, and the fourth 
where the gluon is taken from the insertion 
of the (leading power) soft Lagrangian 
${\cal L}^{(0)}_{s} (y) = 
\bar q_s(y) g_s \slashed{A}_s (y) q_s(y)$.
The soft function is obtained according to
\Eqn{softoperator_scalar} where the complex 
conjugate matrix element can be derived from 
\Eqn{qgSoftTwoEmissions}. In Fig. \ref{fig:SQG} 
we list the diagrams which give a non-zero 
contribution. We obtain
\bea\label{soft-function-momentum-2real} \nn
S^{(2)2r0v}_{g\bar{q}}(\Omega,\omega,\omega') 
&=& (2\pi)^2 g_s^4\, T_F \,
\int [dk_1] \int [dk_2]\, 
\delta(\Omega - 2k_1^0- 2k_2^0) \\ \nn
&&\hspace{-3.0cm}\times\, 
\delta(k_1^2) \theta(k_1^0) \, 
\delta(k_2^2) \theta(k_2^0) \,
\bigg\{ C_F \bigg[ 
-\frac{4}{n_- k_1\, n_-k_2\, n_+ k_2}
\delta(\omega-n_- k_1)\delta(\omega'-n_- k_1) \\ \nn
&&\hspace{2.0cm}
-\frac{1}{(k_1+k_2)^2} 
\bigg(\frac{4 n_- k_1}{n_-k_2 \, n_-(k_1+k_2)}
+ \frac{2(1-\eps) \, n_-k_2}{[n_-(k_1+k_2)]^2}\bigg) \\ \nn
&&\hspace{3.0cm}\times \, 
\delta(\omega-n_- k_1-n_- k_2)
\delta(\omega'-n_- k_1-n_- k_2) \bigg] \\ \nn 
&&\hspace{0.0cm}+\, 
\bigg(C_F - \frac{C_A}{2}\bigg) \bigg[ 
\frac{2}{n_- (k_1+k_2)\, n_-k_2\, n_+ k_2} 
+\frac{1}{n_-k_1 \, n_-(k_1+k_2)\, n_+k_2} \\ \nn
&&\hspace{2.0cm}
-\frac{n_+k_1}{n_-k_1\, n_+k_2\,(k_1+k_2)^2} 
-\frac{n_+(k_1 + k_2)}{n_-(k_1+k_2)\,
n_+k_2 \, (k_1+k_2)^2} \\ \nn 
&&\hspace{2.0cm} + \, 
\frac{2}{(k_1+k_2)^2 \, n_-k_2} \bigg] 
\Big[\delta(\omega-n_- k_1)
\delta(\omega'-n_- k_1-n_- k_2) \\ 
&&\hspace{3.0cm} +\,
\delta(\omega'-n_- k_1)
\delta(\omega-n_- k_1-n_- k_2) \Big]\bigg\}.
\eea
As for the virtual-real contribution, 
we evaluate the integrals over $k_1$
and $k_2$ both by directly integrating 
the expression in \Eqn{soft-function-momentum-2real},
and by reducing such expression to a 
basis of master integrals, in which case 
\Eqn{soft-function-momentum-2real} becomes 
\newpage
\begin{eqnarray}
\label{soft-function-momentum-2realb} \nonumber
S^{(2)2r0v}_{g\bar{q}}(\Omega,\omega,\omega') && 
\\ \nonumber 
&&\hspace{-2.0cm}= 
\,\frac{\alpha_s^2}{(4\pi)^2} T_F
\bigg[ C_F \bigg(\frac{(4-\epsilon)(1-\epsilon) 
(1-2 \epsilon)}{\epsilon^2 \omega^2  
(\Omega -\omega)}\, \hat{I}_6
+\frac{4 (2-3\epsilon)(1-3\epsilon)}{\epsilon^2 
\omega (\Omega -\omega)^2} \, \hat{I}_5 \bigg) 
\delta(\omega - \omega') \\ \nonumber 
&& \hspace{-0.5cm}+\, (C_A-2 C_F) 
\bigg( \frac{(1-2 \epsilon)  
(\omega+\omega')}{\epsilon\,
\omega \omega' (\Omega -\omega)
(\omega-\omega')} \,\hat{I}_3
- \frac{(1-2 \epsilon)
(\omega+\omega')}{\epsilon\,
\omega \omega' (\Omega -\omega')
(\omega-\omega')} \,\hat{I}_1 \\ 
&&\hspace{2.0cm}
+\, \frac{(\Omega -\omega)
(\omega+\omega')}{2 \omega \omega'} \hat{I}_4
+ \frac{(\Omega -\omega')
(\omega+\omega')}{2 \omega \omega'}\hat{I}_2
\bigg)\bigg].
\end{eqnarray}
The master integrals $\hat{I}_i$ are defined 
and evaluated in appendix \ref{app-RR-MIs}.
After substituting the expressions for the 
master integrals in \Eqnss{eq:MISRR1}{eq:MISRR6} 
we finally obtain 
\begin{eqnarray} 
\label{soft-function-momentum-2realc} \nonumber 
S^{(2)2r0v}_{g\bar{q}}(\Omega,\omega,\omega') 
&=& \frac{\alpha_s^2 T_F}{(4\pi)^2} 
\bigg\{ C_F \frac{e^{2 \epsilon \gamma_E} 
\Gamma[1-\eps]}{\eps^2} \frac{1}{\omega}
\bigg[\frac{4}{\Gamma[1-3\eps]}
\bigg(\frac{\mu^4}{\omega(\Omega-\omega)^3}\bigg)^{\eps} 
\\ \nonumber
&& \hspace{0.0cm}
+\,\frac{(4-\eps)\Gamma[2-\eps]}{(1-2\eps)\Gamma[1-2\eps]^2}
\bigg(\frac{\mu^4}{\omega^2(\Omega-\omega)^2}\bigg)^{\eps}
\bigg] \delta(\omega - \omega')
\theta(\Omega-\omega)\theta(\omega) \\ 
&& \hspace{-1.0cm}
+\,\left(C_A - 2 C_F\right) 
\frac{2e^{2 \epsilon \gamma_E}}{
\eps \Gamma[1-2\eps]} \frac{\omega
+\omega'}{\omega \omega'(\omega'-\omega)}
\bigg(\frac{\mu^4}{\omega(\omega'-\omega)
(\Omega-\omega')^2}\bigg)^{\eps} \\ \nonumber 
&& \hspace{0.0cm} \times \, 
\bigg[ \,_2F_1\Big(1,-\epsilon,1-\epsilon, 
\frac{\omega}{\omega-\omega'} \Big) -1 \bigg] 
\theta(\omega)\theta(\omega')
\theta(\omega'-\omega) \theta(\Omega-\omega')\bigg\}.
\end{eqnarray}
\Eqns{soft-function-momentum-1r1v-NNLOd}{soft-function-momentum-2realc}
provide the complete result for the two-loop soft function.


\subsection{Asymptotic limits}
\label{sec:endpoint}

With the results of the NNLO soft function calculation 
at hand, we now have the opportunity to study the 
asymptotic limits of this object.  
This is useful, because endpoint divergences arise 
exactly in these limits, thus understanding the 
emergent asymptotic structure of the soft function 
is useful for any future study of endpoint 
refactorization. Indeed, as shown for instance 
for the off-diagonal ``gluon'' thrust in 
\cite{Beneke:2022obx}, the asymptotic limits of 
the soft (and collinear) functions are the objects 
one needs to subtract from the full functions, in 
order to define matrix elements free of endpoint 
divergences. After this procedure has been completed, 
it is then possible to implement the standard resummation 
procedure, based on the renormalization-group evolution 
of the subtracted functions. 

At NLO the soft function 
is relatively simple, and inspecting 
\Eqn{eq:nlo-matrix} we see that an endpoint 
divergence arises for $\omega, \omega' \to 0$: 
\be
S(\Omega, \omega, \omega',\mu)|_{\omega,\omega' \to 0}
= \frac{\alpha_s \, T_F}{4\pi}  
\frac{e^{\epsilon \gamma_E}}{
	\Gamma[1-\epsilon]} \frac{1}{\omega} 
\bigg(\frac{\mu^{2}}{\omega\,\Omega}\bigg)^{\eps}
\delta(\omega- \omega') \,
\theta(\Omega) \theta(\omega).
\ee
In particular, notice that, because of the 
factor $\delta(\omega- \omega')$, one has 
effectively a dependence on a single variable 
$\omega$. This is analogous to the situation 
in \cite{Beneke:2022obx} where the relevant 
asymptotic limit is $\omega,\omega' \to \infty$, 
however identical simplification occurs, namely, 
the presence of a Dirac Delta function reduces 
the functional dependence of the soft function 
to a single $\omega$ variable (see e.g. Eq. (27) 
in \cite{Beneke:2022obx}).

\begin{figure}[t]
\centering
\includegraphics[width=0.42\textwidth]{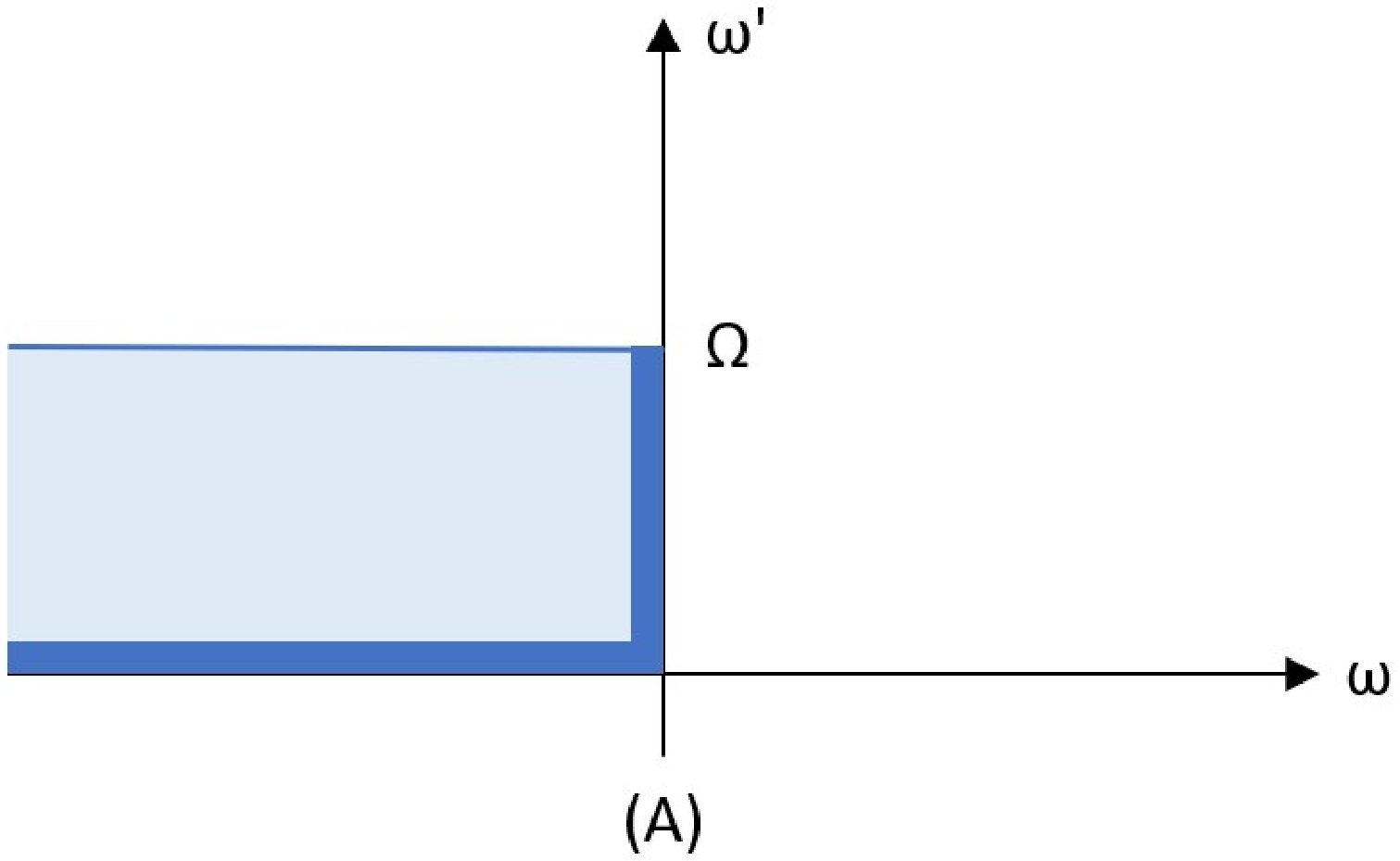} \qquad
\includegraphics[width=0.42\textwidth]{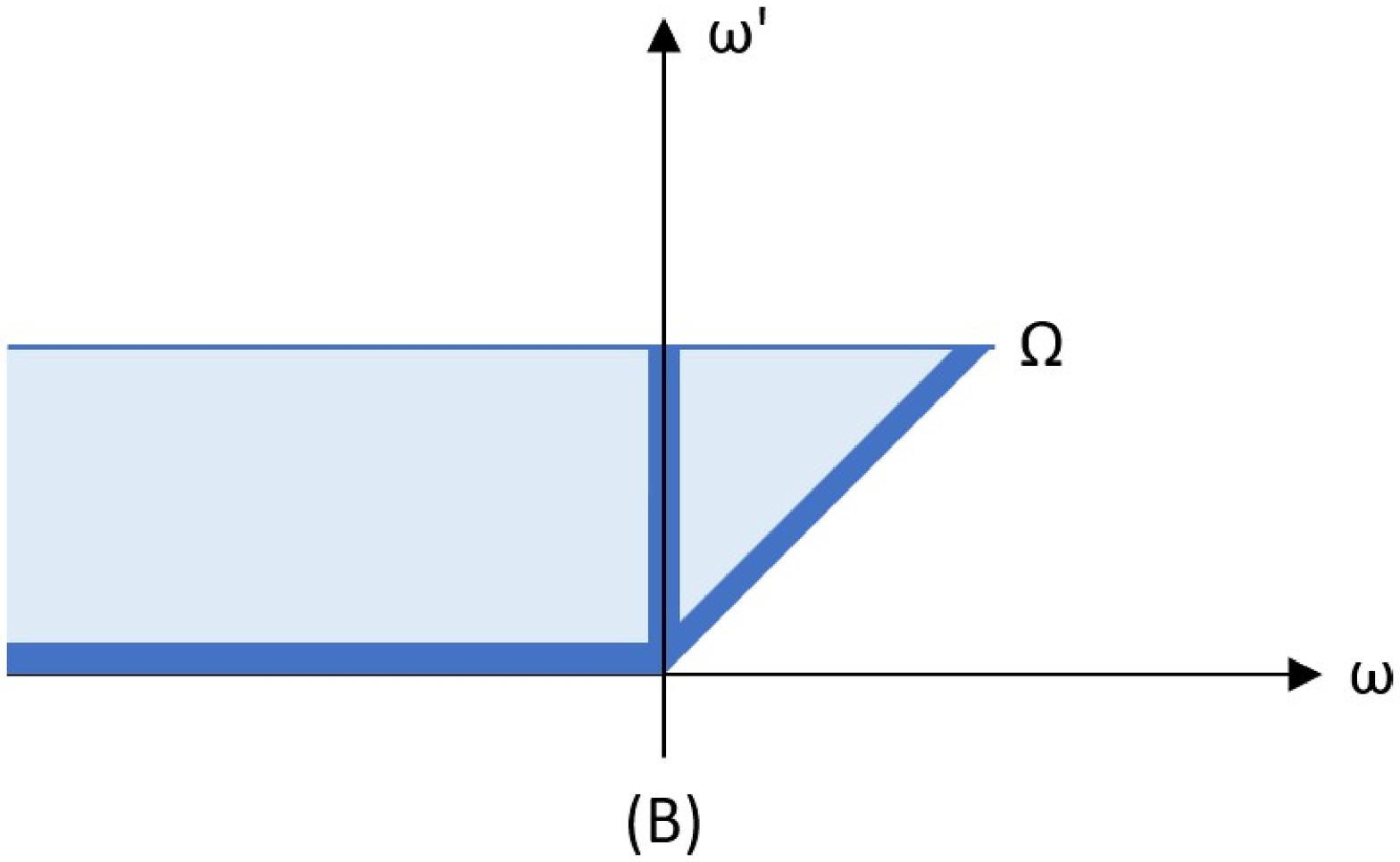} \\[0.2cm]
\includegraphics[width=0.42\textwidth]{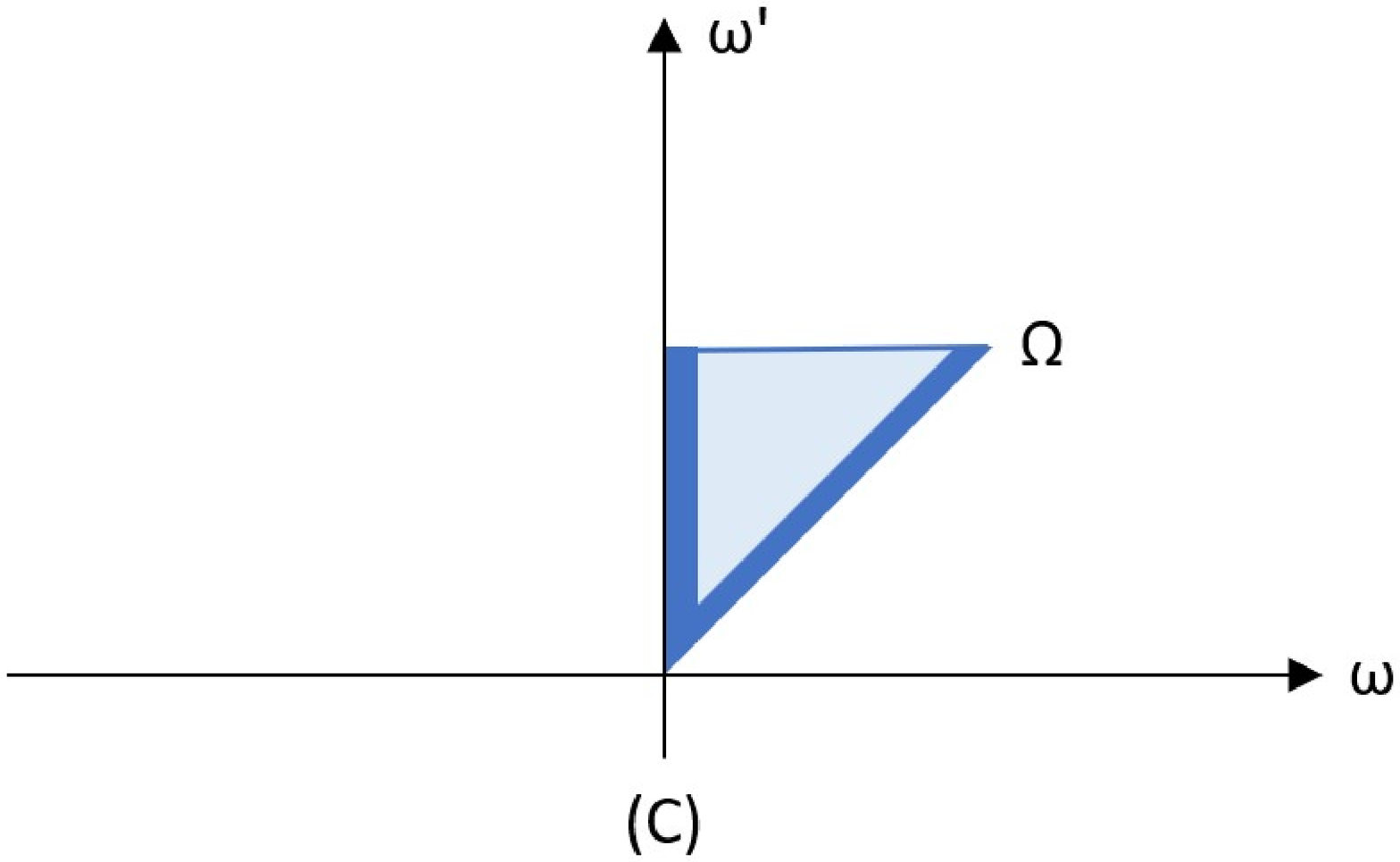}
\caption{Integration domain for the 
	three terms of the two-loop 
	soft functions, defined (from 
	left to right) respectively as 
	$S^{(2A)}(\omega, \omega') $, 
	$S^{(2B)}(\omega, \omega') $ and 
	$S^{(2C)}(\omega, \omega') $ in 
	\Eqn{Stwoloopintdomain}. The shaded blue
	areas represent endpoint regions where 
	the integral of the soft function is 
	divergent.}
\label{fig-domains}
\end{figure}
The calculation of the soft function 
at two loops in sections \ref{sec:twoloopsoftvirtualreal}, \ref{sec:twoloopsoftrealreal} gives us the 
opportunity to explore its asymptotic limits 
for the first time beyond NLO. Indeed, we 
find a more involved structure of endpoint 
singularities, which arise for $\omega \to 0$, 
$\omega' \to 0$ separately, as well as for 
$\omega'-\omega \to0$. To be more specific, 
let us write the soft function as follows: 
\bea\label{Stwoloopintdomain} \nn
S^{(2)}(\omega, \omega') &=& 
\hat S^{(2)}(\omega) \, \delta(\omega - \omega')
\, \theta(\Omega-\omega)\theta(\omega) \\ \nn
&&+\, S^{(2A)}(\omega, \omega') 
\, \theta(-\omega) \theta(\omega') 
\theta(\Omega-\omega')  \\ \nn
&&+\, S^{(2B)}(\omega, \omega') \, 
\theta(\omega'-\omega)
\theta(\omega')\theta(\Omega-\omega') \\
&&+\, S^{(2C)}(\omega, \omega') \, 
\theta(\omega)\theta(\omega')
\theta(\omega'-\omega) 
\theta(\Omega-\omega'), 
\eea
where for simplicity we have dropped 
the dependence on $\Omega$ and the 
scale $\mu$. In \Eqn{Stwoloopintdomain}
we classify the various contributions 
in terms of their domain of integration.
The term $\hat S^{(2)}(\omega)$ corresponds 
to the first contribution of the double-real 
correction in \Eqn{soft-function-momentum-2realc}:
\bea \nn
\hat S^{(2)}(\omega) &=&
\frac{\alpha_s^2 T_F}{(4\pi)^2}\,
C_F \, \frac{e^{2 \epsilon \gamma_E} 
	\Gamma[1-\eps]}{\eps^2} \frac{1}{\omega}
\bigg[\frac{4}{\Gamma[1-3\eps]}
\bigg(\frac{\mu^4}{\omega(\Omega-\omega)^3}\bigg)^{\eps} \\ 
&& \hspace{2.0cm}
+\,\frac{(4-\eps)\Gamma[2-\eps]}{(1-2\eps)\Gamma[1-2\eps]^2}
\bigg(\frac{\mu^4}{\omega^2(\Omega-\omega)^2}\bigg)^{\eps}
\bigg],
\eea
and the domain of integration for this term is the same
as the one of the NLO soft function. We  then have
$S^{(2A)}(\omega, \omega')$ and 
$S^{(2B)}(\omega, \omega')$, which 
correspond respectively to the first 
and second term of the virtual-real 
contribution in \Eqn{soft-function-momentum-1r1v-NNLOd}:
\bea \nonumber
\label{soft-function-momentum-1r1v-A}
S^{(2A)}(\omega, \omega')  
&=& \frac{\alpha_s^2\, T_F}{(4\pi)^2} (2C_F-C_A) 
\frac{e^{2\eps \gamma_E}\,\Gamma[1+\eps]}{\eps\,
	\Gamma[1-\eps]} \\ \nn
&&\hspace{-2.0cm} \times \,
{\rm Re}\bigg\{ \frac{1}{(-\omega)\omega'}
\bigg[\frac{\omega+\omega'}{\omega'}
\, _2F_1 \bigg(1,1+\eps,1-\eps,\frac{\omega}{\omega'}\bigg)
- 1 \bigg] 
\bigg(\frac{\mu^4}{(-\omega) 
	\omega'(\Omega - \omega')^2}\bigg)^{\eps} \bigg\}, \\ \nn
\label{soft-function-momentum-1r1v-B}
S^{(2B)}(\omega, \omega')  
&=& \frac{\alpha_s^2\, T_F}{(4\pi)^2} (2C_F-C_A) 
\frac{e^{2\eps \gamma_E}\,\Gamma[1+\eps]}{\eps\,
	\Gamma[1-\eps]} \\ 
&&\hspace{0.0cm} \times \,
{\rm Re}\bigg\{\frac{2
	(\omega+\omega')}{\omega\omega'(\omega'-\omega)}
\bigg(\frac{\mu^4}{(\omega'-\omega)^2
	(\Omega - \omega')^2}\bigg)^{\eps}
\frac{\Gamma[1-\eps]^2}{\Gamma[1-2\eps]}\bigg\}.
\eea
Last, we have the double real contribution 
of \Eqn{soft-function-momentum-2realc} 
\emph{not} proportional to $\delta(\omega 
- \omega')$:
\begin{eqnarray} 
\label{soft-function-momentum-2realcCHANGED} \nonumber 
S^{(2C)}(\omega, \omega')  
&=& \frac{\alpha_s^2\, T_F}{(4\pi)^2}
\left(2 C_F - C_A \right) 
\frac{2e^{2 \epsilon \gamma_E}}{
	\eps \Gamma[1-2\eps]} \frac{\omega
	+\omega'}{\omega \omega'(\omega'-\omega)} \\  
&& \hspace{0.0cm} \times \, 
\bigg(\frac{\mu^4}{\omega(\omega'-\omega)
	(\Omega-\omega')^2}\bigg)^{\eps}
\bigg[ 1- \,_2F_1\Big(1,-\epsilon,1-\epsilon, 
\frac{\omega}{\omega-\omega'} \Big) \bigg] .
\end{eqnarray}
The integration domains of the three
functions are represented in Fig.~\ref{fig-domains}, 
with the blue shaded areas representing 
the endpoint regions where singularities 
occur. In this regard, let us notice that 
a further divergence for $\omega \to \infty$
would be present for the factors
$S^{(2A)}(\omega,\omega')$ and 
$S^{(2B)}(\omega,\omega')$ taken separately, 
but it cancels in their sum. In the
boundary regions of Fig.~\ref{fig-domains},
taking the limit $\omega,\omega' \to 0$ with 
$\omega \sim \omega'$ we have
\bea\label{asymptotic-soft-functionNNLOdelta} \nn
\hat S^{(2)}(\omega)|_{\omega,\omega' \to 0} &=&
\frac{\alpha_s^2 T_F}{(4\pi)^2}\,
C_F \, \frac{e^{2 \epsilon \gamma_E} 
	\Gamma[1-\eps]}{\eps^2} \frac{1}{\omega}
\bigg[\frac{4}{\Gamma[1-3\eps]}
\bigg(\frac{\mu^4}{\omega\,\Omega^3}\bigg)^{\eps} \\ 
&& \hspace{3.0cm}
+\,\frac{(4-\eps)\Gamma[2-\eps]}{(1-2\eps)\Gamma[1-2\eps]^2}
\bigg(\frac{\mu^4}{\omega^2\,\Omega^2}\bigg)^{\eps}
\bigg], \\ \nn 
\label{asymptotic-soft-functionNNLOA}
S^{(2A)}(\omega, \omega')|_{\omega,\omega' \to 0}
&=& \frac{\alpha_s^2\, T_F}{(4\pi)^2} (2C_F-C_A) 
\frac{e^{2\eps \gamma_E}\,\Gamma[1+\eps]}{\eps\,
	\Gamma[1-\eps]} \\ 
&&\hspace{-2.0cm} \times \,
{\rm Re}\bigg\{ \frac{1}{(-\omega)\omega'}
\bigg[\frac{\omega+\omega'}{\omega'}
\, _2F_1 \bigg(1,1+\eps,1-\eps,\frac{\omega}{\omega'}\bigg)
- 1 \bigg] 
\bigg(\frac{\mu^4}{(-\omega) 
	\omega'\,\Omega^2}\bigg)^{\eps} \bigg\}, 
\\ \nn
\label{asymptotic-soft-functionNNLOB}
S^{(2B)}(\omega, \omega')|_{\omega,\omega' \to 0}
&=& \frac{\alpha_s^2\, T_F}{(4\pi)^2} (2C_F-C_A) 
\frac{e^{2\eps \gamma_E}\,\Gamma[1+\eps]}{\eps\,
	\Gamma[1-\eps]} \\ 
&&\hspace{0.0cm} \times \,
{\rm Re}\bigg\{\frac{2
	(\omega+\omega')}{\omega\omega'(\omega'-\omega)}
\bigg(\frac{\mu^4}{(\omega'-\omega)^2
	\Omega^2}\bigg)^{\eps}
\frac{\Gamma[1-\eps]^2}{\Gamma[1-2\eps]}\bigg\}, 
\\ \nn
\label{asymptotic-soft-functionNNLOC}
S^{(2C)}(\omega, \omega')|_{\omega,\omega' \to 0}
&=& \frac{\alpha_s^2\, T_F}{(4\pi)^2}
\left(2 C_F - C_A \right) 
\frac{2e^{2 \epsilon \gamma_E}}{
	\eps \Gamma[1-2\eps]} \frac{\omega
	+\omega'}{\omega \omega'(\omega'-\omega)} \\  
&& \hspace{0.0cm} \times \, 
\bigg(\frac{\mu^4}{\omega(\omega'-\omega)
	\Omega^2}\bigg)^{\eps}
\bigg[ 1- \,_2F_1\Big(1,-\epsilon,1-\epsilon, 
\frac{\omega}{\omega-\omega'} \Big) \bigg].
\eea
These limits provide valuable information
that will be useful to set up a refactorization  
procedure such as the one devised in 
\cite{Beneke:2022obx}. 
This analysis goes beyond the scope 
of this paper, and we leave it for 
future work.

\section{Comparison to fixed order results}
\label{sec:Validation}

We can now use the one loop collinear 
and the two loop soft functions, which appear
in the factorization theorem \Eqn{eq:QGnlpfact},
to evaluate the partonic cross section up to the 
second order in perturbation theory.  We check 
the result by comparing the bare cross section 
with an in-house calculation of the NLP NNLO 
partonic cross section obtained with the method 
of regions \cite{Beneke:1997zp}. Next we remove 
the initial-state collinear singularities via 
PDF renormalization and compare with the finite 
cross section available in the literature 
\cite{Hamberg:1990np}. 

\subsection{First order}

The partonic cross section in the $g \bar q$
channel starts at first order in perturbation 
theory. At this order, taking into account 
\Eqn{eq:qgtreecollfuncscalar}, and the fact 
that $H(Q^2) = 1+\ord(\alpha_s)$, we have 
\be\label{eq:QGnlpfactNLOb}
\Delta^{(1)}_{g\bar{q}}(z)|_{\rm NLP} = 
2\,\int {d\omega}\, d\omega' 
\, S^{(1)}(\Omega,\omega, \omega') \,.
\ee
We replace the soft function in 
\Eqn{eq:nlo-matrix} and integrate
over $\omega$, $\omega'$. The exact 
result reads 
\be\label{DeltaqgNLO-AllOeps}
\Delta^{(1)}_{g\bar{q}}(z)|_{\rm NLP} =
- \frac{\alpha_s T_F}{4\pi} 
\bigg(\frac{\mu^2}{\Omega^2}\bigg)^{\eps}
\frac{2e^{\eps \gamma_E}\Gamma[1-\eps]}{\eps
\Gamma[1-2\eps]}.
\ee
Expanding in $\eps$, identifying
$\Omega = Q(1-z)$ and setting for 
simplicity $\mu = Q$ we get  
\be\label{DeltaqgNLO}
\Delta^{(1)}_{g\bar{q}}(z)|_{\rm NLP} =
\frac{\alpha_s T_F}{4\pi} \bigg[
-\frac{2}{\eps} +4{\cal L}_1(z) +\eps 
\Big(3\zeta_2 -4 {\cal L}_2(z) \Big) 
+ \ord(\eps^2) \bigg],
\ee
where, in order to keep equations 
compact, we introduced the 
notation
\be
\Lo_n(z) \equiv \ln^n (1-z).
\ee
Let us notice here that the 
one loop soft function in 
\Eqn{eq:nlo-matrix} is finite 
for $\omega \neq 0$, therefore 
the single pole in \Eqn{DeltaqgNLO} 
arises from the integration over 
$\omega$, for $\omega \to 0$:
we see explicitly that the expansion 
in powers of $\eps$ and renormalization 
before taking the convolution in $\omega$, 
$\omega'$ would result in an endpoint 
divergence.

\subsection{Second order}
\label{sec:secondordervalidation}

At second order we need to take 
into account three contributions.
The first involves the hard 
function at one loop: 
\be\label{eq:QGnlpfactNNLOhb}
\Delta^{(2)}_{g\bar{q}}(z)|_{\rm NLP,h} = 
2 H^{(1)}(Q^2)\,\int {d\omega}\, d\omega' 
\, S^{(1)}(\Omega,\omega, \omega') \,.
\ee
The integration over the soft function 
is the same as the one occurring at NLO, 
and the one loop hard function 
$H^{(1)}(Q^2)$, where superscript (1) indicates order $\alpha_s$ result, can be found in Eq.~(5.6) 
of \cite{Beneke:2019oqx}\footnote{
At higher orders the hard function 
can be obtained from the standard 
definition $H(Q^2) = |C^{A0}(Q^2)|^2$  as given above \Eqn{softoperator2},
where the two-loop Wilson coefficient
can be found e.g. in \cite{Becher:2007ty}, 
and the three loop coefficient has been 
given in \cite{Gehrmann:2010ue}.}. 
Integrating over $\omega$, $\omega'$,
the result with exact scale dependence 
can be expressed as:
\bea\label{Delta1r1vH-AllOeps} \nn
\Delta^{(2)}_{g\bar{q}}(z)|_{\rm NLP,h} &=& 
\bigg(\frac{\alpha_s}{4\pi}\bigg)^2 T_F \,C_F\, 
\bigg(\frac{\mu^2}{\Omega^2}\bigg)^{\eps}
\bigg(\frac{\mu^2}{Q^2}\bigg)^{\eps}
\frac{2e^{\eps \gamma_E}\Gamma[1-\eps]}{\eps
\Gamma[1-2\eps]} 
\bigg[\frac{4}{\eps^2} 
+ \frac{6}{\eps} + 16 + 14 \zeta_2  \\ 
&&\hspace{-1.0cm}+\,\eps \bigg(32 
- 21 \zeta_2 - \frac{28 \zeta_3}{3}\bigg) 
+ \eps^2 \bigg(64 - 56 \zeta_2 
- 14 \zeta_3 + \frac{73 \zeta_4}{4}
\bigg) + \ord(\eps^2) \bigg].
\eea
Expanding in powers of $\eps$, 
identifying $\Omega = Q(1-z)$ 
and setting $\mu = Q$ we get  
\bea\label{Delta1r1vH} \nn
\Delta^{(2)}_{g\bar{q}}(z)|_{\rm NLP,h} &=& 
\bigg(\frac{\alpha_s}{4\pi}\bigg)^2 T_F \,C_F\, 
\bigg\{ \frac{8}{\eps^3} + \frac{12 - 16 \Lo_1(z)}{\eps^2} 
+ \frac{32 - 24 \Lo_1(z) + 16 \Lo_2(z) - 40 \zeta_2}{\eps} \\ 
&&\hspace{-1.0cm}+\, 64 - \frac{112 \zeta_3}{3} 
+ \Lo_1(z) (80 \zeta_2 -64) + 24 \Lo_2(z) 
- \frac{32}{3} \Lo_3(z) - 60 \zeta_2 
+ \ord(\eps) \bigg\}.
\eea
This result is in agreement with the in-house 
method of regions calculation where the virtual 
gluon is hard and the emitted quark is soft. 

The second contribution involves 
the collinear function at one loop:
\bea\label{eq:QGnlpfactNNLOccb} \nonumber
\Delta^{(2)}_{g\bar{q}}(z)|_{\rm NLP,c} &=& 
8 H^{(0)}(Q^2)\,\int {d\omega}\, d\omega' 
\big[
G^{(0)*}_{\xi q}(x_an_+p_A;\omega') \,
G^{(1)}_{\xi q}(x_an_+p_A;\omega)  \\ 
&&\hspace{2.0cm} +\,
G^{(1)*}_{\xi q}(x_an_+p_A;\omega') \,
G^{(0)}_{\xi q}(x_an_+p_A;\omega)\big]
\, S^{(1)}(\Omega,\omega, \omega') \,,
\eea
which, taking into account the result for 
the tree level jet function, simplifies to
\be\label{eq:QGnlpfactNNLOccbb} 
\Delta^{(2)}_{g\bar{q}}(z)|_{\rm NLP,c} =
-4 \int {d\omega}\, d\omega' \,
\Big( G^{(1)}_{\xi q}(x_an_+p_A;\omega)
+ G^{(1)*}_{\xi q}(x_an_+p_A;\omega') \Big)
S^{(1)}(\Omega,\omega, \omega') \,.
\ee
Inserting the one loop jet function 
from \Eqn{eq:qgoneloopcollfuncscalar},
the one loop soft function 
\Eqn{eq:nlo-matrix} and 
integrating we get
\bea\label{Delta1r1vC-AllOeps} \nn
\Delta^{(2)}_{g\bar{q}}(z)|_{\rm NLP,c} &=&
-\bigg(\frac{\alpha_s}{4\pi}\bigg)^2 T_F (C_F-C_A)
\bigg(\frac{\mu^2}{\Omega^2}\bigg)^{\eps}
\bigg(\frac{\mu^2}{Q \Omega}\bigg)^{\eps} \\
&&\hspace{1.0cm}\times\,
\frac{2(2-\eps(4+\eps))}{\eps^3(1-2\eps)} 
\frac{e^{2\eps \gamma_E}\Gamma^2[1-\eps]
\Gamma[1+\eps]}{\Gamma[1-3\eps]}.
\eea
Expanding in powers of $\eps$, 
identifying $\Omega = Q(1-z)$ 
and setting $\mu = Q$ we get 
\bea\label{Delta1r1vC} \nn
\Delta^{(2)}_{g\bar{q}}(z)|_{\rm NLP,c} &=& 
\bigg(\frac{\alpha_s}{4\pi}\bigg)^2 T_F (C_F-C_A)
\bigg\{
- \frac{4}{\eps^3} + \frac{12 \Lo_1(z)}{\eps^2}  
+ \frac{2 -18 \Lo_2(z) + 12 \zeta_2}{\eps}  \\ 
&&\hspace{1.5cm}+\,4 + 18 \Lo_3(z) 
- \Lo_1(z) (6 + 36\zeta_2) 
+ \frac{104 \zeta_3}{3} + \ord(\eps) \bigg\}.
\eea 
This result also agrees with the in-house 
method of regions calculation, where the virtual 
gluon is collinear and the emitted quark is soft. 
As discussed after \Eqn{eq:qgoneloopcollfuncExpandedscalar},
let us highlight that, in contrast to the $q\bar q$
channel, where the collinear function starts 
contributing only at NLL accuracy, in this case 
it contributes already at LL level.
Comparing \Eqns{eq:qgoneloopcollfuncExpandedscalar}{Delta1r1vC} 
we see more in detail how this happens. Focusing 
on the highest pole contribution, which is in direct 
correspondence with the LL, we see that the collinear 
function itself contains a $1/\eps^2$ pole, which 
becomes a $1/\eps^3$ leading pole by means of the 
convolution with the one loop soft function. In this 
respect, the collinear contribution in the $q\bar q$ 
channel contains a $1/\eps$ pole (see Eqs. 
(4.30) -- (4.34) in \cite{Beneke:2019oqx}), which 
is then raised to a $1/\eps^2$ subleading pole by 
the convolution integration.
A physical interpretation of this phenomenon 
has been put forward in \cite{Moult:2019uhz,Beneke:2020ibj}. 
In general, the idea is that in a splitting
$1 \to 2+3$ with soft 3, the leading pole is related 
to the difference of the Casimir charge of the
energetic particles 1 and 2. In the diagonal $q\bar q$
channel the collinear quark 1 emits a soft gluon 3, 
continuing as the collinear quark 2, thus there is no 
change of the Casimir charge along the collinear 
direction. In the $g\bar q$ channel, a collinear 
gluon 1 emits a soft quark 3, continuing as a 
collinear quark, which has a different Casimir charge. 
Indeed, the collinear function in 
\Eqn{eq:qgoneloopcollfuncExpandedscalar} is 
proportional to the difference of the Casimir 
charges $C_F - C_A$.

Last, the third contribution involves the 
soft function at two loops: 
\be\label{eq:QGnlpfactNNLOsb}
\Delta^{(2)}_{g\bar{q}}(z)|_{\rm NLP,s} = 
2 \int {d\omega}\, d\omega' 
\, S^{(2)}(\Omega,\omega, \omega') \,.
\ee
As discussed in section~\ref{sec:SoftFunctions},
the soft function at two loops receives two 
contributions, where the additional soft gluon 
is respectively virtual or real. These two 
contributions have been given respectively 
in \Eqns{soft-function-momentum-1r1v-NNLOd}{soft-function-momentum-2realc}.
For a better comparison with the method of regions, 
it proves useful to evaluate the two contributions 
separately. Furthermore, the integration of 
the term involving the virtual gluon, 
\Eqn{soft-function-momentum-1r1v-NNLOd}, 
is comparatively more involved. In 
particular, the term in the last line 
of \Eqn{soft-function-momentum-1r1v-NNLOd}
requires some care, because of the 
singularity at $\omega = 0$, which 
can be integrated by considering the 
standard identity $1/{(\omega - i 0_+)}
= P(1/\omega) + i \pi \delta(\omega)$, 
where $P$ indicates the principal value prescription. 
The term in the second line of 
\Eqn{soft-function-momentum-1r1v-NNLOd}
instead can be integrated over $\omega$ 
and $\omega'$ without particular issues. 
In the end we obtain a result for 
\Eqn{eq:QGnlpfactNNLOsb} valid to 
all orders in $\eps$: 
\bea\label{Soft1r1v-AllOeps} \nn
\Delta^{(2)}_{g\bar{q}}(z)|_{\rm NLP,s,1r1v} &=& 
- \bigg(\frac{\alpha_s}{4\pi}\bigg)^2 
T_F \big(C_A - 2C_F\big)
\bigg(\frac{\mu^2}{\Omega^2}\bigg)^{2\eps} \\
&&\hspace{1.0cm}\times \,
\frac{2 {\rm Re}[e^{-i \eps \pi}]
e^{2\eps \gamma_E}\Gamma[1-2\eps]
\Gamma[1-\eps]^2\Gamma[1+\eps]^2}{\eps^3
\Gamma[1-4\eps]}.
\eea
Expansion in powers of $\eps$,
identifying $\Omega = Q(1-z)$ 
and setting $\mu = Q$ gives 
\bea \nn
\Delta^{(2)}_{g\bar{q}}(z)|_{\rm NLP,s,1r1v} &=& 
\bigg(\frac{\alpha_s}{4\pi}\bigg)^2 
T_F \bigg(C_F-\frac{C_A}{2}\bigg)
\bigg\{\frac{4}{\eps^3} - \frac{16\Lo_1(z)}{\eps^2}  
+ \frac{32 \Lo_2(z) - 28 \zeta_2 }{\eps} \\
&&\hspace{2.0cm}-\, \frac{128}{3} \Lo_3(z) + 112 \zeta_2 \Lo_1(z)  
- \frac{224 \zeta_3}{3} + \ord(\eps) \bigg\}.
\eea
The integration of the double real contribution, 
\Eqn{soft-function-momentum-2realc} is straightforward.
Again, we obtain a result valid to all order in $\eps$:
\bea\label{Soft2r-AllOeps} \nn
\Delta^{(2)}_{g\bar{q}}(z)|_{\rm NLP,s,2r} &=& 
- \bigg(\frac{\alpha_s}{4\pi}\bigg)^2 
T_F \bigg(\frac{\mu^2}{\Omega^2}\bigg)^{2\eps} 
\frac{e^{2\eps \gamma_E}\Gamma[1-\eps]^2}{\eps^3
(1-2\eps)\Gamma[1-4\eps]} \bigg\{ C_F 
\Big[12 - (21-\eps)\eps \Big] \\ \nn
&&\hspace{1.0cm}-\,(C_A - 2C_F) 
\bigg[\frac{1}{\eps}-9+14\eps
- \frac{\,_2F_1 \big[2,2,3-2\eps,1\big]}{1-\eps} \\
&&\hspace{2.0cm}
+4(1-2\eps) \, _3F_2 \big[\{1,1,-\eps\},\{1-\eps,-2\eps\},1\big]
\bigg]\bigg\},
\eea
which upon expansion in $\eps$,
identifying $\Omega = Q(1-z)$ 
and setting $\mu = Q$ gives 
\bea \nn
\Delta^{(2)}_{g\bar{q}}(z)|_{\rm NLP,s,2r} &=& 
\bigg(\frac{\alpha_s}{4\pi}\bigg)^2 
T_F \bigg\{ C_F \bigg[ -\frac{8}{\eps^3} 
+ \frac{32 \Lo_1(z)-3}{\eps^2}  
+ \frac{12 \Lo_1(z)- 64 \Lo_2(z)-7+60\zeta_2}{\eps} \\ \nn
&&\hspace{0.5cm} +\, \frac{256}{3} \Lo_3(z)-24 \Lo_2(z)
+ \Lo_1(z) (28 -240 \zeta_2) - 14
+ 21\zeta_2 + \frac{556\zeta_3}{3} \bigg] \\ \nn
&&\hspace{0.0cm} +\, C_A \bigg[ -\frac{2}{\eps^3} 
+ \frac{8 \Lo_1(z)}{\eps^2} 
+ \frac{12 \zeta_2 - 16 \Lo_2(z)}{\eps}
+ \frac{64}{3} \Lo_3(z) \\ 
&&\hspace{0.5cm}-\, 48 \zeta_2 \Lo_1(z) 
+ \frac{94 \zeta_3}{3} \bigg] + \ord(\eps) \bigg\}
\eea

Summing together all terms contributing at NNLO we obtain
\bea\label{bare-partonic-xs} \nn
\Delta^{(2)}_{g\bar{q}}(z)|_{\rm NLP} &=& 
\bigg(\frac{\alpha_s}{4\pi}\bigg)^2 
T_F \bigg\{ C_F \bigg[ \frac{9 + 12\Lo_1(z)}{\eps^2} 
+ \frac{27 - 12\Lo_1(z) - 34 \Lo_2(z) + 4 \zeta_2}{\eps} \\ \nn
&& \hspace{2.0cm}+\,  54 + 50 \Lo_3(z)
- \Lo_1(z) (42 + 84\zeta_2) - 39\zeta_2+ 108 \zeta_3 \bigg\} \\ \nn
&&\hspace{1.0cm} +\, C_A \bigg[ 
\frac{4\Lo_1(z)}{\eps^2} + \frac{14 \zeta_2 -2 - 14\Lo_2(z)}{\eps} \\
&& \hspace{2.0cm}-\, 4 + \frac{74}{3} \Lo_3(z)
+ \Lo_1(z) (6 - 68 \zeta_2) + 34 \zeta_3
\bigg] + \ord(\eps) \bigg\}.
\eea
This gives our result for the bare partonic 
cross section at NNLO in perturbation theory.
The result agrees with an in-house computation
of the bare partonic cross section obtained with 
the method of regions. This provides us with a first 
confirmation that the (bare) factorization 
theorem in \Eqn{eq:QGnlpfact} is correct,
and is not missing any contribution  
up to $\mathcal{O}(\alpha^2_s)$.

\subsection{Renormalization}
\label{NNLO-qg}

As a final check, we need to compare with the 
known result in literature \cite{Hamberg:1990np}, 
for which we need the finite partonic cross 
section. As discussed in appendix \ref{appPDFr}, 
the latter is obtained by applying 
PDF renormalization to \Eqn{bare-partonic-xs},
which relies on the finiteness of the hadronic
cross section:
\begin{equation}
\frac{d\sigma_{\rm DY}}{dQ^2} = 
\sigma_0
\sum_{a,b} \int_{\tau}^1\, \frac{dz}{z}\, 
{\cal L}^{\rm bare}_{ab}\bigg(\frac{\tau}{z}\bigg)\, 
\Delta^{\rm bare}_{ab}(z)
= \sigma_0
\sum_{a,b} \int_{\tau}^1\, \frac{dz}{z}\, 
{\cal L}^{\rm ren}_{ab}\bigg(\frac{\tau}{z}\bigg)\, 
\Delta^{\rm ren}_{ab}(z),
\label{eq:dsigsqDeltaR}
\end{equation}
where by ``ren'' we indicate the finite, 
renormalized functions. In practice, the 
finite cross section is obtained by applying 
a counterterm to the bare cross section. We 
provide these counterterms as a function of 
the Altarelli-Parisi splitting kernels in 
\Eqn{RenToBarePerturbative}. At first order
in perturbation theory, the explicit 
calculation gives
\bea\label{DeltaqgNLO-fin-eps} 
\Delta_{g\bar q}^{\rm ren\, (1)}(z) &=&
\frac{\alpha_s T_F}{4\pi} \bigg[
4{\cal L}_1(z) + \eps \bigg( 3\zeta_2 
- 4{\cal L}_2(z) \bigg)+ \ord(\eps^2) \bigg],
\eea
which agree respectively with Eqs.~(2.19) and~(2.21) 
of \cite{Hamberg:1990np} upon expansion to first order 
in $(1-z)$. The pole in \Eqn{DeltaqgNLO} is correctly 
removed by the term $P^{(0)}_{qg}/(\eps)$ in the 
second line of \Eqn{RenToBarePerturbative}. Proceeding 
in a similar way and evaluating explicitly the counterterm 
in the third equality of \Eqn{RenToBarePerturbative} 
we have:
\bea\nn
\Delta_{g\bar q}^{(2)}(z)|_{\rm NLP}^{\rm ren} &=& 
\Delta_{g\bar q}^{(2)}(z)|_{\rm NLP} \\ \nn
&&\hspace{-2.0cm}
+\, \bigg(\frac{\alpha_s}{4\pi}\bigg)^2 
T_F \bigg\{ C_F \bigg[ 
-\frac{9 + 12\Lo_1(z)}{\eps^2}  
+ \frac{ 34 \Lo_2(z) + 12\Lo_1(z)  
-27 - 4 \zeta_2}{\eps} \\ 
&& \hspace{1.0cm}-\, 64 - 12 \Lo_2(z) - \frac{80}{3} \Lo_3(z)
+ 51\zeta_2 + 68 \zeta_2\Lo_1(z) - 32 \zeta_3 \bigg] \\ \nn
&&\hspace{0.0cm} +\, C_A \bigg[-\frac{4\Lo_1(z)}{\eps^2}  
+ \frac{2 + 14 \Lo_2(z) - 14 \zeta_2}{\eps}
- 16 \Lo_3(z) + 44 \zeta_2 \Lo_1(z) -32 \zeta_3 
\bigg]\bigg\},
\eea
and in the end we get
\bea \nn
\Delta_{g\bar q}^{(2)}(z)|_{\rm NLP}^{\rm ren} &=& 
\bigg(\frac{\alpha_s}{4\pi}\bigg)^2 
T_F \bigg\{ C_F \bigg[\frac{26}{3} \Lo_3(z)
+ \Lo_1(z) (6 - 24 \zeta_2) -4 + 2\zeta_3 \bigg]  \\  
&&\hspace{-2.5cm}+\, C_A \bigg[\frac{70}{3} \Lo_3(z) 
-12 \Lo_2(z) - \Lo_1(z) (42 + 16 \zeta_2) -10 
+ 12 \zeta_2 + 76 \zeta_3 \bigg] + \ord(\eps) \bigg\},
\eea
which reproduces Eq.~(B.38) + (B.39) of 
\cite{Hamberg:1990np}. This concludes 
our validation of the factorization 
theorem in \Eqn{eq:QGnlpfact}.


\section{Summary}
\label{sec:summary}

In this work, we derived the factorization formula
for the $g\bar{q}$-channel of the Drell-Yan production
process at general powers in the threshold expansion.
Specifying the result to the first non-trivial power, 
which is at next-to-leading power in $(1-z)$ compared
to the leading $q\bar q$ channel, we arrived at a compact 
formula comprising of a LP hard function, two NLP collinear 
functions on each side of the cut and one generalized soft 
function. The simplicity of this result is in contrast with 
the more involved structure of the $q\bar{q}$ case, where 
four separate soft functions contribute along with 
their corresponding collinear functions \cite{Beneke:2019oqx}. 
The first main result of this work is the NLP 
factorization formula given in \Eqn{eq:QGnlpfact}.

In sections~\ref{sec:CollinearFunctions} 
and~\ref{sec:SoftFunctions} we calculated the 
collinear and soft functions which appear in the 
factorization formula. In order to validate this 
formula to NNLO accuracy, hard and collinear 
functions are needed up to $\mathcal{O}(\alpha_s)$ 
while the soft function up to $\mathcal{O}(\alpha_s^2)$. 
The hard function which appears in \Eqn{eq:QGnlpfact} 
is the LP hard function which is known in the literature.
In section~\ref{sec:CollinearFunctions}, we calculated 
the NLP collinear function with an external PDF-collinear 
gluon which we have defined in an  analogous way to 
the collinear functions with an external 
PDF-collinear quark appearing in \cite{Beneke:2019oqx}.  
Due to its relation to the radiative jet functions 
appearing in $h\to gg$ amplitude,  the radiative 
corrections of this object are known up to 
$\mathcal{O}(\alpha_s^2)$ \cite{Liu:2021mac}. 
We found agreement for the $\mathcal{O}(\alpha_s)$ 
contribution which is given in \Eqn{eq:qgoneloopcollfuncscalar}.

In section \ref{sec:SoftFunctions}, we calculated 
the soft function which contains soft quark 
insertions on both sides of the cut. We carried 
out the computation by utilising state-of-the-art 
fixed-order loop integral methods such as 
reduction to master integrals and application 
of differential equations. The $\mathcal{O}(\alpha_s)$ 
result for the soft function 
is given in 
\Eqn{eq:nlo-matrix}, and two-loop soft function 
results are presented in \Eqn{soft-function-momentum-1r1v-NNLOd} 
and \Eqn{soft-function-momentum-2realc} for 
the one-real one-virtual and double-real 
contributions, respectively. We have retained 
all-order $\epsilon$ dependence throughout the 
calculation of the NLO and NNLO contributions 
to the soft function, which 
 enabled us to study the structure of its
asymptotic limits in section~\ref{sec:endpoint}. 
In particular, we noticed a 
relatively simple structure at NLO, where 
singularities arise for $\omega \to 0$. 
More involved structure emerges at NNLO 
which has implications for resummation 
beyond LL using refactorization procedures 
such as the ones developed in \cite{Beneke:2022obx,Liu:2020wbn}.

Using the calculations in sections~\ref{sec:CollinearFunctions} 
and~\ref{sec:SoftFunctions}, we tested the 
validity of the factorization formula derived 
in section~\ref{sec:NLPfactorization} through 
a comparison of our results against in-house 
expansion-by-regions calculations and known 
results in the literature. We evaluated the 
convolutions over the $\omega,\omega'$ 
variables obtaining results valid in exact 
$d$-dimensions at $\mathcal{O}(\alpha_s)$ 
and $\mathcal{O}(\alpha_s^2)$. 
We then expanded the final results to $d\to4$, 
carried out PDF-factorization of the collinear 
divergences appearing in initial states, which 
is standard for fixed order calculations, and 
we compared with \cite{Hamberg:1990np}, 
finding agreement.

In summary, the bare factorization theorem derived 
here, along with the higher-order calculations of the functions 
appearing in our formula and  the study of the asymptotic limits of the soft function, will serve as a spring 
board for future investigations of this channel and an 
eventual desirable four dimensional renormalization 
prescription.


\subsubsection*{Acknowledgments} 
The authors would like to thank Martin Beneke, Mathias Garny, 
Robert Szafron, Jian Wang for collaboration and discussions at 
early stages of this work. The authors would also like to express 
special thanks to the Mainz Institute for Theoretical Physics 
(MITP) of the Cluster of Excellence PRISMA+ (Project ID 39083149) 
for hospitality and support during the program ``Power Expansions 
on the Lightcone: From Theory to Phenomenology'', where part of 
this work has been conducted. S.J. would like to thank the University of
Milano-Bicocca, INFN, T-2 group at Los Alamos National Laboratory, the EIC Theory Institute and the physics department at Brookhaven National Laboratory (under contract DE-SC0012704), and the theory department at Jefferson Laboratory for kind hospitality 
while parts of this work were completed.

The work of A.B. was supported in part by the ERC 
Starting Grant REINVENT-714788. S.J. is supported 
by STFC under grant ST/T001011/1 and the Royal Society 
University Research Fellowship  (URF/R1/201268).
L.V. is supported by Fellini - Fellowship for Innovation 
at INFN, funded by the European Union's Horizon 2020 
research programme under the Marie Sk\l{}odowska-Curie 
Cofund Action, grant agreement no. 754496 and by 
Compagnia di San Paolo through grant 
TORP\_S1921\_EX-POST\_21\_01. Figures were drawn 
with \texttt{Jaxodraw}~\cite{Binosi:2008ig}. 
Calculations were done in part with 
\texttt{FORM}~\cite{Vermaseren:2000nd} 
and expansions with \texttt{HypExp} 
\cite{Huber:2005yg}.

\begin{appendix}

\section{Soft function master integrals}
\label{appMIs}

\subsection{Virtual-real MIs} \label{app-RV-MIs}

The integrals appearing in the calculation of the virtual-real contribution to the $\mathcal{O}(\alpha^2_s)$ soft function are written as 
\begin{align} \label{MI1r1vdef}
\hat{J}_{\mathcal{T}}(\alpha_1,\alpha_2,
\alpha_3,\alpha_4, \alpha_5,\alpha_6,\alpha_7) 
= (4\pi)^4\bigg( \frac{e^{\gamma_E} 
\mu^2}{4 \pi}\bigg)^{2 \epsilon}
\int \frac{d^d k}{(2 \pi)^{d}} \, 
\frac{d^d k_1}{(2 \pi)^{d-1}} 
\prod_{i=1}^7\, \frac{1}{P^{\alpha_i}_i}\,,
\end{align}
where $k$ identifies the loop momentum while $k_1$ identifies the real emission momentum. The set of propagators $P_i$ are identified by the single topology $\mathcal{A}$ which is defined as 
\begin{align}\label{eq:topologyVR}
P_1&= k^2,\quad P_2=(k+k_1)^2,\quad P_3=n_+k,
\quad P_4=k_1^2,  \\ \nonumber
P_5&= \big(\Omega - n_-k_1   -n_+k_1   \big),
\quad P_6 =\big(\omega-n_-k-n_-k_1\big),
\quad P_7 = \big(\omega'-n_-k_1\big)\,,
\end{align}
where the propagators $\{P_4,P_5,P_6,P_7\}$ are cut.
The virtual-real contribution to the 
soft function in \Eqn{soft-function-momentum-1r1v-NNLOc} 
is given in terms of the two master integrals of topology $\mathcal{A}$ 
\begin{align}
&\hat{J}_1(\Omega,\omega,\omega')\equiv \hat{J}_{\mathcal{A}}(0, 1, 1, 1, 1, 1, 1),\quad  \hat{J}_2(\Omega,\omega,\omega')\equiv \hat{J}_{\mathcal{A}}(1, 1, 1, 1, 1, 1, 1), \label{eq:MISVR}
\end{align}
which can be calculated by direct integration 
by combining the propagators in terms of Feynman 
parameters, or by the differential equation method.
We explicitly obtain the following expressions for these integrals
\be \label{eq:MISVR1}
\hat{J}_1(\Omega,\omega,\omega')
= i \,\frac{e^{2\epsilon\gamma_E}
\Gamma[1+\epsilon]}{\epsilon\,\Gamma[1-\epsilon]}\,
\bigg[\frac{\mu^4}{(-\omega)\omega'
(\Omega-\omega')^2}\bigg]^{\epsilon}  
\,\theta(-\omega) \,\theta(\omega')\,
\theta(\Omega-\omega')\,,
\ee
and
\begin{eqnarray} \label{eq:MISVR2}
\hat{J}_2(\Omega,\omega,\omega')&=& 2 i
\frac{e^{2\epsilon\gamma_E}\Gamma[1-\epsilon]
\Gamma[1+\epsilon]}{\epsilon\,\Gamma[1-2\epsilon]}
\bigg[\frac{\mu^4}{(\omega'-\omega)^2
(\Omega-\omega')^2}\bigg]^{\epsilon}  
\frac{\theta(\omega')\,\theta(\omega'-\omega)\,
\theta(\Omega-\omega')}{(\omega'-\omega)
(\Omega-\omega')}  \nonumber \\ \nn
&& \hspace{-2.0cm}  
-\, i \frac{ e^{2\epsilon \gamma_E}
\Gamma[1+\epsilon]}{\epsilon\,
\Gamma[1-\epsilon]} 
\bigg[\frac{\mu^4}{(-\omega)\omega'
(\Omega-\omega')^2}\bigg]^{\epsilon}
\frac{\theta(-\omega)\,\theta(\omega')\,
\theta(\Omega-\omega')}{\omega'\,
(\Omega-\omega')} 
\, _2F_1\left(1,1+\epsilon;1-\epsilon;
\frac{\omega}{\omega'}\right). \\
\end{eqnarray} 

\subsection{Double real MIs} \label{app-RR-MIs}

The integrals appearing in the soft double real contribution are expressed as
\begin{align}
\hat{I}_{\mathcal{T}}(\alpha_1,\alpha_2,\alpha_3,\alpha_4, 
\alpha_5,\alpha_6,\alpha_7) 
= (4\pi)^4\bigg( \frac{e^{\gamma_E} 
\mu^2}{4 \pi}\bigg)^{2 \epsilon}
\int \frac{d^d k_1}{(2 \pi)^{d-1}} \, 
\frac{d^d k_2}{(2 \pi)^{d-1}} 
\prod_{i=1}^7\, \frac{1}{P^{\alpha_i}_i}\,,
\end{align}
where the set of seven propagators $P_i$ defines a corresponding topology $\mathcal{T}$. In total, for this contribution we need to introduce four different topologies.
The first, $\mathcal{A}$, is defined by the 
following set of propagators:
\begin{align}\label{eq:topologySA} \nonumber
P_1&=(k_1+k_2)^2,\quad P_2=n_+k_2,\quad P_3=k_1^2,
\quad P_4=k_2^2,  \\ \nonumber
P_5&= \big(\Omega - n_-k_1 - n_-k_2 -n_+k_1 - n_+k_2 \big),
\quad P_6 =\big(\omega-n_-k_1\big), \\ 
\quad P_7&= \big(\omega'-n_-k_1-n_-k_2\big)\, ,
\end{align}
where the last five propagators are cut. Similarly, the second topology, $\mathcal{B}$, is defined by the following propagators:
\begin{align}\label{eq:topologySB} \nonumber
P_1&=(k_1+k_2)^2,\quad P_2=n_+k_2,\quad P_3=k_1^2,
\quad P_4=k_2^2,  \\ \nonumber
P_5&= \big(\Omega - n_-k_1 - n_-k_2 -n_+k_1 - n_+k_2 \big),
\quad P_6 =\big(\omega-n_-k_1-n_-k_2\big), \\  
\quad P_7&= \big(\omega'-n_-k_1\big)\,,
\end{align}
where the last five are again cut. Topology $\mathcal{A}$ and $\mathcal{B}$ are related to topology $\mathcal{H}$ of \cite{Broggio:2021fnr} by the following relabellings $\{\omega\to \omega_1,\,\omega^\prime \to  \omega_1+\omega_2 \}$ and $\{\omega\to \omega_1+\omega_2,\,\omega^\prime \to \omega_2\}$, respectively.
As a consequence we can extract the results for the master integrals of topologies $\mathcal{A}$ and $\mathcal{B}$ from the computations carried out for topology $\mathcal{H}$ in \cite{Broggio:2021fnr}, after applying the relevant substitutions.
We have then the topology $\mathcal{C}$, which is equivalent to topology $\mathcal{A}$ of \cite{Broggio:2021fnr},
defined by the following set of propagators:
\begin{align}\label{eq:topologySC} \nonumber
P_1&=(k_1+k_2)^2,\quad P_2=n_+k_2, \quad P_3=n_-k_2,
\quad P_4=k_1^2,\quad P_5=k_2^2,  \\
P_6&= \big(\Omega - n_-k_1 - n_-k_2 -n_+k_1 - n_+k_2 \big), 
\quad P_7 = \big(\omega-n_-k_1\big)\, ,
\end{align}
where the last four propagators are cut.
The last topology, $\mathcal{D}$, is equivalent to topology $\mathcal{D}$ in \cite{Broggio:2021fnr} and it is defined by the following set of propagators:
\begin{align}\label{eq:topologySD} \nonumber
P_1&=(k_1+k_2)^2,\quad P_2=n_+k_2, \quad P_3=n_-k_2,
\quad P_4=k_1^2,\quad P_5=k_2^2,  \\
P_6&= \big(\Omega - n_-k_1 - n_-k_2 -n_+k_1 - n_+k_2 \big),
\quad P_7 =\big(\omega-n_-k_1-n_-k_2\big)\,,
\end{align}
where the last four propagators are cut. Please note that for topologies $\mathcal{A}$ and $\mathcal{B}$ we exclude the $\delta(\omega-\omega')$ factor from the topology definition and we include it explicitly in \Eqn{soft-function-momentum-2realb}.

The double real emission contribution of the soft function 
in \Eqn{soft-function-momentum-2realb}
is expressed in terms of six master integrals belonging to the above topologies.
In particular, two master integrals belong to topology $\mathcal{A}$
\begin{align}
&\hat{I}_1(\Omega,\omega,\omega')\equiv 
\hat{I}_{\mathcal{A}}(0, 0, 1, 1, 1, 1, 1),\quad  
\hat{I}_2(\Omega,\omega,\omega')\equiv 
\hat{I}_{\mathcal{A}}(1, 1, 1, 1, 1, 1, 1) , \label{eq:MISSA}
\end{align}
two belong to topology $\mathcal{B}$ 
\begin{align}
&\hat{I}_3(\Omega,\omega,\omega')\equiv 
\hat{I}_{\mathcal{B}}(0, 0, 1, 1, 1, 1, 1),\quad  
\hat{I}_4(\Omega,\omega,\omega')\equiv 
\hat{I}_{\mathcal{B}}(1, 1, 1, 1, 1, 1, 1) , \label{eq:MISSB}
\end{align}
and the last two belong to topology $\mathcal{C}$ and $\mathcal{D}$
\begin{align}
&\hat{I}_5(\Omega,\omega)\equiv 
\hat{I}_{\mathcal{C}}(0, 0, 0, 1, 1, 1, 1),\quad 
\hat{I}_6(\Omega,\omega)\equiv 
\hat{I}_{\mathcal{D}}(0, 0, 0, 1, 1, 1, 1), \label{eq:MISSCD}
\end{align}
respectively. $\hat{I}_1$ can be easily extracted from the results in \cite{Broggio:2021fnr}, we find 
\begin{eqnarray} \label{eq:MISRR1} \nonumber
\hat{I}_1(\Omega,\omega,\omega') &=& 
\frac{e^{2 \epsilon \gamma_E}}{\Gamma[2-2\epsilon]}
\bigg[\frac{\mu^4}{\omega(\omega'-\omega)
(\Omega-\omega')^2}\bigg]^{\eps}
(\Omega-\omega')
\theta(\omega)\theta(\omega')
\theta(\omega'-\omega)
\theta(\Omega-\omega'). \\
\end{eqnarray}
$\hat{I}_2$ was not explicitly required for the calculations carried out in \cite{Broggio:2021fnr}, hence we evaluate it for the first time here by employing the differential equation method.
$\hat{I}_2$ appears in a system with $\hat{I}_1$ and the differential equation can be solved to all orders in $\epsilon$.
The $\epsilon$-dependent integration constant can be fixed by matching to the version of this integral where $\omega$ and $\omega^\prime$ are integrated over first. We find that the constant is zero to all-orders in $\epsilon$ and the final integral reads
\begin{eqnarray} \label{eq:MISRR2} \nonumber
\hat{I}_2(\Omega,\omega,\omega') &=& 
-\frac{2 e^{2 \epsilon \gamma_E}}{\epsilon
\,\Gamma[1-2\epsilon]}
\bigg[\frac{\mu^4}{\omega(\omega'-\omega)
(\Omega-\omega')^2}\bigg]^{\eps}
\frac{\theta(\omega)\theta(\omega')
\theta(\omega'-\omega)
\theta(\Omega-\omega')}{(\omega'-\omega)
(\Omega-\omega')}  \\ 
&& \times \,\,
_2F_1\bigg(1,-\epsilon,1-\epsilon,
\frac{\omega}{\omega-\omega'}\bigg).
\end{eqnarray}
By exploiting exchange symmetries among topology $\mathcal{A}$ and $\mathcal{B}$ we find the following relations among integrals $\hat{I}_3(\Omega,\omega,\omega') = 
\hat{I}_1(\Omega,\omega',\omega)$
and $\hat{I}_4(\Omega,\omega,\omega') = 
\hat{I}_2(\Omega,\omega',\omega)$.
Finally the last two master integrals read
\begin{align}
\label{eq:MISRR5}
\hat{I}_5(\Omega,\omega) 
&= \frac{e^{2 \epsilon \gamma_E} 
\Gamma[1-\epsilon]}{\Gamma[3-3\epsilon]}
\bigg[\frac{\mu^4}{\omega
(\Omega-\omega)^3}\bigg]^{\eps}
(\Omega-\omega)^2\, \theta(\omega)
\theta(\Omega-\omega),\\[0.2cm]
\label{eq:MISRR6}
\hat{I}_6(\Omega,\omega) 
&= \frac{e^{2 \epsilon \gamma_E} 
\Gamma[1-\epsilon]^2}{\Gamma[2-2\epsilon]^2}
\bigg[\frac{\mu^4}{\omega^2
(\Omega-\omega)^2}\bigg]^{\eps}
\omega (\Omega-\omega)\,
\theta(\omega)
\theta(\Omega-\omega).
\end{align}

\section{PDF renormalization}
\label{appPDFr}

In this appendix we briefly set our notation for PDF 
renormalization. Let us start from \Eqn{eq:dsigsqDelta}, 
and write it in the form 
\begin{equation}
\frac{d\sigma_{\rm DY}}{dQ^2} = \sigma_0
\sum_{a,b} \int_{\tau}^1 dz \int d x_a \, dx_b \, \delta(\tau - x_a x_b z) \,
f^{\rm bare}_{a/A}(x_a,\mu)\, f^{\rm bare}_{b/B}(x_b,\mu) 
\, \Delta^{\rm bare}_{ab}(z,\mu),
\label{eq:dsigsqDeltaApp}
\end{equation}
where we added a superscript to indicate that  
all functions are unrenormalized\footnote{However, we 
implicitly assume that UV renormalization has 
been taken into account, by expressing the 
functions in terms of the renormalized coupling 
constant $\alpha_s(\mu)$, related to the bare 
coupling constant by $\alpha_s^{\rm b} (4\pi)^{\eps} 
e^{-\eps \gamma_{E}} = \alpha_s(\mu) \mu^{\eps} 
\big[1-\alpha_s/(4\pi) \beta_0/\eps\big]$,
with $\beta_0 = 11/3 \, C_A - 2/3 \, n_f$.}.
Due to renormalization invariance of the 
differential cross section, \Eqn{eq:dsigsqDeltaApp}
can be expressed as well in terms of the 
corresponding renormalized function: 
\begin{equation}	
\frac{d\sigma_{\rm DY}}{dQ^2} = \sigma_0
\sum_{a,b} \int_{\tau}^1 dz \int d x_a \, dx_b \, \delta(\tau - x_a x_b z) \,
f^{\rm ren}_{a/A}(x_a,\mu)\, f^{\rm ren}_{b/B}(x_b,\mu)
\, \Delta^{\rm ren}_{ab}(z,\mu).
\label{eq:dsigsqDeltaAppR}
\end{equation}
The relation between the bare and 
renormalized partonic cross section can be 
obtained by considering the relation between 
bare and renormalized PDFs:
\be\label{PDFsRtoB}
f^{\rm ren}_{a/A} = f^{\rm bare}_{b/A} \otimes \Gamma_{ab},
\ee
where the convolution explicitly reads 
\be\label{PDFsRtoBexplicit}
f^{\rm ren}_{a/A}(x_a) = \int_0^1 dy_1 dy_2\, 
\delta(x_a - y_1 y_2)\, f^{\rm bare}(y_1)_{b/A} 
\, \Gamma_{ab}(y_2).
\ee
In turn, the function $\Gamma_{ab}(x)$ has the following
perturbative expansion 
\bea\label{GammaExpansion}
\Gamma_{ab}(x) = \delta_{ab}\, \delta(1-x) 
+ \sum_{n = 0}^{\infty} 
\bigg(\frac{\alpha_s}{4\pi}\bigg)^{n+1} \, 
\Gamma^{(n)}_{ab}(x),
\eea
and for our analysis we need the 
first two orders: 
\bea \label{GammaToAP} \nn
\Gamma^{(0)}_{ab}(x) &=& -\frac{P^{(0)}_{ab}(x)}{\eps}, \\  
\Gamma^{(1)}_{ab}(x) &=& \frac{1}{2\eps^2} 
\bigg[\Big(P^{(0)}_{ac} \otimes P^{(0)}_{cb}\Big)(x) 
+ \beta_0 \, P^{(0)}_{ab}(x) \bigg] - \frac{1}{2\eps} P^{(1)}_{ab}(x).  
\eea
inserting \Eqn{PDFsRtoB} into 
\Eqn{eq:dsigsqDeltaAppR} we get 
the relation 
\be\label{BareToRen}
\Delta^{\rm bare}_{cd}(z) 
= \Big[\Gamma_{ac} \otimes \Gamma_{bd} \otimes 
\Delta^{\rm ren}_{ab} \Big](z).
\ee
This equation can then be solved order
by order in $\alpha_s$ to obtain the 
renormalized cross section coefficients 
in terms of the bare ones. Taking into 
account the UV renormalization and the 
explicit form of \Eqns{GammaExpansion}{GammaToAP}, 
for the $g\bar q$ channel one has\footnote{
Notice that by symmetry we take $\bar q = q$ 
in the indices of $\Gamma_{ab}$ and $P_{ab}$.} 
\bea\label{RenToBarePerturbative} \nn
\Delta^{\rm ren\,(0)}_{g\bar q}(z) 
&=& \Delta^{\rm bare\,(0)}_{g\bar q}(z), \\ \nn
\Delta^{\rm ren\,(1)}_{g\bar q}(z) 
&=& \Delta^{\rm bare\,(1)}_{g\bar q}(z) 
+ \frac{P^{(0)}_{ab}(z)}{\eps}, \\ \nn
\Delta^{\rm ren\,(2)}_{g\bar q}(z) 
&=& \Delta^{\rm bare\,(2)}_{g\bar q}(z) 
+\frac{1}{\eps^2} \bigg[
\frac{3}{2} \Big(P_{qq}^{(0)} \otimes P_{qg}^{(0)}\Big)(z)
+\frac{1}{2} \Big(P_{gg}^{(0)} \otimes P_{qg}^{(0)}\Big)(z)
-\frac{\beta_0}{2} P_{qg}^{(0)}(z) \bigg] \\ \nn
&&\hspace{0.0cm}
+\,\frac{1}{\eps}\bigg[ \frac{1}{2} P_{qg}^{(1)}(z)
+\Big(P_{qq}^{(0)} \otimes \Delta^{\rm bare\,(1)}_{g\bar q}\Big)(z)
+\Big(P_{gg}^{(0)} \otimes \Delta^{\rm bare\,(1)}_{g\bar q}\Big)(z) \\
&&\hspace{1.0cm}
+\Big(P_{qg}^{(0)} \otimes \Delta^{\rm bare\,(1)}_{q\bar q}\Big)(z)
-\beta_0 \, \Delta^{\rm bare\,(1)}_{g\bar q}(z) \bigg],
\eea
where the last term originates from expressing the 
bare cross section as an expansion in terms of the 
renormalized strong coupling constant. Near $x\to 1$
we can expand the Altarelli-Parisi splitting kernels
in powers of $1-x$, and up to NLP we need
\bea\label{AP-explict1L-zto1} \nn
P^{(0)}_{qq}(z) &=& C_F \big[4 {\cal D}_0(z) 
+ 3 \delta(1-z) - 4 \big] + \ord(1-z), \\ \nn
P^{(0)}_{qg}(z) &=& 2 T_F  + \ord(1-z) \\ 
P^{(0)}_{gg}(z) &=& C_A \big[4 {\cal D}_0(z) -4\big] 
+ \beta_0 \,\delta(1-z) + \ord(1-z),
\eea
where we introduced the notation 
\be
 {\cal D}_n(z) = \frac{ln^n(1-z)}{1-z}\bigg|_{+},
\ee
and 
\be \label{AP-explict2L-zto1}
P^{(1)}_{qg}(z) =
C_A T_F \bigg[4 +\frac{2 \pi^2}{3} - 4 {\cal L}_2(z)\bigg] 
+ C_F T_F \bigg[10 - \frac{4 \pi^2}{3} + 4 {\cal L}_2(z) \bigg]
+ \ord(1-z).
\ee

\bibliography{NLP}

\providecommand{\href}[2]{#2}\begingroup\raggedright\begin{thebibliography}{10}

\bibitem{Sterman:1986aj}
G.~Sterman, \emph{{Summation of Large Corrections to Short Distance Hadronic
  Cross-Sections}}, {\emph{Nucl. Phys.} {\bfseries B281} (1987) 310}.

\bibitem{Catani:1989ne}
S.~Catani and L.~Trentadue, \emph{{Resummation of the QCD Perturbative Series
  for Hard Processes}},
  \href{https://doi.org/10.1016/0550-3213(89)90273-3}{\emph{Nucl. Phys.}
  {\bfseries B327} (1989) 323--352}.

\bibitem{Idilbi:2005ky}
A.~Idilbi and X.-d. Ji, \emph{{Threshold resummation for Drell-Yan process in
  soft-collinear effective theory}},
  \href{https://doi.org/10.1103/PhysRevD.72.054016}{\emph{Phys. Rev.}
  {\bfseries D72} (2005) 054016},
  [\href{https://arxiv.org/abs/hep-ph/0501006}{{\ttfamily hep-ph/0501006}}].

\bibitem{Idilbi:2006dg}
A.~Idilbi, X.-d. Ji and F.~Yuan, \emph{{Resummation of threshold logarithms in
  effective field theory for DIS, Drell-Yan and Higgs production}},
  \href{https://doi.org/10.1016/j.nuclphysb.2006.07.002}{\emph{Nucl. Phys.}
  {\bfseries B753} (2006) 42--68},
  [\href{https://arxiv.org/abs/hep-ph/0605068}{{\ttfamily hep-ph/0605068}}].

\bibitem{Becher:2006nr}
T.~Becher and M.~Neubert, \emph{{Threshold resummation in momentum space from
  effective field theory}},
  \href{https://doi.org/10.1103/PhysRevLett.97.082001}{\emph{Phys. Rev. Lett.}
  {\bfseries 97} (2006) 082001},
  [\href{https://arxiv.org/abs/hep-ph/0605050}{{\ttfamily hep-ph/0605050}}].

\bibitem{Moch:2005ky}
S.~Moch and A.~Vogt, \emph{{Higher-order soft corrections to lepton pair and
  Higgs boson production}},
  \href{https://doi.org/10.1016/j.physletb.2005.09.061}{\emph{Phys. Lett.}
  {\bfseries B631} (2005) 48--57},
  [\href{https://arxiv.org/abs/hep-ph/0508265}{{\ttfamily hep-ph/0508265}}].

\bibitem{Becher:2007ty}
T.~Becher, M.~Neubert and G.~Xu, \emph{{Dynamical Threshold Enhancement and
  Resummation in Drell- Yan Production}},
  \href{https://doi.org/10.1088/1126-6708/2008/07/030}{\emph{JHEP} {\bfseries
  07} (2008) 030}, [\href{https://arxiv.org/abs/0710.0680}{{\ttfamily
  0710.0680}}].

\bibitem{Catani:2014uta}
S.~Catani, L.~Cieri, D.~de~Florian, G.~Ferrera and M.~Grazzini,
  \emph{{Threshold resummation at N$^3$LL accuracy and soft-virtual cross
  sections at N$^3$LO}},
  \href{https://doi.org/10.1016/j.nuclphysb.2014.09.012}{\emph{Nucl. Phys. B}
  {\bfseries 888} (2014) 75--91},
  [\href{https://arxiv.org/abs/1405.4827}{{\ttfamily 1405.4827}}].

\bibitem{Ajjath:2020rci}
A.~H. Ajjath, G.~Das, M.~C. Kumar, P.~Mukherjee, V.~Ravindran and K.~Samanta,
  \emph{{Resummed Drell-Yan cross-section at N$^{3}$LL}},
  \href{https://doi.org/10.1007/JHEP10(2020)153}{\emph{JHEP} {\bfseries 10}
  (2020) 153}, [\href{https://arxiv.org/abs/2001.11377}{{\ttfamily
  2001.11377}}].

\bibitem{Low:1958sn}
F.~E. Low, \emph{{Bremsstrahlung of very low-energy quanta in elementary
  particle collisions}},
  \href{https://doi.org/10.1103/PhysRev.110.974}{\emph{Phys. Rev.} {\bfseries
  110} (1958) 974--977}.

\bibitem{Burnett:1967km}
T.~H. Burnett and N.~M. Kroll, \emph{{Extension of the low soft photon
  theorem}}, \href{https://doi.org/10.1103/PhysRevLett.20.86}{\emph{Phys. Rev.
  Lett.} {\bfseries 20} (1968) 86}.

\bibitem{DelDuca:1990gz}
V.~Del~Duca, \emph{{High-energy Bremsstrahlung Theorems for Soft Photons}},
  \href{https://doi.org/10.1016/0550-3213(90)90392-Q}{\emph{Nucl. Phys.}
  {\bfseries B345} (1990) 369--388}.

\bibitem{Bonocore:2014wua}
D.~Bonocore, E.~Laenen, L.~Magnea, L.~Vernazza and C.~D. White, \emph{{The
  method of regions and next-to-soft corrections in Drell-Yan production}},
  \href{https://doi.org/10.1016/j.physletb.2015.02.008}{\emph{Phys. Lett.}
  {\bfseries B742} (2015) 375--382},
  [\href{https://arxiv.org/abs/1410.6406}{{\ttfamily 1410.6406}}].

\bibitem{Bahjat-Abbas:2018hpv}
N.~Bahjat-Abbas, J.~Sinninghe~Damsté, L.~Vernazza and C.~D. White, \emph{{On
  next-to-leading power threshold corrections in Drell-Yan production at
  N$^3$LO}}, \href{https://doi.org/10.1007/JHEP10(2018)144}{\emph{JHEP}
  {\bfseries 10} (2018) 144},
  [\href{https://arxiv.org/abs/1807.09246}{{\ttfamily 1807.09246}}].

\bibitem{Laenen:2008ux}
E.~Laenen, L.~Magnea and G.~Stavenga, \emph{{On next-to-eikonal corrections to
  threshold resummation for the Drell-Yan and DIS cross sections}},
  \href{https://doi.org/10.1016/j.physletb.2008.09.037}{\emph{Phys. Lett.}
  {\bfseries B669} (2008) 173--179},
  [\href{https://arxiv.org/abs/0807.4412}{{\ttfamily 0807.4412}}].

\bibitem{Laenen:2008gt}
E.~Laenen, G.~Stavenga and C.~D. White, \emph{{Path integral approach to
  eikonal and next-to-eikonal exponentiation}},
  \href{https://doi.org/10.1088/1126-6708/2009/03/054}{\emph{JHEP} {\bfseries
  03} (2009) 054}, [\href{https://arxiv.org/abs/0811.2067}{{\ttfamily
  0811.2067}}].

\bibitem{Laenen:2010uz}
E.~Laenen, L.~Magnea, G.~Stavenga and C.~D. White, \emph{{Next-to-eikonal
  corrections to soft gluon radiation: a diagrammatic approach}},
  \href{https://doi.org/10.1007/JHEP01(2011)141}{\emph{JHEP} {\bfseries 01}
  (2011) 141}, [\href{https://arxiv.org/abs/1010.1860}{{\ttfamily 1010.1860}}].

\bibitem{Bonocore:2015esa}
D.~Bonocore, E.~Laenen, L.~Magnea, S.~Melville, L.~Vernazza and C.~D. White,
  \emph{{A factorization approach to next-to-leading-power threshold
  logarithms}}, \href{https://doi.org/10.1007/JHEP06(2015)008}{\emph{JHEP}
  {\bfseries 06} (2015) 008},
  [\href{https://arxiv.org/abs/1503.05156}{{\ttfamily 1503.05156}}].

\bibitem{Bonocore:2016awd}
D.~Bonocore, E.~Laenen, L.~Magnea, L.~Vernazza and C.~D. White,
  \emph{{Non-abelian factorisation for next-to-leading-power threshold
  logarithms}}, \href{https://doi.org/10.1007/JHEP12(2016)121}{\emph{JHEP}
  {\bfseries 12} (2016) 121},
  [\href{https://arxiv.org/abs/1610.06842}{{\ttfamily 1610.06842}}].

\bibitem{Beneke:2018gvs}
M.~Beneke, A.~Broggio, M.~Garny, S.~Jaskiewicz, R.~Szafron, L.~Vernazza et~al.,
  \emph{{Leading-logarithmic threshold resummation of the Drell-Yan process at
  next-to-leading power}},
  \href{https://doi.org/10.1007/JHEP03(2019)043}{\emph{JHEP} {\bfseries 03}
  (2019) 043}, [\href{https://arxiv.org/abs/1809.10631}{{\ttfamily
  1809.10631}}].

\bibitem{Beneke:2019oqx}
M.~Beneke, A.~Broggio, S.~Jaskiewicz and L.~Vernazza, \emph{{Threshold
  factorization of the Drell-Yan process at next-to-leading power}},
  \href{https://doi.org/10.1007/JHEP07(2020)078}{\emph{JHEP} {\bfseries 07}
  (2020) 078}, [\href{https://arxiv.org/abs/1912.01585}{{\ttfamily
  1912.01585}}].

\bibitem{Moult:2018jjd}
I.~Moult, I.~W. Stewart, G.~Vita and H.~X. Zhu, \emph{{First Subleading Power
  Resummation for Event Shapes}},
  \href{https://doi.org/10.1007/JHEP08(2018)013}{\emph{JHEP} {\bfseries 08}
  (2018) 013}, [\href{https://arxiv.org/abs/1804.04665}{{\ttfamily
  1804.04665}}].

\bibitem{Moult:2019mog}
I.~Moult, I.~W. Stewart and G.~Vita, \emph{{Subleading Power Factorization with
  Radiative Functions}},
  \href{https://doi.org/10.1007/JHEP11(2019)153}{\emph{JHEP} {\bfseries 11}
  (2019) 153}, [\href{https://arxiv.org/abs/1905.07411}{{\ttfamily
  1905.07411}}].

\bibitem{Bahjat-Abbas:2019fqa}
N.~Bahjat-Abbas, D.~Bonocore, J.~Sinninghe~Damsté, E.~Laenen, L.~Magnea,
  L.~Vernazza et~al., \emph{{Diagrammatic resummation of leading-logarithmic
  threshold effects at next-to-leading power}},
  \href{https://doi.org/10.1007/JHEP11(2019)002}{\emph{JHEP} {\bfseries 11}
  (2019) 002}, [\href{https://arxiv.org/abs/1905.13710}{{\ttfamily
  1905.13710}}].

\bibitem{Beneke:2019mua}
M.~Beneke, M.~Garny, S.~Jaskiewicz, R.~Szafron, L.~Vernazza and J.~Wang,
  \emph{{Leading-logarithmic threshold resummation of Higgs production in gluon
  fusion at next-to-leading power}},
  \href{https://doi.org/10.1007/JHEP01(2020)094}{\emph{JHEP} {\bfseries 01}
  (2020) 094}, [\href{https://arxiv.org/abs/1910.12685}{{\ttfamily
  1910.12685}}].

\bibitem{Moult:2019uhz}
I.~Moult, I.~W. Stewart, G.~Vita and H.~X. Zhu, \emph{{The Soft Quark
  Sudakov}}, \href{https://doi.org/10.1007/JHEP05(2020)089}{\emph{JHEP}
  {\bfseries 05} (2020) 089},
  [\href{https://arxiv.org/abs/1910.14038}{{\ttfamily 1910.14038}}].

\bibitem{Moult:2019vou}
I.~Moult, G.~Vita and K.~Yan, \emph{{Subleading power resummation of rapidity
  logarithms: the energy-energy correlator in $ \mathcal{N} $ = 4 SYM}},
  \href{https://doi.org/10.1007/JHEP07(2020)005}{\emph{JHEP} {\bfseries 07}
  (2020) 005}, [\href{https://arxiv.org/abs/1912.02188}{{\ttfamily
  1912.02188}}].

\bibitem{Ajjath:2020ulr}
A.~H. Ajjath, P.~Mukherjee and V.~Ravindran, \emph{{Next to soft corrections to
  Drell-Yan and Higgs boson productions}},
  \href{https://doi.org/10.1103/PhysRevD.105.094035}{\emph{Phys. Rev. D}
  {\bfseries 105} (2022) 094035},
  [\href{https://arxiv.org/abs/2006.06726}{{\ttfamily 2006.06726}}].

\bibitem{Beneke:2020ibj}
M.~Beneke, M.~Garny, S.~Jaskiewicz, R.~Szafron, L.~Vernazza and J.~Wang,
  \emph{{Large-x resummation of off-diagonal deep-inelastic parton scattering
  from d-dimensional refactorization}},
  \href{https://doi.org/10.1007/JHEP10(2020)196}{\emph{JHEP} {\bfseries 10}
  (2020) 196}, [\href{https://arxiv.org/abs/2008.04943}{{\ttfamily
  2008.04943}}].

\bibitem{Ajjath:2020sjk}
A.~H. Ajjath, P.~Mukherjee, V.~Ravindran, A.~Sankar and S.~Tiwari, \emph{{On
  next to soft threshold corrections to DIS and SIA processes}},
  \href{https://doi.org/10.1007/JHEP04(2021)131}{\emph{JHEP} {\bfseries 04}
  (2021) 131}, [\href{https://arxiv.org/abs/2007.12214}{{\ttfamily
  2007.12214}}].

\bibitem{vanBeekveld:2021mxn}
M.~van Beekveld, L.~Vernazza and C.~D. White, \emph{{Threshold resummation of
  new partonic channels at next-to-leading power}},
  \href{https://doi.org/10.1007/JHEP12(2021)087}{\emph{JHEP} {\bfseries 12}
  (2021) 087}, [\href{https://arxiv.org/abs/2109.09752}{{\ttfamily
  2109.09752}}].

\bibitem{Beneke:2022obx}
M.~Beneke, M.~Garny, S.~Jaskiewicz, J.~Strohm, R.~Szafron, L.~Vernazza et~al.,
  \emph{{Next-to-leading power endpoint factorization and resummation for
  off-diagonal \textquotedblleft{}gluon\textquotedblright{} thrust}},
  \href{https://doi.org/10.1007/JHEP07(2022)144}{\emph{JHEP} {\bfseries 07}
  (2022) 144}, [\href{https://arxiv.org/abs/2205.04479}{{\ttfamily
  2205.04479}}].

\bibitem{Boughezal:2016zws}
R.~Boughezal, X.~Liu and F.~Petriello, \emph{{Power Corrections in the
  N-jettiness Subtraction Scheme}},
  \href{https://doi.org/10.1007/JHEP03(2017)160}{\emph{JHEP} {\bfseries 03}
  (2017) 160}, [\href{https://arxiv.org/abs/1612.02911}{{\ttfamily
  1612.02911}}].

\bibitem{Moult:2016fqy}
I.~Moult, L.~Rothen, I.~W. Stewart, F.~J. Tackmann and H.~X. Zhu,
  \emph{{Subleading Power Corrections for N-Jettiness Subtractions}},
  \href{https://doi.org/10.1103/PhysRevD.95.074023}{\emph{Phys. Rev.}
  {\bfseries D95} (2017) 074023},
  [\href{https://arxiv.org/abs/1612.00450}{{\ttfamily 1612.00450}}].

\bibitem{Moult:2017jsg}
I.~Moult, L.~Rothen, I.~W. Stewart, F.~J. Tackmann and H.~X. Zhu, \emph{{N
  -jettiness subtractions for $gg\to H$ at subleading power}},
  \href{https://doi.org/10.1103/PhysRevD.97.014013}{\emph{Phys. Rev.}
  {\bfseries D97} (2018) 014013},
  [\href{https://arxiv.org/abs/1710.03227}{{\ttfamily 1710.03227}}].

\bibitem{Ebert:2018lzn}
M.~A. Ebert, I.~Moult, I.~W. Stewart, F.~J. Tackmann, G.~Vita and H.~X. Zhu,
  \emph{{Power Corrections for N-Jettiness Subtractions at ${\cal
  O}(\alpha_s)$}}, \href{https://doi.org/10.1007/JHEP12(2018)084}{\emph{JHEP}
  {\bfseries 12} (2018) 084},
  [\href{https://arxiv.org/abs/1807.10764}{{\ttfamily 1807.10764}}].

\bibitem{Boughezal:2018mvf}
R.~Boughezal, A.~Isgr\'o and F.~Petriello, \emph{{Next-to-leading-logarithmic
  power corrections for $N$-jettiness subtraction in color-singlet
  production}}, \href{https://doi.org/10.1103/PhysRevD.97.076006}{\emph{Phys.
  Rev.} {\bfseries D97} (2018) 076006},
  [\href{https://arxiv.org/abs/1802.00456}{{\ttfamily 1802.00456}}].

\bibitem{Boughezal:2019ggi}
R.~Boughezal, A.~Isgr\'o and F.~Petriello, \emph{{Next-to-leading power
  corrections to $V+1$ jet production in $N$-jettiness subtraction}},
  \href{https://doi.org/10.1103/PhysRevD.101.016005}{\emph{Phys. Rev.}
  {\bfseries D101} (2020) 016005},
  [\href{https://arxiv.org/abs/1907.12213}{{\ttfamily 1907.12213}}].

\bibitem{Ebert:2018gsn}
M.~A. Ebert, I.~Moult, I.~W. Stewart, F.~J. Tackmann, G.~Vita and H.~X. Zhu,
  \emph{{Subleading power rapidity divergences and power corrections for
  q$_{T}$}}, \href{https://doi.org/10.1007/JHEP04(2019)123}{\emph{JHEP}
  {\bfseries 04} (2019) 123},
  [\href{https://arxiv.org/abs/1812.08189}{{\ttfamily 1812.08189}}].

\bibitem{Cieri:2019tfv}
L.~Cieri, C.~Oleari and M.~Rocco, \emph{{Higher-order power corrections in a
  transverse-momentum cut for colour-singlet production at NLO}},
  \href{https://doi.org/10.1140/epjc/s10052-019-7361-8}{\emph{Eur. Phys. J.}
  {\bfseries C79} (2019) 852},
  [\href{https://arxiv.org/abs/1906.09044}{{\ttfamily 1906.09044}}].

\bibitem{Oleari:2020wvt}
C.~Oleari and M.~Rocco, \emph{{Power corrections in a transverse-momentum cut
  for vector-boson production at NNLO: the $qg$-initiated real-virtual
  contribution}},
  \href{https://doi.org/10.1140/epjc/s10052-021-08878-3}{\emph{Eur. Phys. J. C}
  {\bfseries 81} (2021) 183},
  [\href{https://arxiv.org/abs/2012.10538}{{\ttfamily 2012.10538}}].

\bibitem{Liu:2019oav}
Z.~L. Liu and M.~Neubert, \emph{{Factorization at subleading power and
  endpoint-divergent convolutions in $h\to\gamma\gamma$ decay}},
  \href{https://doi.org/10.1007/JHEP04(2020)033}{\emph{JHEP} {\bfseries 04}
  (2020) 033}, [\href{https://arxiv.org/abs/1912.08818}{{\ttfamily
  1912.08818}}].

\bibitem{Liu:2020tzd}
Z.~L. Liu, B.~Mecaj, M.~Neubert and X.~Wang, \emph{{Factorization at subleading
  power, Sudakov resummation, and endpoint divergences in soft-collinear
  effective theory}},
  \href{https://doi.org/10.1103/PhysRevD.104.014004}{\emph{Phys. Rev. D}
  {\bfseries 104} (2021) 014004},
  [\href{https://arxiv.org/abs/2009.04456}{{\ttfamily 2009.04456}}].

\bibitem{Liu:2020wbn}
Z.~L. Liu, B.~Mecaj, M.~Neubert and X.~Wang, \emph{{Factorization at subleading
  power and endpoint divergences in $h\to\gamma\gamma$ decay. Part II.
  Renormalization and scale evolution}},
  \href{https://doi.org/10.1007/JHEP01(2021)077}{\emph{JHEP} {\bfseries 01}
  (2021) 077}, [\href{https://arxiv.org/abs/2009.06779}{{\ttfamily
  2009.06779}}].

\bibitem{Liu:2022ajh}
Z.~L. Liu, M.~Neubert, M.~Schnubel and X.~Wang, \emph{{Factorization at
  next-to-leading power and endpoint divergences in gg \textrightarrow{} h
  production}}, \href{https://doi.org/10.1007/JHEP06(2023)183}{\emph{JHEP}
  {\bfseries 06} (2023) 183},
  [\href{https://arxiv.org/abs/2212.10447}{{\ttfamily 2212.10447}}].

\bibitem{Bell:2022ott}
G.~Bell, P.~B\"oer and T.~Feldmann, \emph{{Muon-electron backward scattering: a
  prime example for endpoint singularities in SCET}},
  \href{https://doi.org/10.1007/JHEP09(2022)183}{\emph{JHEP} {\bfseries 09}
  (2022) 183}, [\href{https://arxiv.org/abs/2205.06021}{{\ttfamily
  2205.06021}}].

\bibitem{Feldmann:2022ixt}
T.~Feldmann, N.~Gubernari, T.~Huber and N.~Seitz, \emph{{Contribution of the
  electromagnetic dipole operator ${\cal O}_7$ to the $\bar B_s \to \mu^+\mu^-$
  decay amplitude}},
  \href{https://doi.org/10.1103/PhysRevD.107.013007}{\emph{Phys. Rev. D}
  {\bfseries 107} (2023) 013007},
  [\href{https://arxiv.org/abs/2211.04209}{{\ttfamily 2211.04209}}].

\bibitem{Cornella:2022ubo}
C.~Cornella, M.~K\"onig and M.~Neubert, \emph{{Structure-dependent QED effects
  in exclusive B decays at subleading power}},
  \href{https://doi.org/10.1103/PhysRevD.108.L031502}{\emph{Phys. Rev. D}
  {\bfseries 108} (2023) L031502},
  [\href{https://arxiv.org/abs/2212.14430}{{\ttfamily 2212.14430}}].

\bibitem{Hurth:2023paz}
T.~Hurth and R.~Szafron, \emph{{Refactorisation in subleading $\bar B \to X_s
  \gamma$}}, \href{https://doi.org/10.1016/j.nuclphysb.2023.116200}{\emph{Nucl.
  Phys. B} {\bfseries 991} (2023) 116200},
  [\href{https://arxiv.org/abs/2301.01739}{{\ttfamily 2301.01739}}].

\bibitem{Beneke:2020ibjB}
M.~Beneke, M.~Garny, S.~Jaskiewicz, R.~Szafron, L.~Vernazza and J.~Wang,
  \emph{{Unpublished, 2020}}, .

\bibitem{Broggio:2021fnr}
A.~Broggio, S.~Jaskiewicz and L.~Vernazza, \emph{{Next-to-leading power
  two-loop soft functions for the Drell-Yan process at threshold}},
  \href{https://doi.org/10.1007/JHEP10(2021)061}{\emph{JHEP} {\bfseries 10}
  (2021) 061}, [\href{https://arxiv.org/abs/2107.07353}{{\ttfamily
  2107.07353}}].

\bibitem{Liu:2021mac}
Z.~L. Liu, M.~Neubert, M.~Schnubel and X.~Wang, \emph{{Radiative quark jet
  function with an external gluon}},
  \href{https://doi.org/10.1007/JHEP02(2022)075}{\emph{JHEP} {\bfseries 02}
  (2022) 075}, [\href{https://arxiv.org/abs/2112.00018}{{\ttfamily
  2112.00018}}].

\bibitem{Liu:2020eqe}
Z.~L. Liu, B.~Mecaj, M.~Neubert, X.~Wang and S.~Fleming, \emph{{Renormalization
  and Scale Evolution of the Soft-Quark Soft Function}},
  \href{https://doi.org/10.1007/JHEP07(2020)104}{\emph{JHEP} {\bfseries 07}
  (2020) 104}, [\href{https://arxiv.org/abs/2005.03013}{{\ttfamily
  2005.03013}}].

\bibitem{Bodwin:2021cpx}
G.~T. Bodwin, J.-H. Ee, J.~Lee and X.-P. Wang, \emph{{Analyticity,
  renormalization, and evolution of the soft-quark function}},
  \href{https://doi.org/10.1103/PhysRevD.104.016010}{\emph{Phys. Rev. D}
  {\bfseries 104} (2021) 016010},
  [\href{https://arxiv.org/abs/2101.04872}{{\ttfamily 2101.04872}}].

\bibitem{Beneke:2002ph}
M.~Beneke, A.~P. Chapovsky, M.~Diehl and T.~Feldmann, \emph{{Soft collinear
  effective theory and heavy to light currents beyond leading power}},
  \href{https://doi.org/10.1016/S0550-3213(02)00687-9}{\emph{Nucl. Phys.}
  {\bfseries B643} (2002) 431--476},
  [\href{https://arxiv.org/abs/hep-ph/0206152}{{\ttfamily hep-ph/0206152}}].

\bibitem{Beneke:2002ni}
M.~Beneke and T.~Feldmann, \emph{{Multipole expanded soft collinear effective
  theory with non-abelian gauge symmetry}},
  \href{https://doi.org/10.1016/S0370-2693(02)03204-5}{\emph{Phys. Lett.}
  {\bfseries B553} (2003) 267--276},
  [\href{https://arxiv.org/abs/hep-ph/0211358}{{\ttfamily hep-ph/0211358}}].

\bibitem{Jaskiewicz:2021cfw}
S.~E. Jaskiewicz, \emph{{Factorization and Resummation at Subleading Powers}},
  Ph.D. thesis, Technical University of Munich, May, 2021.

\bibitem{Beneke:2017ztn}
M.~Beneke, M.~Garny, R.~Szafron and J.~Wang, \emph{{Anomalous dimension of
  subleading-power N-jet operators}},
  \href{https://doi.org/10.1007/JHEP03(2018)001}{\emph{JHEP} {\bfseries 03}
  (2018) 001}, [\href{https://arxiv.org/abs/1712.04416}{{\ttfamily
  1712.04416}}].

\bibitem{Beneke:2018rbh}
M.~Beneke, M.~Garny, R.~Szafron and J.~Wang, \emph{{Anomalous dimension of
  subleading-power $N$-jet operators. Part II}},
  \href{https://doi.org/10.1007/JHEP11(2018)112}{\emph{JHEP} {\bfseries 11}
  (2018) 112}, [\href{https://arxiv.org/abs/1808.04742}{{\ttfamily
  1808.04742}}].

\bibitem{Beneke:2019kgv}
M.~Beneke, M.~Garny, R.~Szafron and J.~Wang, \emph{{Violation of the
  Kluberg-Stern-Zuber theorem in SCET}},
  \href{https://doi.org/10.1007/JHEP09(2019)101}{\emph{JHEP} {\bfseries 09}
  (2019) 101}, [\href{https://arxiv.org/abs/1907.05463}{{\ttfamily
  1907.05463}}].

\bibitem{Marcantonini:2008qn}
C.~Marcantonini and I.~W. Stewart, \emph{{Reparameterization Invariant
  Collinear Operators}},
  \href{https://doi.org/10.1103/PhysRevD.79.065028}{\emph{Phys. Rev.}
  {\bfseries D79} (2009) 065028},
  [\href{https://arxiv.org/abs/0809.1093}{{\ttfamily 0809.1093}}].

\bibitem{Kolodrubetz:2016uim}
D.~W. Kolodrubetz, I.~Moult and I.~W. Stewart, \emph{{Building Blocks for
  Subleading Helicity Operators}},
  \href{https://doi.org/10.1007/JHEP05(2016)139}{\emph{JHEP} {\bfseries 05}
  (2016) 139}, [\href{https://arxiv.org/abs/1601.02607}{{\ttfamily
  1601.02607}}].

\bibitem{Feige:2017zci}
I.~Feige, D.~W. Kolodrubetz, I.~Moult and I.~W. Stewart, \emph{{A Complete
  Basis of Helicity Operators for Subleading Factorization}},
  \href{https://doi.org/10.1007/JHEP11(2017)142}{\emph{JHEP} {\bfseries 11}
  (2017) 142}, [\href{https://arxiv.org/abs/1703.03411}{{\ttfamily
  1703.03411}}].

\bibitem{Moult:2017rpl}
I.~Moult, I.~W. Stewart and G.~Vita, \emph{{A subleading operator basis and
  matching for gg $\to$ H}},
  \href{https://doi.org/10.1007/JHEP07(2017)067}{\emph{JHEP} {\bfseries 07}
  (2017) 067}, [\href{https://arxiv.org/abs/1703.03408}{{\ttfamily
  1703.03408}}].

\bibitem{Beneke:2010da}
M.~Beneke, P.~Falgari and C.~Schwinn, \emph{{Threshold resummation for pair
  production of coloured heavy (s)particles at hadron colliders}},
  \href{https://doi.org/10.1016/j.nuclphysb.2010.09.009}{\emph{Nucl. Phys.}
  {\bfseries B842} (2011) 414--474},
  [\href{https://arxiv.org/abs/1007.5414}{{\ttfamily 1007.5414}}].

\bibitem{Vogt:2010cv}
A.~Vogt, \emph{{Leading logarithmic large-x resummation of off-diagonal
  splitting functions and coefficient functions}},
  \href{https://doi.org/10.1016/j.physletb.2010.06.010}{\emph{Phys. Lett. B}
  {\bfseries 691} (2010) 77--81},
  [\href{https://arxiv.org/abs/1005.1606}{{\ttfamily 1005.1606}}].

\bibitem{Liu:2017vkm}
T.~Liu and A.~A. Penin, \emph{{High-Energy Limit of QCD beyond the Sudakov
  Approximation}},
  \href{https://doi.org/10.1103/PhysRevLett.119.262001}{\emph{Phys. Rev. Lett.}
  {\bfseries 119} (2017) 262001},
  [\href{https://arxiv.org/abs/1709.01092}{{\ttfamily 1709.01092}}].

\bibitem{Liu:2018czl}
T.~Liu and A.~Penin, \emph{{High-Energy Limit of Mass-Suppressed Amplitudes in
  Gauge Theories}}, \href{https://doi.org/10.1007/JHEP11(2018)158}{\emph{JHEP}
  {\bfseries 11} (2018) 158},
  [\href{https://arxiv.org/abs/1809.04950}{{\ttfamily 1809.04950}}].

\bibitem{Beneke:1997zp}
M.~Beneke and V.~A. Smirnov, \emph{{Asymptotic expansion of Feynman integrals
  near threshold}},
  \href{https://doi.org/10.1016/S0550-3213(98)00138-2}{\emph{Nucl. Phys.}
  {\bfseries B522} (1998) 321--344},
  [\href{https://arxiv.org/abs/hep-ph/9711391}{{\ttfamily hep-ph/9711391}}].

\bibitem{Hamberg:1990np}
R.~Hamberg, W.~L. van Neerven and T.~Matsuura, \emph{{A complete calculation of
  the order $\alpha_s^{2}$ correction to the Drell-Yan $K$ factor}},
  \href{https://doi.org/10.1016/0550-3213(91)90064-5}{\emph{Nucl. Phys.}
  {\bfseries B359} (1991) 343--405}.

\bibitem{Gehrmann:2010ue}
T.~Gehrmann, E.~W.~N. Glover, T.~Huber, N.~Ikizlerli and C.~Studerus,
  \emph{{Calculation of the quark and gluon form factors to three loops in
  QCD}}, \href{https://doi.org/10.1007/JHEP06(2010)094}{\emph{JHEP} {\bfseries
  06} (2010) 094}, [\href{https://arxiv.org/abs/1004.3653}{{\ttfamily
  1004.3653}}].

\bibitem{Binosi:2008ig}
D.~Binosi, J.~Collins, C.~Kaufhold and L.~Theussl, \emph{{JaxoDraw: A Graphical
  user interface for drawing Feynman diagrams. Version 2.0 release notes}},
  \href{https://doi.org/10.1016/j.cpc.2009.02.020}{\emph{Comput. Phys. Commun.}
  {\bfseries 180} (2009) 1709--1715},
  [\href{https://arxiv.org/abs/0811.4113}{{\ttfamily 0811.4113}}].

\bibitem{Vermaseren:2000nd}
J.~A.~M. Vermaseren, \emph{{New features of FORM}},
  \href{https://arxiv.org/abs/math-ph/0010025}{{\ttfamily math-ph/0010025}}.

\bibitem{Huber:2005yg}
T.~Huber and D.~Maitre, \emph{{HypExp: A Mathematica package for expanding
  hypergeometric functions around integer-valued parameters}},
  \href{https://doi.org/10.1016/j.cpc.2006.01.007}{\emph{Comput. Phys. Commun.}
  {\bfseries 175} (2006) 122--144},
  [\href{https://arxiv.org/abs/hep-ph/0507094}{{\ttfamily hep-ph/0507094}}].

\end{thebibliography}\endgroup

\end{appendix}

\end{document}